\RequirePackage{fix-cm}

\documentclass[smallextended]{svjour3}
\smartqed  

\usepackage{graphicx}
\usepackage{hyperref}
\usepackage{amsmath}
\usepackage{cite}
\usepackage{bm}
\usepackage{amssymb}
\usepackage{comment}
\usepackage{color}
\usepackage{todonotes}

\DeclareMathOperator{\csch}{cosech}

\numberwithin{equation}{section}

\allowdisplaybreaks

\begin{document}

\title{Nonminimal coupling, quantum scalar field stress-energy tensor and energy conditions on global anti-de Sitter space-time}
\titlerunning{Nonminimal coupling, RSET and energy conditions on adS}

\author{ Sivakumar Namasivayam \and 
        Elizabeth Winstanley 
}

\institute{School of Mathematical and Physical Sciences, The University of Sheffield, \\ Hicks Building, Hounsfield Road, Sheffield. S3 7RH United Kingdom
       \\       \email{SNamasivayam1@sheffield.ac.uk, E.Winstanley@sheffield.ac.uk}           
}
\date{\today}

\maketitle

\abstract{
We compute the renormalized expectation value of the stress-energy tensor operator for a quantum scalar field propagating on three-dimensional global anti-de Sitter space-time.
The scalar field has general mass and nonminimal coupling to the Ricci scalar curvature, and is subject to Dirichlet, Neumann or Robin boundary conditions at the space-time boundary.
We consider both vacuum and thermal states, and explore whether the weak and null energy conditions are satisfied by the quantum stress-energy tensor. 
We uncover a rather complicated picture: compliance with these two energy conditions depends strongly on the mass and coupling of the scalar field to the curvature, and the boundary conditions applied.
}

\maketitle

\section{Introduction}
\label{sec:intro}

In classical general relativity, the Einstein equations relate the curvature of the space-time geometry to the matter content of the space-time:
\begin{equation}
     R_{\alpha \gamma }-\frac{1}{2} R g_{\alpha \gamma } + \Lambda  g_{\alpha \gamma } = 8\pi T_{\alpha \gamma },
     \label{eq:classEE}
\end{equation}
where $R_{\alpha \gamma }$ is the Ricci tensor, $R$ is the Ricci scalar, $\Lambda $ is the cosmological constant,
$g_{\alpha \gamma}$  is the metric tensor, $T_{\alpha \gamma }$ is the stress-energy tensor (SET) and we are using units in which $G = c= \hbar = k_{{\rm {B}}}=1$. 
In principle, given any SET, one could then solve the Einstein equations (\ref{eq:classEE}) to find possible space-time geometries sourced by that SET. 
Similarly, given any space-time, one can use (\ref{eq:classEE}) to give the corresponding matter SET.
In practice, the form of the SET $T_{\alpha \gamma}$ is constrained by physical principles. 
For example, in order to satisfy (\ref{eq:classEE}), it must be the case that the SET is conserved: 
$\nabla _{\alpha  }T^{\alpha \gamma }=0$, where $\nabla _{\alpha }$ is the usual covariant derivative. 
Energy conditions (see \cite{Kontou:2020bta,Curiel:2014zba,Martin-Moruno:2017exc} for reviews) play an important role in constraining the form of the matter SET. 
The idea is that energy conditions encode the general behaviour of SETs for ``physically reasonable matter'', and that by insisting that the matter content of Einstein's equations (\ref{eq:classEE}) satisfies these energy conditions, one can rule out ``unphysical'' space-times such as wormholes or those with closed time-like curves.

Of the pointwise energy conditions considered in the literature, in this paper we focus on just two: the weak energy condition (WEC) and null energy condition (NEC). 
The WEC stipulates that for any future-directed time-like vector $V^\alpha $, we have $T_{\alpha \gamma} V^{\alpha } V^{\gamma } \ge 0$.
Put simply, this requires that for all observers, the energy density measured by an observer is never negative.
The NEC imposes a similar condition on the SET, but for null vectors; in particular, the NEC requires that for a null vector $k^\alpha $, we have $T_{\alpha \gamma} k^{\alpha } k^{\gamma } \ge 0$. 
If we assume that the SET takes the perfect fluid form
\begin{equation}
    T^{\alpha \gamma } = \left(  E + P \right) U^{\alpha } U^{\gamma } + P g^{\alpha \gamma },
    \label{eq:fluid}
\end{equation}
where $E$ is the energy density, $P$ the pressure density and $U^{\alpha }$ is the four-velocity of the fluid, then the WEC is equivalent to $E\ge 0$ and $E+P\ge 0$, while the NEC is equivalent to $E+P\ge 0$.
Thus the WEC implies the NEC, but not the converse; the NEC is a weaker condition than the WEC.

Let us consider the mathematically simplest type of classical matter field, a scalar field $\Phi $ having mass $m$ and satisfying the wave equation
\begin{equation}
    \left[ \nabla ^{\alpha }\nabla _{\alpha } - \left( m^{2}+\xi R \right) \right] \Phi = 0,
    \label{eq:KG}
\end{equation}
where $\xi $ is a coupling constant. 
When $\xi = 0$, the scalar field is minimally coupled to the space-time geometry, and both the WEC and NEC are satisfied \cite{Kontou:2020bta}.
On the other hand, when $\xi \neq 0 $ and the field is nonminimally coupled to the Ricci curvature scalar, 
both the WEC and NEC can be violated by classical configurations satisfying the Einstein equations (see, for example, \cite{Kontou:2020bta,Flanagan:1996gw,Barcelo:1999hq,Barcelo:2000zf}), where the SET of the classical scalar field is given by \cite{Decanini:2005eg}
\begin{multline}
    T_{\alpha \gamma } = \left( 1 - 2\xi  \right) \left( \nabla _{\alpha } \Phi \right)  \left( \nabla _{\gamma } \Phi \right) - \frac{1}{2}\left( 1 - 4\xi \right)  g_{\alpha \gamma }
    \left[ \left( \nabla _{\lambda }\Phi \right)  \left( \nabla ^{\lambda }\Phi \right) + \left( m^{2} + \xi R \right) \Phi ^{2}\right]
    \\
     - 2\xi \left[ \Phi \nabla _{\alpha }\nabla _{\gamma } \Phi + \frac{1}{2}R_{\alpha \gamma }\Phi ^{2} \right] . 
     \label{eq:SET}
\end{multline}
It is possible, via a conformal transformation of the metric and a field redefinition
\begin{equation}
    g_{\alpha \gamma } \rightarrow \Omega (\Phi) ^{2} g_{\alpha \gamma } , \qquad \Phi \rightarrow F(\Phi ) ,
\end{equation}
where $\Omega (\Phi )$ is the conformal factor (which depends on the scalar field) and $F(\Phi )$ is a function of the scalar field, 
to transform the classical theory with a nonminimally-coupled scalar field (in what is termed the Jordan frame) to an alternative frame (the Einstein frame) in which the transformed scalar field is minimally coupled to the transformed space-time scalar curvature, but has a nonzero self-interaction potential.
While the NEC (and hence also the WEC) can be violated in the Jordan frame, the NEC is satisfied in the Einstein frame (if the self-interaction potential is positive then the WEC is also satisfied in the Einstein frame).
One could therefore argue that it is the Einstein frame which is the physically relevant one (see, for example, \cite{Magnano:1993bd,Faraoni:1998qx,Fliss:2023rzi} for discussion of this issue and further references on the subject). 

If we consider a quantum rather than classical scalar field, the classical SET on the right-hand-side of the Einstein equations (\ref{eq:classEE}) is replaced by a (renormalized) expectation value of the quantum SET operator:
\begin{equation}
    R_{\alpha \gamma }-\frac{1}{2} R g_{\alpha \gamma } + \Lambda  g_{\alpha \gamma } 
    = 8 \pi  \langle {\hat {T}}_{\alpha \gamma } \rangle .
    \label{eq:SCEE}
\end{equation}
Within the framework of quantum field theory in curved space-time, the RSET (renormalized SET) $\langle {\hat {T}}_{\alpha \gamma } \rangle $ is computed for a suitable state of the quantum field on a fixed, background, classical space-time. 
The semiclassical Einstein equations (\ref{eq:SCEE}) then govern the backreaction of the quantum field on the space-time geometry.
If we are interested in a free quantum scalar field which is nonminimally coupled to the background geometry (that is, in the Jordan frame), we cannot simply make a conformal transformation to the Einstein frame, because we have already 
fixed the space-time background before computing the RSET.
The question of whether the RSET for a nonminimally-coupled quantum scalar field satisfies either the WEC or NEC must therefore be addressed directly in the Jordan frame. 

It is well-known that the RSET for a minimally-coupled quantum scalar field does not always satisfy the NEC, even on Minkowski space-time \cite{Epstein:1965zza}. 
The extent to which the NEC is violated (if at all) depends on the quantum state of the field. 
For example, in Minkowski space-time, a free quantum scalar field (whether minimally or nonminimally coupled) in the global vacuum state has vanishing RSET $\langle {\hat {T}}_{\alpha \gamma } \rangle =0$ and hence the NEC (and also the WEC) is trivially satisfied. 
The WEC (and hence also the NEC) is also satisfied when the scalar field is in a global thermal state in Minkowski space-time.
Our purpose in this paper is to explore whether this remains the case for vacuum and thermal states on global anti-de Sitter (adS) space-time, for both minimally and nonminimally-coupled scalar fields.

While adS space-time is maximally symmetric, which simplifies computations on this background, the properties of the RSET for a quantum scalar field are nonetheless rather nontrivial. 
As well as having constant negative curvature, adS space-time possesses a timelike boundary, on which boundary conditions must be imposed in order to have a well-defined quantum field theory
\cite{Avis:1977yn,Wald:1980jn,Ishibashi:2003jd,Ishibashi:2004wx,Dappiaggi:2016fwc,Benini:2017dfw,Dappiaggi:2017wvj,Dappiaggi:2018pju,Dappiaggi:2018xvw,Gannot:2018jkg,Barroso:2019cwp,Morley:2020ayr,Dappiaggi:2021wtr,Namasivayam:2022bky,Campos:2022byi,Morley:2023exv}. 
As a result, expectation values of quantum operators possess rich properties, depending on the mass and coupling of the scalar field, as well as the boundary conditions applied.
While the vacuum expectation value (v.e.v.) of the RSET when the field is subject to either Dirichlet or Neumann boundary conditions retains the maximal symmetry of the underlying adS space-time \cite{Allen:1985wd,Allen:1986ty,Kent:2014nya} (and thus trivially satisfies the NEC, but not necessarily the WEC), considering either Robin boundary conditions and/or nonzero temperature breaks the maximal symmetry \cite{Allen:1986ty,Ambrus:2018olh,Barroso:2019cwp,Pitelli:2019svx,Morley:2020ayr,Namasivayam:2022bky,Morley:2023exv}. 
Our focus in this paper is studying whether the NEC/WEC are satisfied by the RSET when the quantum scalar field is in a nonmaximally-symmetric state.
While the RSET for a massless, conformally-coupled, scalar field in a thermal state on four-dimensional adS was computed many years ago for Dirichlet and Neumann boundary conditions \cite{Allen:1986ty}, and more recently also for Robin boundary conditions \cite{Morley:2023exv}, the corresponding computation for other masses and couplings to the curvature has yet to be completed.
Following \cite{Namasivayam:2022bky}, in this paper we address this calculation in the simpler setting of three space-time dimensions.

The outline of the paper is as follows.
We begin, in section~\ref{sec:DN}, by considering the effect of the scalar field mass and coupling on the RSET (and thereby the energy conditions) for thermal states with Dirichlet and Neumann boundary conditions. 
We then examine the consequences of generalizing to Robin boundary conditions in section~\ref{sec:Robin}, considering both v.e.v.s and thermal expectation values (t.e.v.s) of the RSET.
Our conclusions can be found in section~\ref{sec:conc}.

\section{RSET with Dirichlet and Neumann boundary conditions}
\label{sec:DN}

We work in three-dimensional anti-de Sitter space-time (adS${}_{3}$) whose metric, in global coordinates, is
\begin{equation}
	ds^{2}=L^{2} \sec ^{2} \rho \, \left[ -dt^{2}+d\rho ^{2} + \sin ^{2} \rho \, d\theta ^{2} \right] ,
	\label{eq:metric}
\end{equation}
where $0\le \rho <\pi/2, \, 0 \le \theta < 2\pi$ and $ -\pi \le t \le \pi$ with $ -\pi$ and $\pi$ identified.  
The time coordinate $t$ is periodic with period $2\pi$, leading to closed time-like curves, but this can be circumvented by considering the covering space of adS, where the time coordinate is ``unwrapped'' to give $-\infty < t < \infty$.  Henceforth we work on this covering space. 
AdS${}_{3}$ is a maximally symmetric space-time with a constant negative curvature where the cosmological constant, $\Lambda$, is related to the inverse radius of curvature, $L$, by $ \Lambda = -1/L^2$.

\begin{figure}
    \centering
    \includegraphics[width=0.8\linewidth]{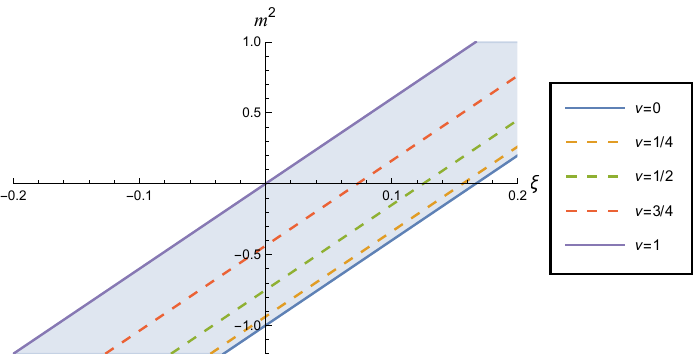}
    \caption{$(\xi, m^{2})$-plane of the parameters in the scalar field equation (\ref{eq:KG}). 
    Lines of constant $\nu $ (\ref{eq:nu}) are diagonal lines. 
    The shaded region corresponds to $0<\nu <1$, in which there is a choice of boundary conditions that can be applied to the scalar field.  
    The dashed lines correspond to three particular values of $\nu $ which will form the focus of our study.}
    \label{fig:nu}
\end{figure}

We study a real, free scalar field, $\Phi$,  with general mass, $m$, and coupling to the background curvature, satisfying the wave equation (\ref{eq:KG}), where $R=-6/L^{2}$ is the Ricci scalar curvature. 
As in~\cite{Namasivayam:2022bky}, to simplify the notation we define the parameter $\nu $ by  
\begin{equation}
	\nu  = {\sqrt {1+  \left( m^{2} + \xi R\right) \,L^{2}}}. 
	\label{eq:nu}
\end{equation}
In three space-time dimensions, the scalar field is conformally coupled if $\xi = 1/8$, and thus a massless, conformally coupled scalar field has $\nu  = 1/2$.
The scalar field equation (\ref{eq:KG}) depends on the mass $m$ and coupling $\xi $ only through the combination $\nu $, however the SET (\ref{eq:SET}) depends on $m$ and $\xi $ separately.
There is thus a degeneracy in this set-up, with different values of $m$ and $\xi $ yielding the same scalar field equation (\ref{eq:KG}) but different SETs (\ref{eq:SET}). 
To make this explicit, in Fig.~\ref{fig:nu} we show the $(\xi , m^{2})$-plane which parameterizes the scalar field theory. 
Lines of constant $\nu $ (\ref{eq:nu}) are diagonal lines in this plane. 
In this paper we focus on values of $\nu $ lying in the interval $(0,1)$, since in this interval we have a choice of boundary conditions which can be applied to the scalar field \cite{Ishibashi:2004wx}.
In section~\ref{sec:Robin} we will consider Robin boundary conditions, but we first study the RSET when Dirichlet or Neumann boundary conditions are applied to the field.

\subsection{Vacuum RSET with Dirichlet and Neumann boundary conditions}
\label{sec:vacDN}

The vacuum Green functions $G_{0}^{{{D}}/{{N}}}(s)$ with Dirichlet ($D$) and Neumann ($N$) boundary conditions applied are~\cite{Namasivayam:2022bky}
\begin{subequations}
\label{eq:G0DN}
\begin{align}
   G_0^{{{D}}} (s) = & ~ - \frac{i \nu }{4 \pi L}  {}_2F_1\left (1+ \nu,1-\nu,\frac{3}{2};-\sinh^2\left[ \frac{s}{2L}\right]\right)
 \nonumber \\ &  \qquad
 + \frac{i}{8 \pi L\, \sinh\left(\frac{s}{2L}\right)} {}_2F_1 \left(\frac{1}{2}+\nu,\frac{1}{2}-\nu,\frac{1}{2};-\sinh^2\left[ \frac{s}{2L}\right]\right),
\label{eq:GFoD} 
\\ 
G_0^{{{N}}} (s) = & ~  \frac{i \nu }{4 \pi L}  {}_2F_1\left (1+ \nu,1-\nu,\frac{3}{2};-\sinh^2\left[ \frac{s}{2L}\right]\right) 
\nonumber \\ & \qquad 
+ \frac{i}{8 \pi L\, \sinh\left(\frac{s}{2L}\right)} {}_2F_1 \left(\frac{1}{2}+\nu,\frac{1}{2}-\nu,\frac{1}{2};-\sinh^2\left[ \frac{s}{2L}\right]\right),
\label{eq:GFoN}
\end{align}
\end{subequations}
where ${}_2F_1 (a,b,c;z)$ are hypergeometric functions and  $s(x,x')$ is the proper distance between the points $x,x'$, given by~\cite{Avis:1977yn}
\begin{equation}
\cosh \left( \frac{s (x,x')}{L}\right) = \frac{\cos \Delta t}{\cos \rho \cos \rho'}-  \tan \rho \tan \rho' \cos \Delta \theta,
\label{eq:cosh_s}
\end{equation}
with $ \Delta t = t'-t$ and  $\Delta \theta = \theta'-\theta$.  
The only difference between the Dirichlet and Neumann vacuum Green functions is the sign of the first term.

To find the RSET, we first subtract from the Green functions (\ref{eq:G0DN}) the singular part $G_{\text{sing}}(x,x')$ of the Hadamard parametrix, leaving the state-dependent biscalar $W(x,x')$:
\begin{equation}
    W(x,x')=-i\left[ G(x,x')-G_{\text{sing}}(x,x')\right] .
    \label{eq:Whad}
\end{equation}
In three space-time dimensions, the RSET is then given, in terms of the coincidence limits of $W(x,x')$ and its derivatives, by \cite{Decanini:2005eg}
\begin{equation}
    \langle \psi |\hat{T}_{\alpha \gamma }| \psi \rangle= -w_{\alpha \gamma } 
    +\frac{1}{2}\left(1-2\xi \right)w_{;\alpha\gamma} +\frac{1}{2}\left(2\xi-\frac{1}{2} \right) g_{\alpha \gamma} \nabla _{\lambda }\nabla ^{\lambda } w +\xi R_{\alpha \gamma }\,w ,
    \label{eq:diff_operator3}
\end{equation}
where
\begin{equation}
    w(x) = \lim_{x' \to x} W(x,x'), \qquad 
    w_{\alpha \gamma }(x) = \lim_{x' \to x} W(x,x')_{;\alpha \gamma }.
    \label{eq:walphagamma}
\end{equation}
The resulting v.e.v.s of the RSET when Dirichlet or Neumann boundary conditions are applied are then \cite{Kent:2014nya}
\begin{subequations}
\label{eq:SETvacDN}
\begin{align}
\langle {\hat {T}}_{\alpha \gamma} \rangle _{0}^{{D}} & = \frac{1}{12 \pi L^3} \left[ \nu^3 + \left( 6 \xi -1 \right)\nu\right] g_{\alpha \gamma } = \frac{m^{2}\nu }{12\pi L}g_{\alpha \gamma}, 
\label{eq:SETvacD}
\\
\langle {\hat {T}}_{\alpha \gamma} \rangle _{0}^{{N}} & = -\frac{1}{12 \pi L^3} \left[ \nu^3 + \left( 6 \xi -1\right)\nu\right] g_{\alpha \gamma}=-\frac{m^{2}\nu }{12\pi L}g_{\alpha \gamma}.
\label{eq:SETvacN}
\end{align}
\end{subequations}
The v.e.v.s (\ref{eq:SETvacDN}) differ only by an overall sign.
Both are proportional to the metric, indicating that the corresponding vacuum states respect the maximal symmetry of the underlying space-time, and hence the v.e.v.~of the RSET is determined entirely by its trace.
For a massless scalar field, we find that $\langle {\hat {T}}_{\alpha \gamma} \rangle _{0}^{{{D}}/{{N}}}=0$, as expected since there is no trace anomaly in three space-time dimensions.

Since $\langle {\hat {T}}_{\alpha \gamma} \rangle _{0}^{{{D}}/{{N}}}$ is proportional to the metric, the NEC is satisfied trivially for both Dirichlet and Neumann boundary conditions, and all scalar field masses and couplings.
Furthermore, $\langle {\hat {T}}_{\alpha \gamma} \rangle _{0}^{{{D}}/{{N}}}$ has the perfect fluid (\ref{eq:fluid}) form with 
\begin{equation}
    E = -P = \mp \frac{m^{2}\nu }{12\pi L},
    \label{eq:DNvevE}
\end{equation}
where the $-$ sign is for Dirichlet boundary conditions and the $+$ sign for Neumann boundary conditions.
Therefore, when $0<\nu <1$, the WEC is satisfied by the v.e.v.~of the RSET for Neumann but not Dirichlet boundary conditions.
However, if we consider the semiclassical Einstein equations (\ref{eq:SCEE}), then the RSETs (\ref{eq:SETvacDN}) correspond to a renormalization of the cosmological constant \cite{Kent:2014nya,Thompson:2024vuj}, so there are no deep consequences of any violation of the WEC.

\subsection{Thermal RSET with Dirichlet and Neumann boundary conditions}
\label{sec:tevDN}

The thermal Green functions, $G_\beta^{D/N}(x,x')$,  for inverse temperature $\beta$, can be expressed as an infinite sum involving the vacuum Green function $G_{0}^{D/N}(x,x')$ \cite{Birrell:1982ix}:
\begin{equation}
G_\beta^{D/N}(t,\rho, \theta;t',\rho ', \theta')=\sum_{j=-\infty}^{\infty} G_{0}^{D/N} (t + i j\beta, \rho, \theta; t',\rho', \theta'),
\label{thermal_vac}
\end{equation}
giving~\cite{Namasivayam:2022bky}
\begin{subequations}
\label{eq:GFthermalDN}
\begin{align}
G_\beta^{D} (s) = & ~ \sum_{j=-\infty}^{\infty} -\frac{i \nu }{4 \pi L}  {}_2F_1\left (1+ \nu,1-\nu,\frac{3}{2};-\sinh^2\left[ \frac{s_\beta}{2L}\right]\right)\nonumber \\
 & \qquad + \frac{i}{8 \pi L\, \sinh\left(\frac{s_\beta}{2L}\right)} {}_2F_1 \left(\frac{1}{2}+\nu,\frac{1}{2}-\nu,\frac{1}{2};-\sinh^2\left[ \frac{s_\beta}{2L}\right]\right),
\label{eq:GFoD_thermal}
\\
G_\beta^{N} (s) = & ~  \sum_{j=-\infty}^{\infty} \frac{i \nu }{4 \pi L}  {}_2F_1\left (1+ \nu,1-\nu,\frac{3}{2};-\sinh^2\left[ \frac{s_\beta}{2L}\right] \right)\nonumber \\
 & \qquad + \frac{i}{8 \pi L\, \sinh\left(\frac{s_\beta}{2L}\right)} {}_2F_1 \left(\frac{1}{2}+\nu,\frac{1}{2}-\nu,\frac{1}{2};-\sinh^2\left[ \frac{s_\beta}{2L}\right]\right),
\label{eq:GFoN_thermal}
\end{align}
\end{subequations}
where $s_\beta$ is given by 
\begin{equation}
\cosh \left( \frac{s_\beta}{L}\right) = \frac{\cos (\Delta t+ ij\beta)}{\cos \rho \cos \rho'}- \tan \rho \tan \rho' \cos \Delta \theta .
\label{eq:cosh_s_new}
\end{equation}
To determine the t.e.v.~of the RSET with Dirichlet and Neumann boundary conditions, we apply  the methodology 
used in \cite{Namasivayam:2022bky}  and consider the difference between  the expectation values of the RSET in the thermal and vacuum states.  
As the vacuum Green functions in (\ref{eq:G0DN}) are given by the $j=0$ contributions to the sums in (\ref{eq:GFthermalDN}), we define a regular biscalar $W_{\beta }^{D/N}(x,x')$ as follows:
 \begin{align}
W_{\beta}^{D/N}(x,x') = & ~ \sum_{j=-\infty,j \ne 0}^{\infty} \mp \frac{i \nu }{4 \pi L}  {}_2F_1\left (1+ \nu,1-\nu,\frac{3}{2};-\sinh^2\left[ \frac{s_\beta}{2L}\right]\right)\nonumber \\
 & \qquad + \frac{i}{8 \pi L\, \sinh\left(\frac{s_\beta}{2L}\right)} {}_2F_1 \left(\frac{1}{2}+\nu,\frac{1}{2}-\nu,\frac{1}{2};-\sinh^2\left[ \frac{s_\beta}{2L}\right]\right) .
\label{eq:GFoD_renthermal}
\end{align}
We then use this biscalar in  (\ref{eq:diff_operator3}) to give the difference between the t.e.v.~and the v.e.v.~of the RSET. 
We can then simply add the relevant v.e.v. to the final answer, to give the t.e.v.. 
Despite the simplicity of the expression (\ref{eq:GFoD_renthermal}), the resulting components of the RSET are algebraically lengthy, these can be found in appendix~\ref{sec:terms_DN}.
We compute the t.e.v.s numerically in {\tt{Mathematica}}, and find that the sum over $j$ converges very rapidly.

 \begin{figure}
     \begin{center}
           \includegraphics[width=0.52\textwidth]{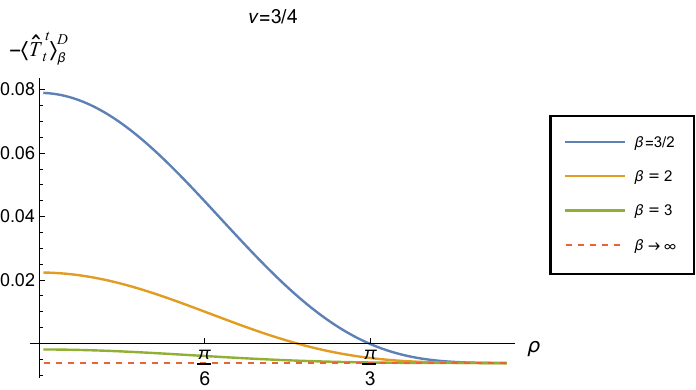} \qquad
       \includegraphics[width=0.39\textwidth]{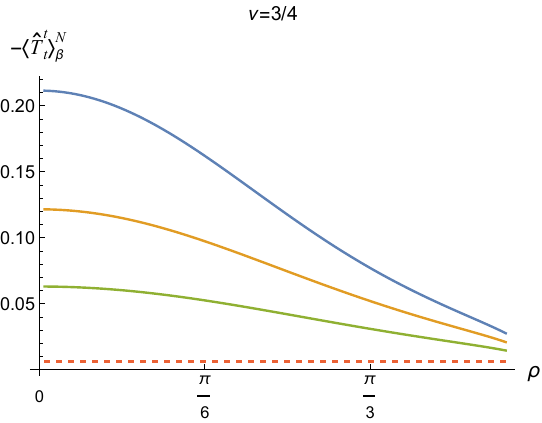}
              \caption{ Renormalized t.e.v.s of the energy density component of the RSET, $-\langle \hat{T}_t^t \rangle _\beta ^{D/N}$, with Dirichlet (left) and Neumann (right) boundary conditions, for  a selection of inverse temperatures, $\beta$~($\beta \to \infty$ corresponds to the vacuum state). For both plots $\nu=3/4$ and $\xi =1/8$.}
       \label{fig:thermDN_vbs}
          \end{center}
   \end{figure} 

In Fig.~\ref{fig:thermDN_vbs} we show the energy density component of the RSET, $-\langle \hat{T}_t^t \rangle _\beta ^{D/N}$, for Dirichlet (left) and Neumann (right) boundary conditions when $\nu = 3/4$, $\xi = 1/8$ and a selection of values of the inverse temperature $\beta $. 
The red dotted lines denote the v.e.v., which corresponds to $\beta \rightarrow \infty $.
For both boundary conditions, we see that the energy density as a function of $\rho $ has a maximum at the space-time origin and is monotonically decreasing as $\rho $ increases. 
This is to be expected, as the local inverse temperature ${\widetilde {\beta }}$ is related to the (fixed) inverse temperature $\beta $ via  \cite{Tolman:1930zza,Tolman:1930ona,Ambrus:2018olh}
\begin{equation}
    {\widetilde {\beta }} = \beta {\sqrt {-g_{tt}}} = \beta L \sec \rho ,
    \label{eq:loctemp}
\end{equation}
and hence the local temperature ${\widetilde {\beta }}^{-1}$ has a maximum at $\rho =0$ and tends to zero as $\rho \rightarrow \pi/2$ and the boundary is approached.
The value of the maximum in the energy density at $\rho =0$, unsurprisingly, increases with increasing temperature $\beta ^{-1}$.

Of more interest is the difference between Dirichlet (left) and Neumann (right) boundary conditions. 
In particular, we see that the energy density is significantly greater when Neumann boundary conditions are applied, compared to Dirichlet boundary conditions.
Similar behaviour is seen for the vacuum polarization on adS in three dimensions \cite{Namasivayam:2022bky}.
We also see that the energy density decreases more rapidly as the space-time boundary is approached for Dirichlet boundary conditions than for Neumann boundary conditions.

For both Dirichlet and Neumann boundary conditions, for the values of $\nu $ and $\xi $ considered in Fig.~\ref{fig:thermDN_vbs} (which correspond to a massive but conformally coupled scalar field), we see that the energy density in the thermal states is always greater than that in the corresponding vacuum state. 
For Dirichlet and Neumann boundary conditions, it is the difference in expectation values between the thermal and vacuum states which affects the backreaction of the quantum field on the space-time background, due to the maximal symmetry of the vacuum states with these boundary conditions applied \cite{Thompson:2024vuj}. To see this,
following \cite{Thompson:2024vuj}, we write (\ref{eq:SCEE}) in the alternative form
\begin{equation}
    R_{\alpha }^{\gamma }- \frac{1}{2} R \delta _{\alpha }^{\gamma } + \Lambda \delta _{\alpha }^{\gamma} 
    = \Delta {\hat {T}}_{\alpha }^{\gamma } + \langle {\hat {T}}_{\alpha }^{\gamma} \rangle _{0}^{{{D}}/{{N}}} ,
    \label{eq:SCEEalt}
\end{equation}
where 
\begin{equation}
 \Delta {\hat {T}}_{\alpha }^{\gamma }= \langle {\hat {T}}_{\alpha }^{\gamma} \rangle _{\beta }^{{{D}}/{{N}}}- \langle {\hat {T}}_{\alpha }^{\gamma} \rangle _{0}^{{{D}}/{{N}}}   
 \label{eq:effectiveRSET}
\end{equation}
is the difference between the thermal and vacuum expectation values.
Using the v.e.v.s (\ref{eq:SETvacDN}), we can write (\ref{eq:SCEEalt}) as
\begin{equation}
     R_{\alpha }^{\gamma }- \frac{1}{2} R \delta _{\alpha }^{\gamma } + {\widetilde {\Lambda }} \delta _{\alpha }^{\gamma} 
    = \Delta {\hat {T}}_{\alpha }^{\gamma } ,
    \label{eq:SCEEalt1}
\end{equation}
where ${\widetilde {\Lambda }}=\Lambda \pm m\nu ^{2}/(12 \pi L)$ is the renormalized cosmological constant.
From Fig.~\ref{fig:thermDN_vbs}, we see that the effective RSET $\Delta {\hat {T}}_{\alpha }^{\gamma }$ for the conformally coupled field has a positive energy density. 

\begin{figure}
     \begin{center}
     \begin{tabular}{cc}
        \includegraphics[width=0.45\textwidth]{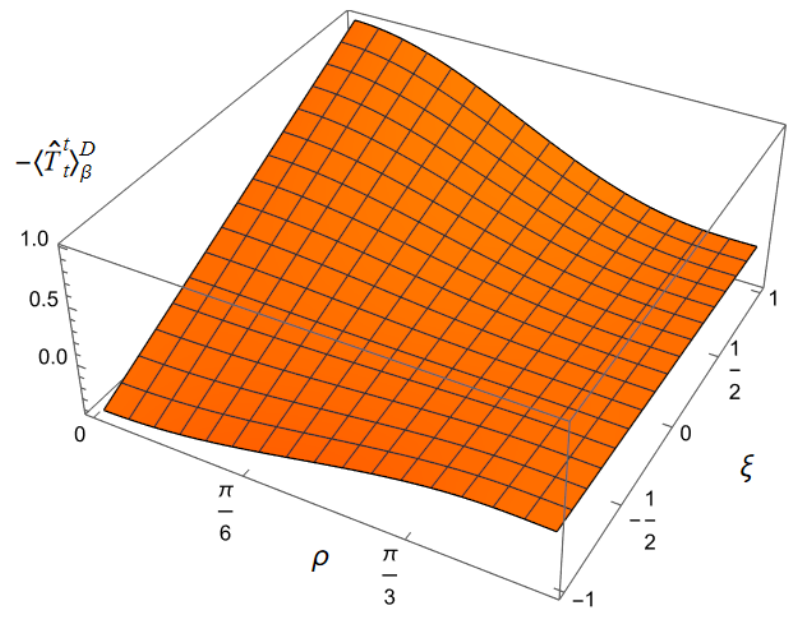} & \includegraphics[width=0.43\textwidth]{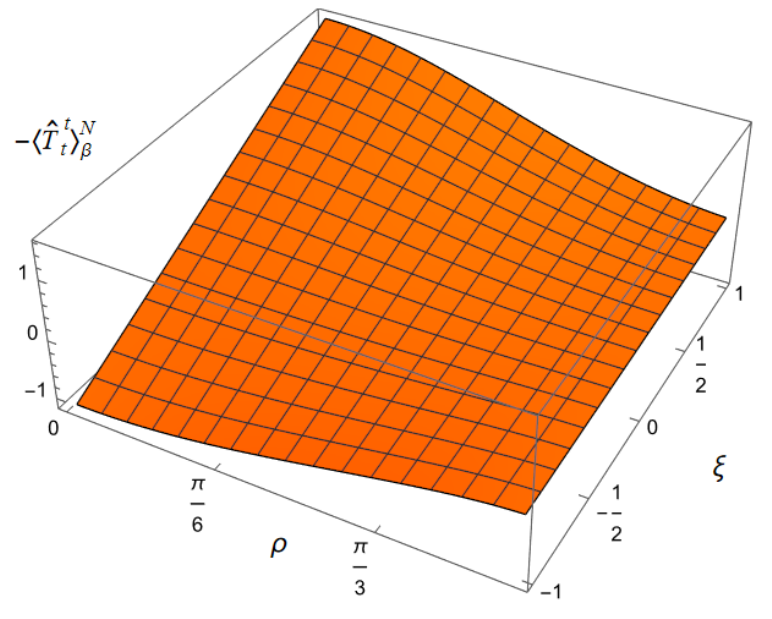}       
        \\[0.3cm]
       \\[0.3cm]
       \includegraphics[width=0.45\textwidth]{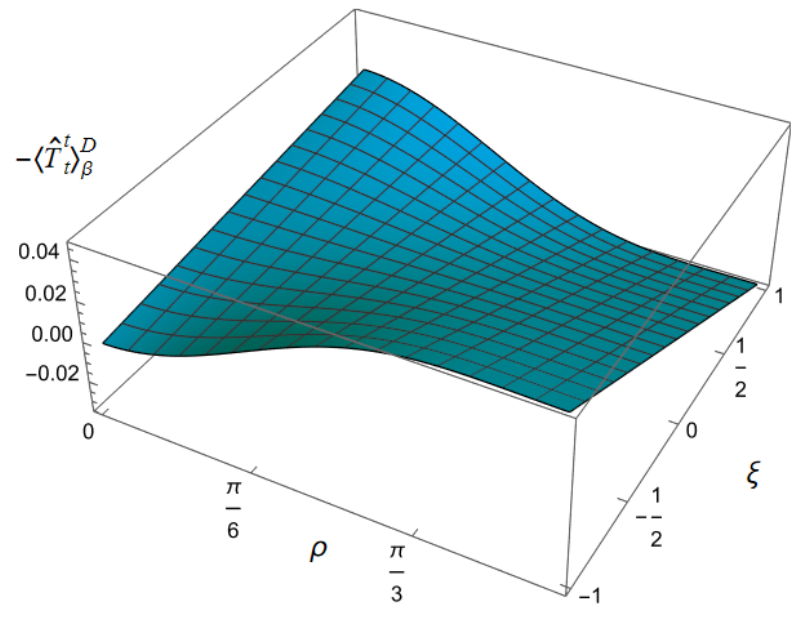} &
       \includegraphics[width=0.43\textwidth]{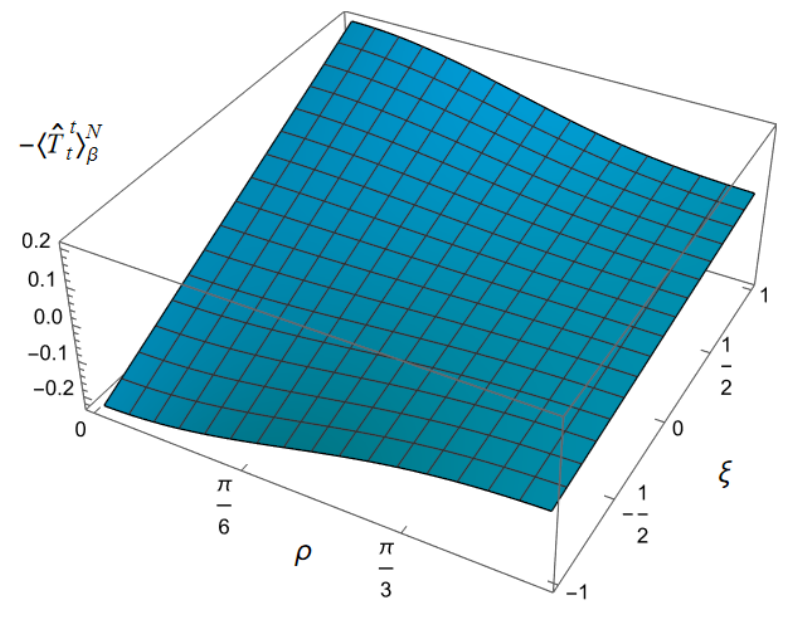}
       \end{tabular}
       \caption{3D surface plots of the energy density component $-\langle T_t^t \rangle _\beta^{D/N}$ of the RSET as functions of the radial coordinate $\rho$ and the coupling constant $\xi$, for different values of the inverse temperature $\beta$. The left column has the Dirichlet boundary condition and the right column has Neumann  boundary condition, with  inverse temperature $\beta=1$ (top) and $\beta=3$ (bottom).  For all plots, the parameter $\nu=1/4$ (\ref{eq:nu}).}
       \label{fig:thermal_3D_DNV14}
          \end{center}
   \end{figure}

To explore whether this remains the case for nonconformally-coupled fields, in Fig.~\ref{fig:thermal_3D_DNV14} 
we show the energy density $-\langle \hat{T}_t^t \rangle _\beta ^{D/N}$ as a function of both the radial coordinate $\rho $ and coupling constant $\xi $, for fixed $\nu = 1/4$, and two values of the inverse temperature $\beta $, considering Dirichlet boundary conditions on the left, and Neumann boundary conditions on the right.
For all values of $\xi$, the t.e.v. energy density approaches the v.e.v. energy density (\ref{eq:DNvevE}) as $\rho \rightarrow \pi/2$ and the space-time boundary is approached.
However, the profiles of the energy density as a function of $\rho $ depend strongly on the value of the coupling constant $\xi $ (bearing in mind that varying $\xi $ for fixed $\nu $ (\ref{eq:nu}) corresponds to varying $m^{2}$, the squared mass of the field). 
For sufficiently large $\xi $, the energy density at the origin is a maximum and decreases as $\rho $ increases towards the boundary, as seen in Fig.~\ref{fig:thermDN_vbs} for a conformally coupled field (albeit with a different value of $\nu $). 
However, as  $\xi $ decreases and becomes negative, the energy density takes its minimum value at the origin and is monotonically increasing as $\rho $ increases. 
In this case the effective RSET (\ref{eq:effectiveRSET}) no longer satisfies the WEC.
The violations of the WEC are maximal at the origin, and tend to zero on the space-time boundary. 

\begin{figure}
     \begin{center}
     \begin{tabular}{cc}
       \includegraphics[width=0.45\textwidth]{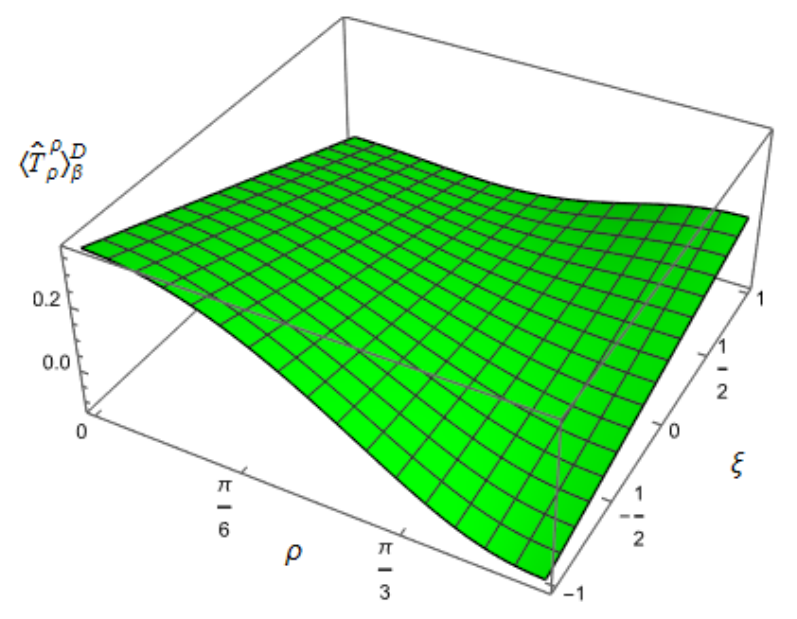} &  \includegraphics[width=0.45\textwidth]{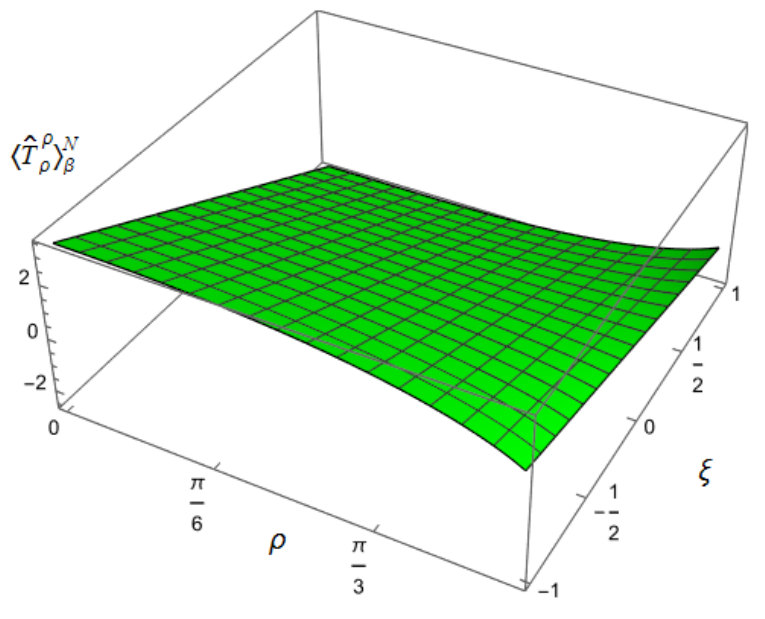}     
       \end{tabular}
       \caption{3D surface plots of the t.e.v.s of the $\langle T_\rho^ \rho \rangle _\beta^{D/N}$  component of the RSET, as functions of the radial coordinate $\rho$ and the coupling constant $\xi$. The left column shows  Dirichlet and the right column Neumann  boundary conditions. For both plots, the parameter $\nu=3/4$ (\ref{eq:nu}) and the inverse temperature is $\beta=1$. }       
       \label{fig:thermal_DNV14}
          \end{center}
   \end{figure}
  
To examine whether the NEC is satisfied even though the WEC is not, we need to consider other components of the RSET.
Fig.~\ref{fig:thermal_DNV14} shows the component $\langle T_\rho^ \rho \rangle _\beta^{D/N}$ of the RSET as a function of $\rho $ and $\xi $, with fixed inverse temperature $\beta = 1$ and $\nu =3/4$. 
Once again we see a strong dependence on the coupling constant $\xi $, but rather different behaviour from that observed for the $-\langle T_t^t \rangle _\beta^{D/N}$ component in Fig.~\ref{fig:thermal_3D_DNV14}.
For sufficiently large and positive $\xi$, the value of $\langle T_\rho^ \rho \rangle _\beta^{D/N}$ has a minimum at the origin and increases towards the boundary, while for sufficiently large and negative $\xi $ the value at the origin is a maximum and $\langle T_\rho^ \rho \rangle _\beta^{D/N}$ decreases as the boundary is approached. 

\begin{figure}[htbp]
     \begin{center}
     \begin{tabular}{cc}     
       \includegraphics[width=0.50\textwidth]{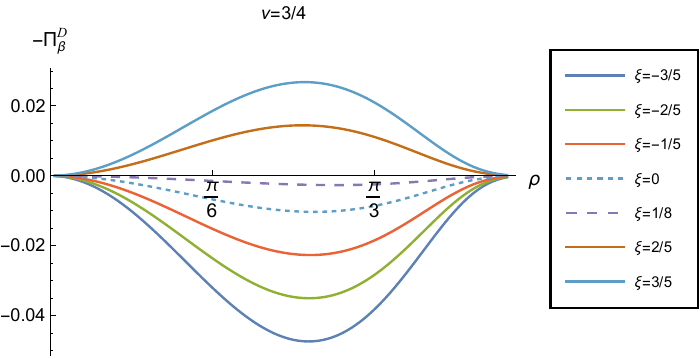} &
       \includegraphics[width=0.40\textwidth]{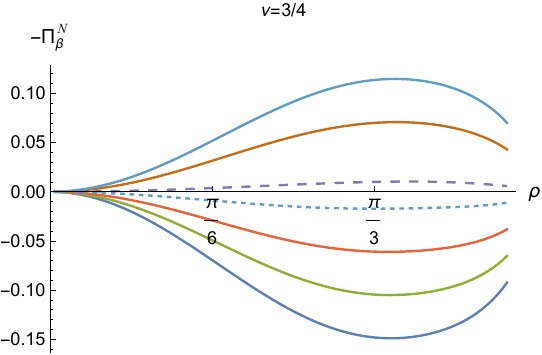}
       \end{tabular}
       \caption{ Negative thermal pressure deviators, $-\Pi_\beta^{D/N}$ for $\nu = 3/4$ with Dirichlet (left) and Neumann (right) boundary conditions, as functions of the radial coordinate $\rho$ for selected values of the coupling constant $\xi $. For both plots, the inverse temperature is $\beta=1$.}
       \label{fig:therm_press_devDN}
          \end{center}
   \end{figure}

The surface plots of the remaining RSET component, $\langle T_\theta^ \theta \rangle _\beta^{D/N}$, are indistinguishable by eye from those for $\langle T_\rho^ \rho \rangle _\beta^{D/N}$.
However, these two components of the RSET are not, in fact, equal.
We therefore employ the Landau decomposition of the RSET in an analogous way to that in four space-time dimensions \cite{Ambrus:2018olh}
\begin{equation}
    \langle {\hat {T}}_\alpha ^\gamma \rangle_{\beta}^{D/N }= \text{Diag} \left\{ -E_{\beta}^{D/N },\, P_{\beta}^{ D/N} +\Pi_{\beta}^{D/N}, \, P_{\beta}^{D/N } -\Pi_{\beta}^{D/N} \right\} ,
    \label{eq:Landau}
\end{equation}
where $E_{\beta}^{D/N }$ is the energy density, $P_{\beta}^{D/N }$ the pressure and  $\Pi_\beta^{D/N}$ is the pressure deviator \cite{Ambrus:2018olh}.
For a classical thermal  gas of particles, the pressure deviator vanishes identically.
In Fig.~\ref{fig:therm_press_devDN} we show
$-\Pi_\beta^{D/N}$ as a function of the radial coordinate $\rho$ . 
For both Dirichlet and Neumann boundary conditions and all values of $\xi$ considered,  the pressure deviators $\Pi_\beta^{D/N}$  attain their maximum magnitude between the space-time origin and boundary and are zero at the space-time origin. 
At the space-time boundary however, the difference between the two boundary conditions is noticeable. 
For Dirichlet boundary conditions, $\Pi_\beta^D$ vanishes on the boundary whereas for Neumann boundary conditions, $\Pi_\beta^N$ is nonzero for all $\rho $ considered, with its magnitude increasing with increasing $|\xi |$. 
However, for both Dirichlet and Neumann boundary conditions, $\Pi_\beta^{D/N}$ have  their minimum magnitudes at fixed $\rho $ for conformal coupling ($\xi =1/8$), suggesting that it is for conformal coupling that the quantum scalar field most closely resembles its classical counterpart.

\begin{figure}
     \begin{center}
     \begin{tabular}{cc} 
     \includegraphics[width=0.46\textwidth]{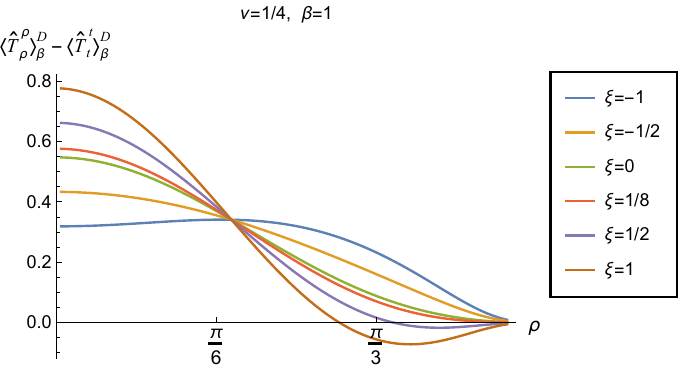} & \includegraphics[width=0.46\textwidth]{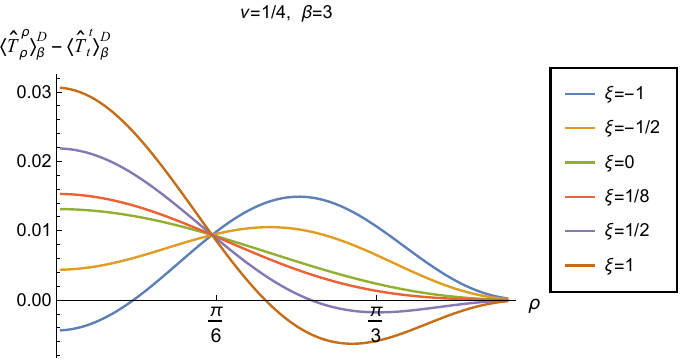}        
        \\[0.3cm]
         \includegraphics[width=0.46\textwidth]{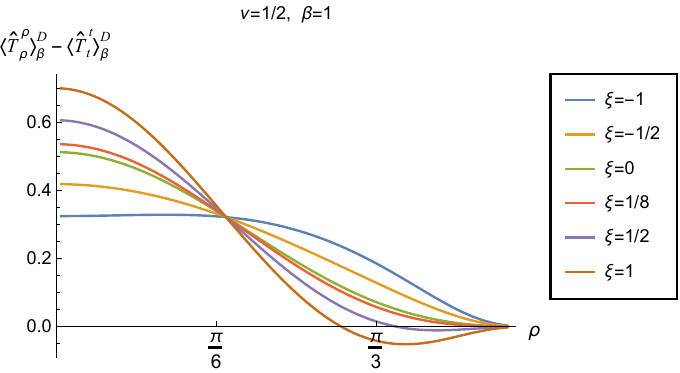} & 
         \includegraphics[width=0.46\textwidth]{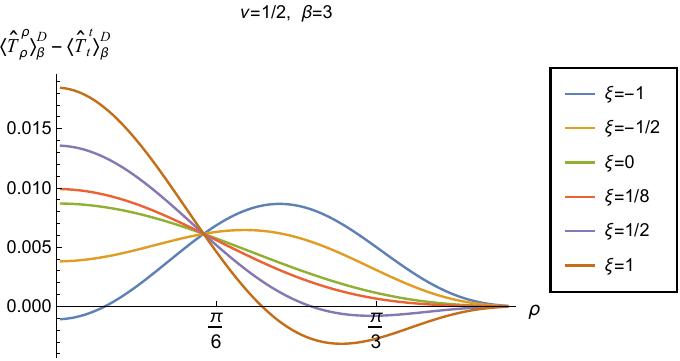}
       \\[0.3cm]
             \includegraphics[width=0.46\textwidth]{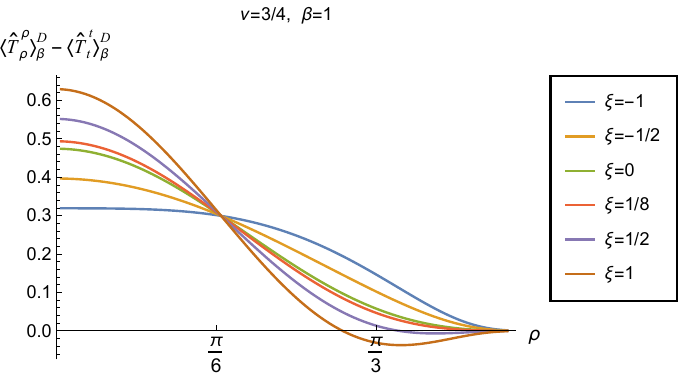}  & 
             \includegraphics[width=0.46\textwidth]{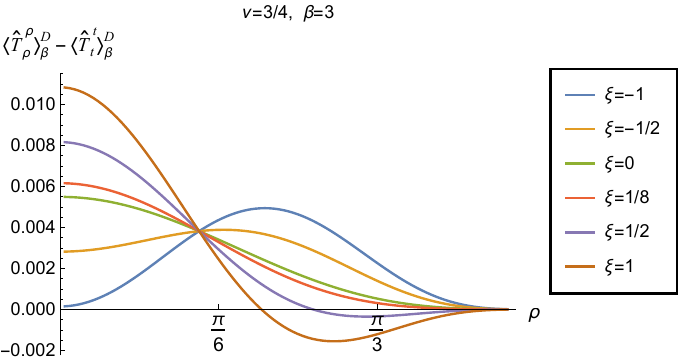}        
       \end{tabular}
       \caption{T.e.v.s of the combination of RSET components $\langle {\hat {T}}_\rho^\rho\rangle_{\beta}^{D}  -\langle {\hat {T}}_t^t\rangle_{\beta}^{D}$  with Dirichlet boundary conditions and a selection of values of the coupling constant $\xi$. In the top row $\nu=1/4$, middle row $\nu=1/2$, whilst the bottom row has $\nu=3/4$. In the left column $\beta=1$ whilst the right column has $\beta=3$.  }
       \label{fig:thermNECD}
          \end{center}
   \end{figure}

\begin{figure}
     \begin{center}
     \begin{tabular}{cc} 
     \includegraphics[width=0.46\textwidth]{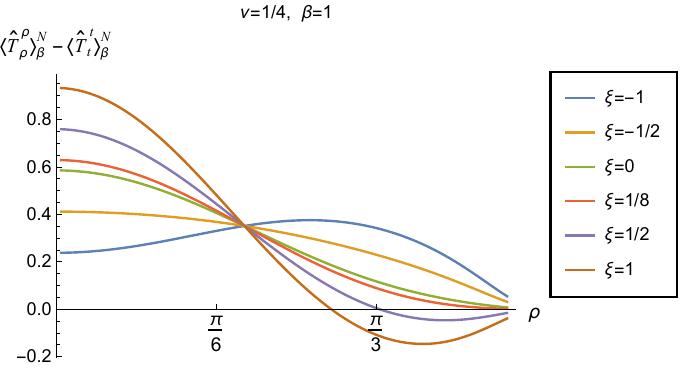} & \includegraphics[width=0.46\textwidth]{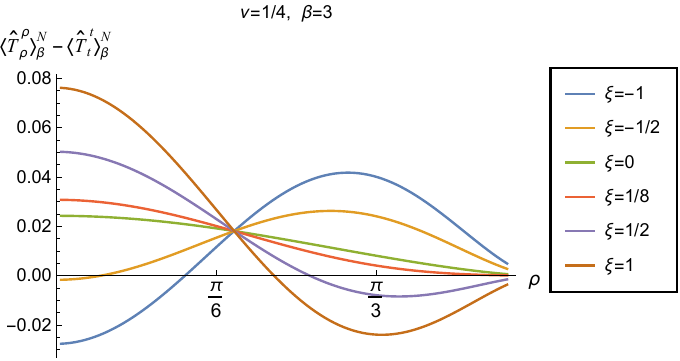}        
        \\[0.3cm]
         \includegraphics[width=0.46\textwidth]{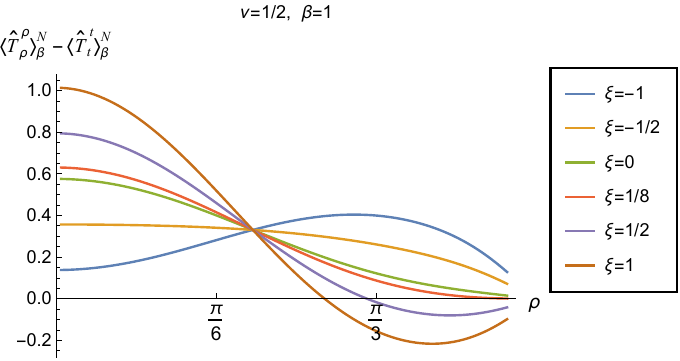} & \includegraphics[width=0.45\textwidth]{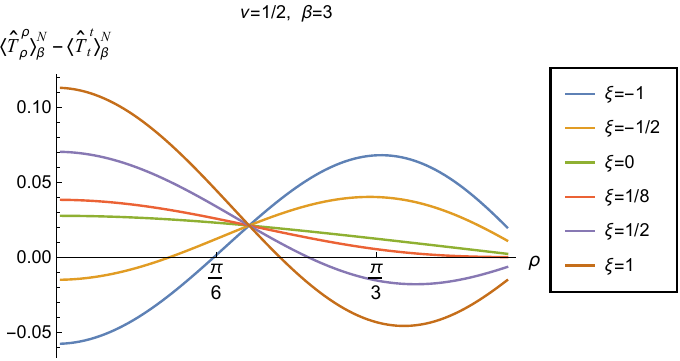}
       \\[0.3cm]
             \includegraphics[width=0.46\textwidth]{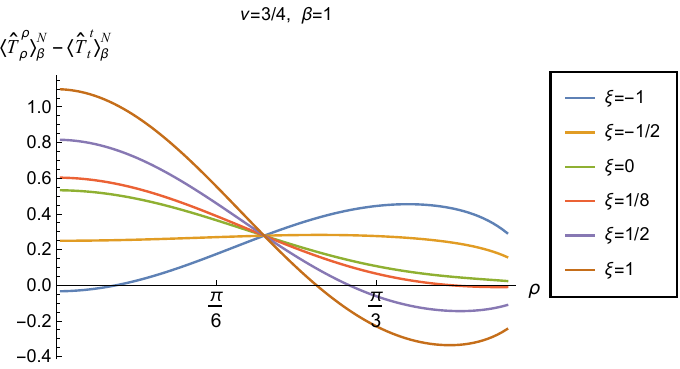}  & \includegraphics[width=0.45\textwidth]{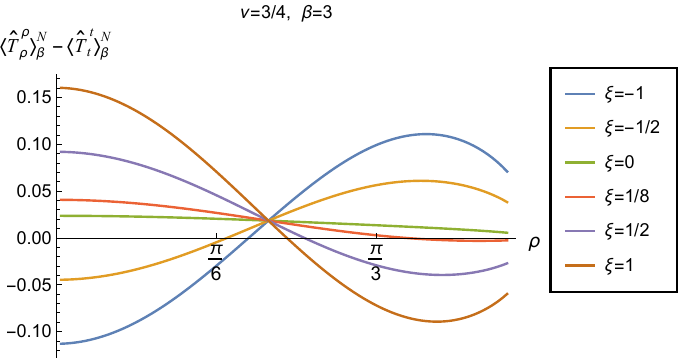}        
       \end{tabular}
       \caption{T.e.v.s of the combination of RSET components $\langle {\hat {T}}_\rho^\rho\rangle_{\beta}^{N }  -\langle {\hat {T}}_t^t\rangle_{\beta}^{N }$  for Neumann boundary conditions and a selection of values of the   coupling constant $\xi$. In the top row $\nu=1/4$, middle row $\nu=1/2$, whilst the bottom row has $\nu=3/4$. In the left column $\beta=1$ whilst the right column had $\beta=3$.  }
       \label{fig:thermNECN}
          \end{center}
   \end{figure}

Although the RSET no longer has the perfect fluid form (\ref{eq:fluid}), the NEC will be satisfied if 
$\langle {\hat {T}}_\rho^\rho\rangle_{\beta}^{D/N }  -\langle {\hat {T}}_t^t\rangle_{\beta}^{D/N }\ge 0$ and $\langle {\hat {T}}_\theta^\theta   \rangle_{\beta}^{D/N }-\langle {\hat {T}}_t^t \rangle_{\beta}^{D/N }\ge 0$.
However, since the pressure deviator $\Pi _{\beta}^{D/N }$ 
is roughly an order of magnitude smaller than the corresponding RSET components, in order to avoid a proliferation of plots displaying qualitatively similar results, in this work we focus on the combination of RSET components $-\langle {\hat {T}}_t^t\rangle_{\beta}^{D/N }+\langle {\hat {T}}_\rho^\rho  \rangle_{\beta}^{D/N }$ (further plots and discussion of the RSET components can be found in \cite{NamasivayamPhD}).
The positivity of this combination is a necessary (but not sufficient) condition for the NEC to be satisfied. 
Since for the v.e.v., $\langle {\hat {T}}_\rho^\rho\rangle_{0}^{D/N } = \langle {\hat {T}}_t^t\rangle_{\beta}^{D/N } $
(\ref{eq:SETvacDN}), when studying the NEC in this section we do not need to consider the effective RSET (\ref{eq:effectiveRSET}). 

In Figs.~\ref{fig:thermNECD} and \ref{fig:thermNECN} respectively, we show the combination of RSET components 
    $\langle {\hat {T}}_\rho^\rho\rangle_{\beta}^{D/N }  -\langle {\hat {T}}_t^t\rangle_{\beta}^{D/N }$ for Dirichlet and Neumann boundary conditions, and selected values of the coupling constant $\xi $, parameter $\nu $ (\ref{eq:nu}) and inverse temperature $\beta $.
    The NEC is violated when this combination of RSET components is less than zero. 

Considering first Dirichlet boundary conditions (Fig.~\ref{fig:thermNECD}), at the higher temperature ($\beta = 1$, left-hand plots) we see that, for the values of the coupling constant $\xi $ considered, there are no violations of the NEC in a neighbourhood of the origin.
However, for sufficiently large positive $\xi$, there are small regions close to the boundary where the NEC is violated. 
At the lower temperature ($\beta = 3$, right-hand plots), the NEC is again violated in a neighbourhood of the boundary for sufficiently large and positive $\xi $, but there are also violations of the NEC close to the origin for sufficiently large and negative $\xi $, at least for some values of the parameter $\nu $ (\ref{eq:nu}). 

The situation for Neumann boundary conditions (Fig.~\ref{fig:thermNECN}) is similar at the lower temperature ($\beta = 3$, right-hand plots), with violations of the NEC in a region including the origin for sufficiently large and negative $\xi $, and violations in a neighbourhood of the boundary for sufficiently large and positive $\xi $.
At the higher temperature ($\beta = 1$, left-hand plots), we again find that the NEC is violated close to the boundary for sufficiently large and positive $\xi $, but, unlike Dirichlet boundary conditions, there are also small violations of the NEC close to the origin for sufficiently large and negative $\xi $ when $\nu = 3/4$.
We conjecture that similar violations of the NEC would occur for Neumann boundary conditions and other values of $\nu $, and for Dirichlet boundary conditions, if we considered coupling constants $\xi $ having larger magnitudes.

For both Dirichlet and Neumann boundary conditions, and for both inverse temperatures considered, we do not find any significant violations of the NEC for either minimally ($\xi = 0$) or conformally coupled ($\xi = 1/8$) fields, and for values of $\xi $ lying in a sufficiently small interval containing these two particular values (there may be very small violations of the NEC very close to the space-time boundary).
 
 All of the plots in Figs.~\ref{fig:thermNECD} and \ref{fig:thermNECN} have an additional feature in common, namely that the curves for different $\xi $ and fixed $\nu $, $\beta $ all intersect at a particular value of $\rho\approx \pi /6$. 
This phenomenon can be understood as follows. 
The quantities $w$ (and its derivatives) and $w_{\alpha \gamma }$ (\ref{eq:walphagamma}) which appear in the RSET (\ref{eq:diff_operator3}) are constructed from the biscalar $W(x,x')$ (\ref{eq:Whad}) which depends on the scalar field mass $m$ and coupling constant $\xi $ only through the combination $\nu $ (\ref{eq:nu}). 
Therefore the terms in the RSET (\ref{eq:diff_operator3}) which are proportional to $\xi $ depend on $\nu $ and not on $\xi$, $m^{2}$ separately.
In all our figures, we consider fixed $\nu $, even when varying the coupling constant $\xi $.
The curves for varying values of $\xi $ intersect when the terms in (\ref{eq:diff_operator3}) proportional to $\xi $ vanish for a particular value of the radial coordinate $\rho$.

\section{RSET  with Robin Boundary conditions}
\label{sec:Robin}

To determine the RSET when Robin boundary conditions are applied to the scalar field we employ Euclidean methods, following \cite{Morley:2020ayr,Namasivayam:2022bky,Morley:2023exv}. 
   
\subsection{Euclidean Green functions and RSET}
\label{sec:GRobin}
   
We perform a Wick rotation, $t \to i \tau$, to give the Euclidean adS${}_{3}$ metric
\begin{equation}
    ds^2 = L^2\sec^2\rho \, \,[d\tau^2 + d\rho^2 +\sin^2\rho \, d\theta^2].
\label{eq:Euclid_metric}
\end{equation}
Suitable ansatze for the vacuum, $ G^E_0(x,x')$, and thermal, $G^E_\beta (x,x')$, Euclidean Green functions are~\cite{Namasivayam:2022bky}
\begin{subequations}
\label{eq:Euclidean_Green}
\begin{align}
G^E_0(x,x') = & ~ \frac{1}{4\pi^2} \int_{-\infty}^{\infty} e^{i\omega \Delta \tau} d\omega \,\sum_{\ell=-\infty}^{\infty} e^{i \ell\Delta \theta} g_{\omega\ell} (\rho,\rho'),
\label{eq:Euclid_vac_Green}
\\
G^E_\beta(x,x') = & ~ \frac{\kappa}{4\pi^2} \sum_{n=-\infty}^{\infty} \,\sum_{\ell=-\infty}^{\infty} e^{i n\kappa \Delta \tau}e^{i \ell\Delta \theta} g_{n\ell} (\rho,\rho'),
\label{eq:Euclid_Thermal_Green}
\end{align}
\end{subequations}
where  $\Delta \tau = \tau - \tau'$ and $ \Delta \theta = \theta - \theta' $ are the separation of the time and angular coordinates respectively, $\kappa=2\pi/\beta$ (with $\beta $ the inverse temperature) and $g_{\omega \ell}(\rho, \rho')$, $g_{nl}(\rho , \rho')$ are, respectively, the vacuum and thermal radial Green functions.

Applying Robin boundary conditions to the scalar field, the vacuum radial Green function $g_{\omega\ell}^\zeta (\rho,\rho')$ takes the form~\cite{Namasivayam:2022bky} 
\begin{align}
g_{\omega\ell}^\zeta (\rho,\rho') = & ~\mathcal{N}_{\omega\ell}^\zeta \,\, [\cos\rho]^{1+\nu}\,\, [\cos\rho']^{1+\nu}[\sin\rho]^{|\ell|}\,\,[\sin\rho']^{|\ell|} \nonumber\\
&  \quad \times {}_{2}F_{1} \Big (\frac{1}{2}[1+|\ell| +\nu - i\omega] ,\frac{1}{2}[1+|\ell| +\nu + i\omega] , 1+|\ell| ; \sin^2\rho _{<}\Big )  \nonumber \\
& \quad \times   \Bigg \{\cos\zeta \,{}_{2}F_{1} \Big (\frac{1}{2}[1+|\ell| +\nu - i\omega],\frac{1}{2}[1+|\ell| +\nu + i\omega], 1+\nu; \cos^2\rho_{>} \Big )   \nonumber\\
& \qquad  +\sin\zeta\,\, [\cos\rho _{>}]^{-2\nu} {}_{2}F_{1} \Big (\frac{1}{2}[1+|\ell| -\nu - i\omega],\frac{1}{2}[1+|\ell| -\nu + i\omega], 1-\nu; \cos^2\rho_{>} \Big ) \Bigg \}, 
\label{eq:gR}
\end{align}
where $\rho _{<}=\min\{ \rho, \rho' \}$ and $\rho _{>}=\max \{ \rho, \rho' \}$.
The normalization constant $\mathcal{N}_{\omega l}^\zeta$ is given by 
\begin{equation}
    \mathcal{N}_{\omega \ell}^\zeta = \left [ 2L\nu\left \{\cos\zeta  \frac{\Gamma(\nu) \Gamma ( 1+\lvert \ell \rvert )}{|\Gamma(\frac{1}{2}[1+\lvert \ell \rvert +\nu -i\omega])|^2}- \sin\zeta \frac{\Gamma(-\nu) \Gamma(1+ \lvert \ell \rvert ) }{\lvert \Gamma(\frac{1}{2}[1+\lvert  \ell \rvert -\nu -i\omega])\rvert ^2}  \right\} \right]^{-1}.
    \label{eq:Nconstantzeta}
\end{equation}
The corresponding thermal radial Green function, $g_{n\ell }^{\zeta}(\rho, \rho')$, is given by substituting $\omega = n\kappa $ into the vacuum radial Green function (\ref{eq:gR}).
In (\ref{eq:gR}, \ref{eq:Nconstantzeta}), the quantity $\zeta $ parameterizes the Robin boundary conditions, with $\zeta =0$ corresponding to Dirichlet boundary conditions, and $\zeta = \pi /2$ corresponding to Neumann boundary conditions.
The denominator of (\ref{eq:Nconstantzeta}) becomes zero if $\zeta$ satisfies the equation
\begin{equation}
\tan\zeta  =  \frac{ \,\Gamma(\nu)\lvert \Gamma(\frac{1}{2}[\lvert \ell \rvert +1 -\nu-i\omega]\rvert ^2}{\Gamma(-\nu)\lvert \Gamma(\frac{1}{2}[\lvert \ell \rvert  +1 +\nu-i\omega])\rvert ^2},
\label{eq:stability}
\end{equation}
in which case the vacuum and thermal Green functions become divergent. 
Denoting the smallest positive root of (\ref{eq:stability}) by $\zeta _{\text{crit}}$ (where $\zeta _{\text{crit}}>\pi/2$), we restrict attention to values of the Robin parameter below this critical value, that is, $ 0 \le \zeta < \zeta_{\text{crit}}$~\cite{Namasivayam:2022bky}.
In our numerical results in the following subsections, we will focus particularly on the values $\nu = 1/4$, $1/2$ and $3/4$ for the parameter $\nu $ (\ref{eq:nu}), for which $\zeta _{\text{crit}}$ is approximately given by
\begin{equation} 
\zeta_{\text{crit}}\approx   
\begin{cases}
0.64 \pi \qquad \text{for} \qquad \nu =1/4,\\
0.57\pi \qquad \text{for} \qquad \nu =1/2,\\
0.52 \pi \qquad \text{for} \qquad \nu =3/4.
    \end{cases}
    \label{eq:v_values}
\end{equation}

Since the divergences in the Green functions are independent of the quantum state under consideration, we follow \cite{Namasivayam:2022bky} and employ a state-subtraction technique to find the RSET when Robin boundary conditions are applied to the quantum scalar field.
Previous studies of the expectation value of the square of the scalar field (the ``vacuum polarization'') in both three \cite{Namasivayam:2022bky} and four space-time dimensions \cite{Morley:2020ayr} and of the RSET for a massless, conformally-coupled scalar field on four-dimensional global adS \cite{Morley:2023exv} have shown that it is Neumann boundary conditions which give the generic behaviour on the space-time boundary.
We therefore follow the methodology of \cite{Namasivayam:2022bky} and consider the difference in RSET between vacuum and thermal states with Robin and Neumann boundary conditions applied.
To implement this approach, we subtract the vacuum/thermal Euclidean Green function with Neumann boundary conditions applied from the corresponding Green function with Robin boundary conditions applied, to give the regular biscalars
\begin{subequations}
\label{eq:Euclid_Green_diff}
\begin{align}
W_{0}^{\zeta }(x,x')= & ~ G^{\zeta }_{0}(x,x') -G^{N}_{0}(x,x')
=\frac{1}{4\pi^2} \int_{-\infty}^{\infty} e^{i\omega \Delta \tau} d\omega \,\sum_{\ell=-\infty}^{\infty} e^{i \ell\Delta \theta} \Big[ g_{\omega\ell}^\zeta (\rho,\rho')-g_{\omega \ell}^N (\rho,\rho')\Big ],
\label{eq:Euclid_vac_Green_diff}
\\
W_{\beta }^{\zeta }(x,x') = & ~G^{\zeta}_{\beta}(x,x') -G^{N}_{\beta}(x,x')
=\frac{\kappa}{4\pi^2} \sum_{n=-\infty}^{\infty} \,\sum_{\ell=-\infty}^{\infty} e^{i n\kappa \Delta \tau}e^{i \ell\Delta \theta} \Big [ g_{n\ell}^\zeta (\rho,\rho')- g_{n \ell}^N (\rho,\rho')\Big],
\label{eq:Euclid_Thermal_Green_diff}
\end{align} 
\end{subequations}
where $G^{\zeta }_{0/\beta }(x,x')$ are the vacuum/thermal Green functions (\ref{eq:Euclidean_Green}) with Robin boundary conditions applied, and $G^{N}_{0/\beta }(x,x')=G^{\frac{\pi }{2}}_{0/\beta }(x,x')$ are the vacuum/thermal Green functions for Neumann boundary conditions.
Similarly, we denote the radial Green functions for Neumann boundary conditions by $g_{\omega \ell}^N (\rho,\rho')$ and $g_{n \ell}^N (\rho,\rho')$ for vacuum and thermal states, respectively.

The biscalars (\ref{eq:Euclid_Green_diff}) are substituted into  (\ref{eq:diff_operator3}) to give the difference in RSET expectation values between the relevant states with Robin and Neumann boundary conditions.
The final v.e.v.s and t.e.v.s of the RSET for Robin boundary conditions are then found by adding the corresponding RSET expectation values for Neumann boundary conditions, computed in section~\ref{sec:DN}.

\subsection{Vacuum RSET with Robin boundary conditions}
\label{sec:vacRobin}

Substituting the biscalar (\ref{eq:Euclid_vac_Green_diff}) into (\ref{eq:diff_operator3}) gives an expression for the difference in v.e.v.s~of the RSET with Robin and Neumann boundary conditions, involving an infinite sum over the quantum number $\ell $ and an integral over the frequency $\omega $. 
The lengthy expressions for the quantities to be summed/integrated over can be found in appendix~\ref{sec:terms_Robin_vac}.
The sums and integrals are performed numerically using {\tt {Mathematica}}, following the methodology of \cite{Namasivayam:2022bky}.
The $\omega $ integral is computed first, followed by the sum over $\ell $.
The integrals converge rapidly for large $|\omega |$, and, for fixed values of the radial coordinate $\rho $, the sum over $\ell $ is also rapidly converging. 
We find that truncating the integral at $|\omega |=100$ and the sum at $|\ell |=100$ gives final results which are sufficiently accurate for our purposes, away from the space-time boundary.
For example, for a representative value of the Robin parameter $\zeta = \pi /10$ with $\nu = 1/2$ and $\xi = 1/8$, the relative error in the $\langle {\hat {T}}_{\tau }^{\tau }\rangle $ component of the RSET at $\rho = 94\pi /200$ is $\sim 4\times 10^{-6}$ comparing to performing the sum and integral over $|\omega |,|\ell | \le 200$.
As observed in \cite{Namasivayam:2022bky}, the sum over $\ell $ is not uniformly convergent as the radial coordinate $\rho $ varies.
Accordingly, the relative error is much smaller for smaller values of the radius $\rho $. 

\begin{figure}
     \begin{center}
     \begin{tabular}{cc}
        \includegraphics[width=0.50\textwidth]{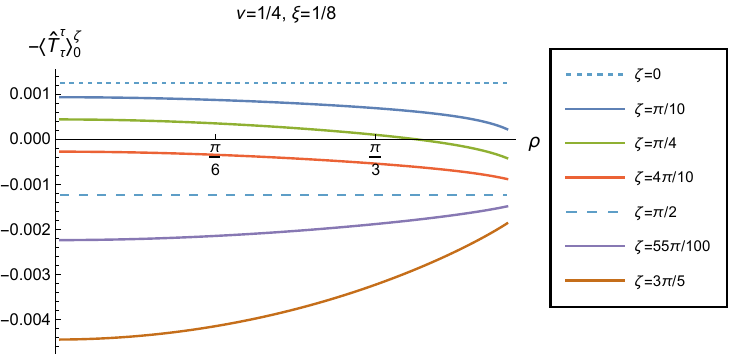} & 
        \includegraphics[width=0.40\textwidth]{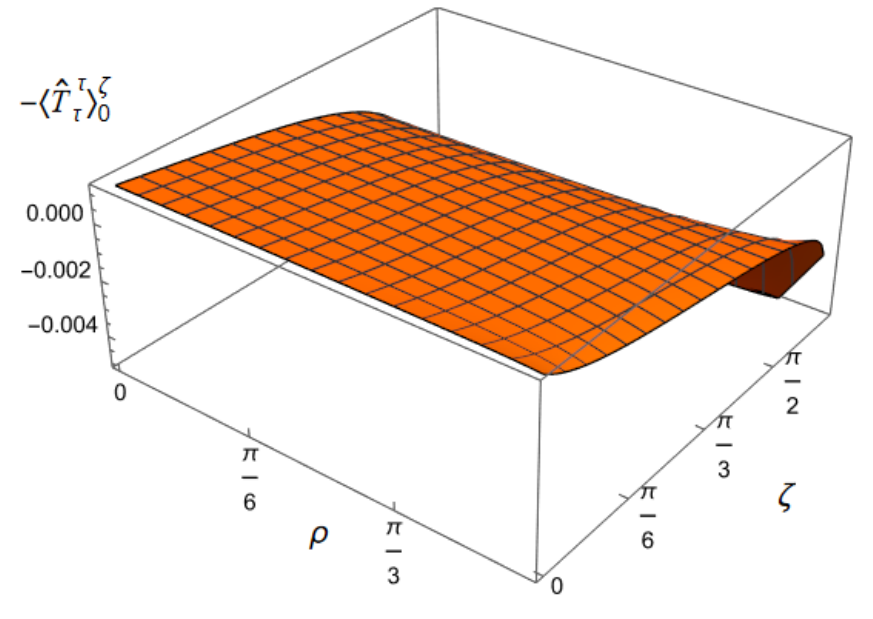}
        \\[0.3cm]
       \includegraphics[width=0.50\textwidth]{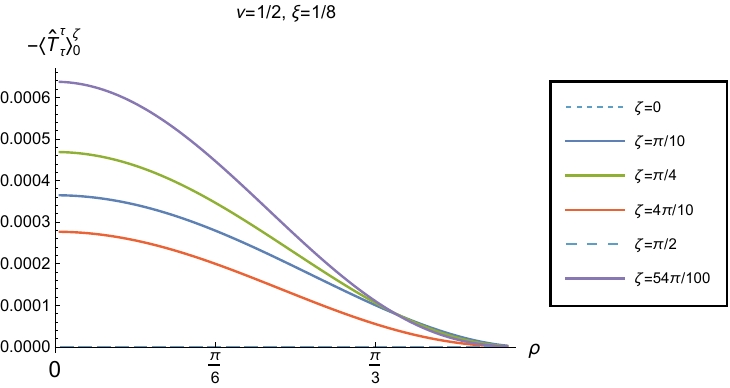} &
       \includegraphics[width=0.40\textwidth]{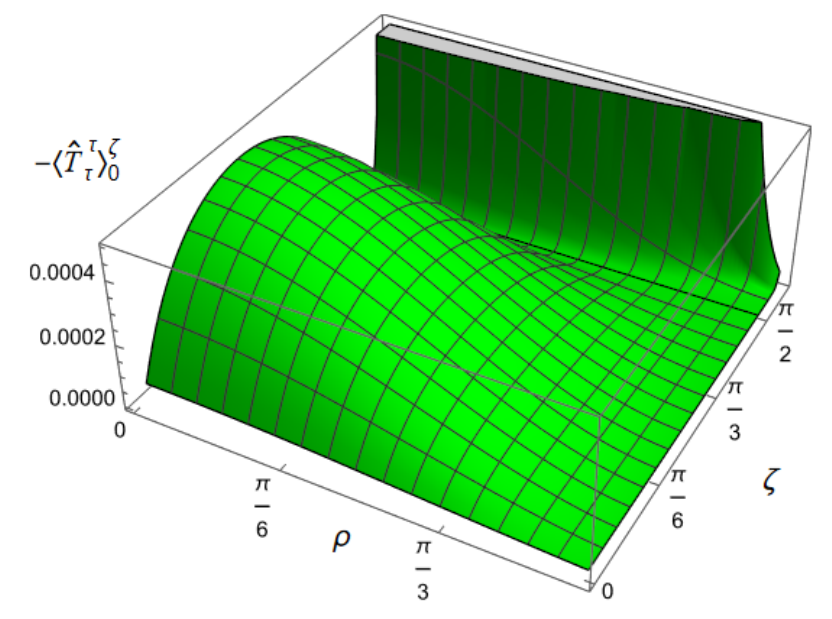}
       \\[0.3cm]
       \includegraphics[width=0.50\textwidth]{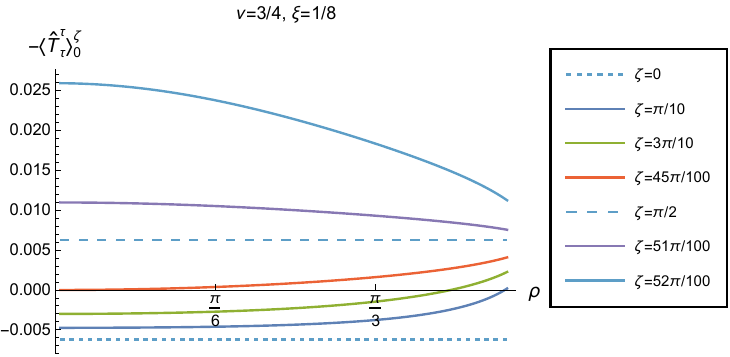} &
       \includegraphics[width=0.40\textwidth]{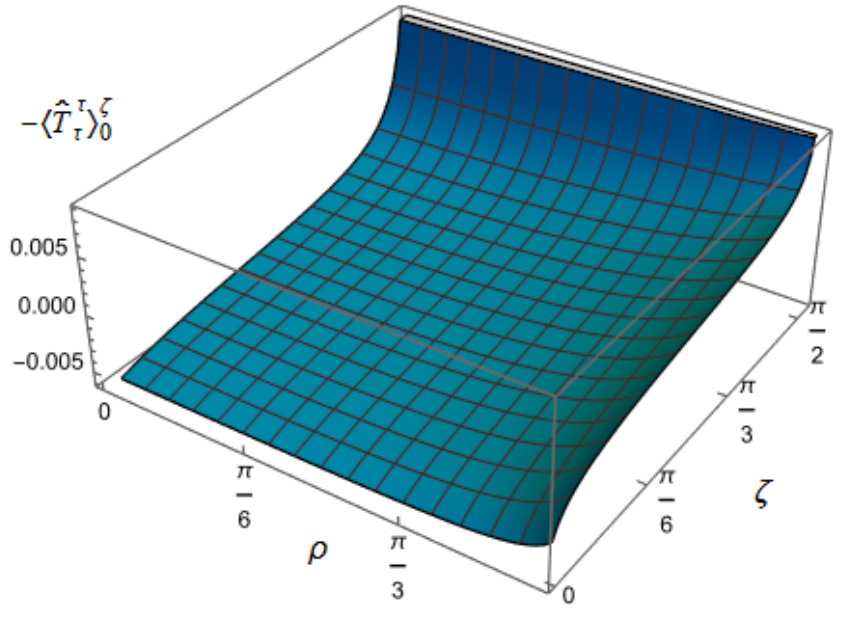}
       \end{tabular}
       \caption{Renormalized v.e.v.s of the energy density component of the RSET, $-\langle \hat{T}_\tau^\tau \rangle_0^\zeta$, with Robin boundary conditions and  different values of $\nu$. Plots in the left column show $-\langle \hat{T}_\tau^\tau \rangle_0^\zeta$ as a function of the radial coordinate $\rho$ for a selection of Robin parameters, $\zeta$. The right column shows the 3D-surface plots of the energy density as functions of  $\rho$ and $\zeta$. For all plots, $\xi=1/8$.}
       \label{fig:vacR_tt_nus}
          \end{center}
   \end{figure}

   \begin{figure}
     \begin{center}
     \begin{tabular}{cc}
        \includegraphics[width=0.50\textwidth]{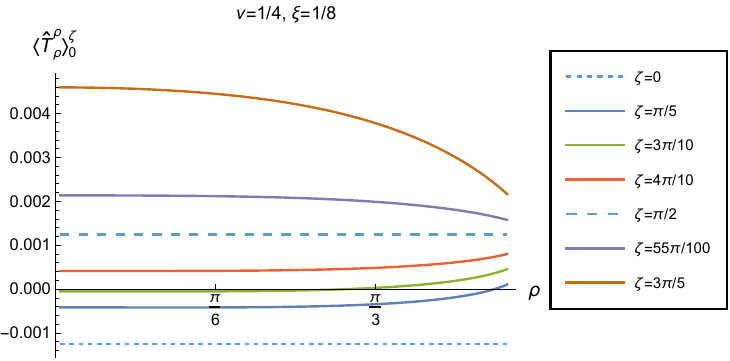} & 
        \includegraphics[width=0.40\textwidth]{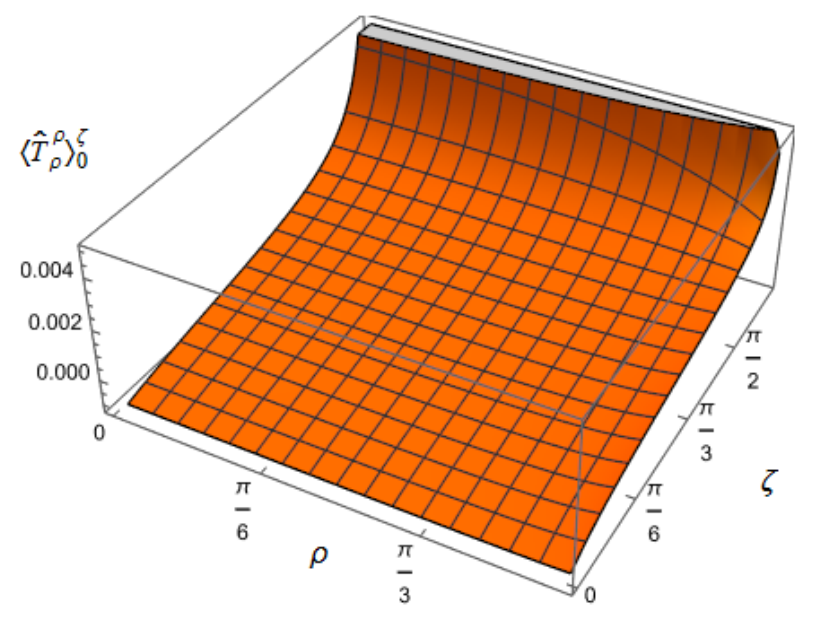}
        \\[0.3cm]
       \includegraphics[width=0.50\textwidth]{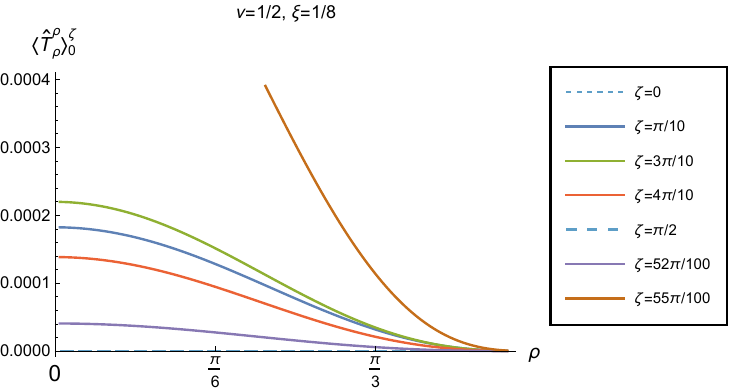} &
       \includegraphics[width=0.40\textwidth]{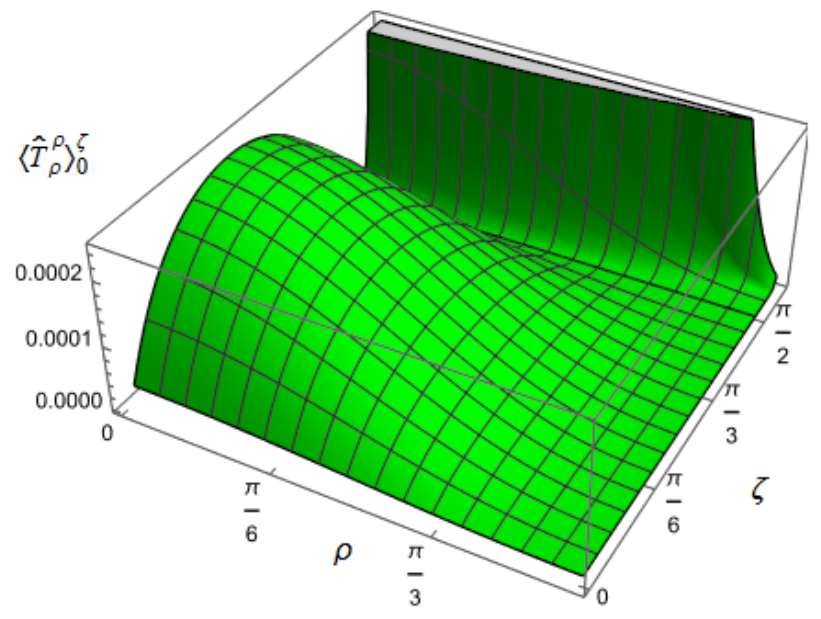}
       \\[0.3cm]
       \includegraphics[width=0.50\textwidth]{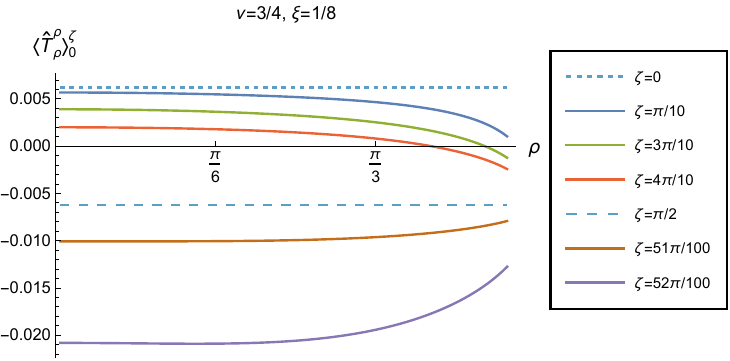} &
       \includegraphics[width=0.40\textwidth]{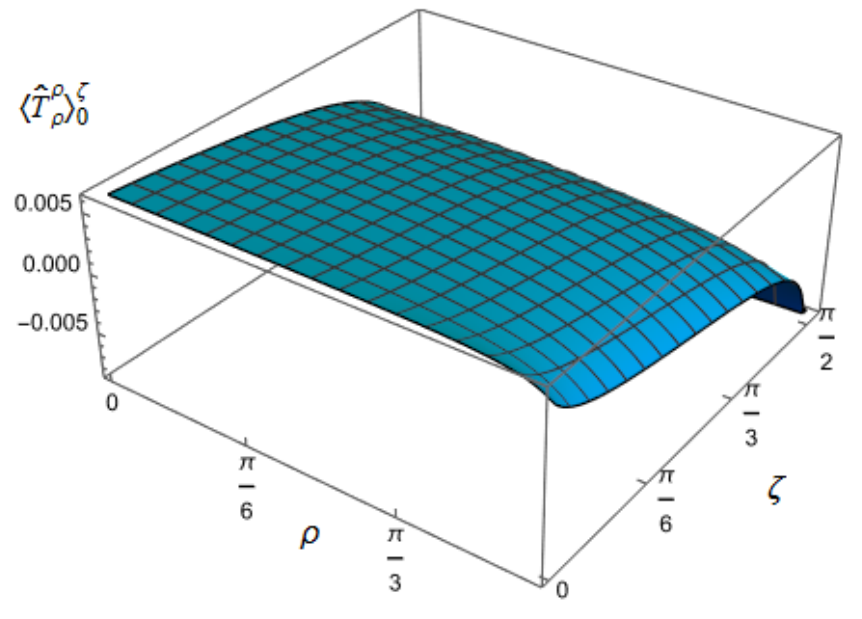}
       \end{tabular}
       \caption{Renormalized v.e.v.s of the $\langle \hat{T}_\rho^{\rho} \rangle _0^\zeta$ component of the RSET  with Robin boundary conditions, for different values of $\nu$. Plots on the left hand column show the $\langle \hat{T}_\rho^{\,\rho} \rangle _0^\zeta$ as a function of the radial coordinate $\rho $ for a selection of  Robin parameters $\zeta$. The right hand column shows the surface plots of $\langle \hat{T}_\rho^{\,\rho} \rangle _0^\zeta$ as functions of $\rho$ and $\zeta$. For all plots, $\xi=1/8$.}
       \label{fig:vacR22}
          \end{center}
   \end{figure}

    \begin{figure}
     \begin{center}
     \begin{tabular}{cc}
        \includegraphics[width=0.50\textwidth]{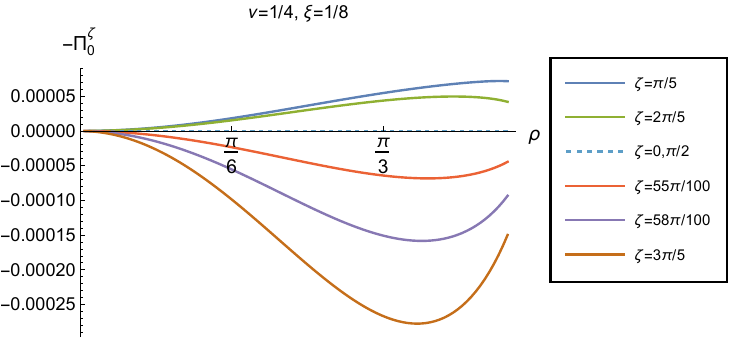} & 
        \includegraphics[width=0.40\textwidth]{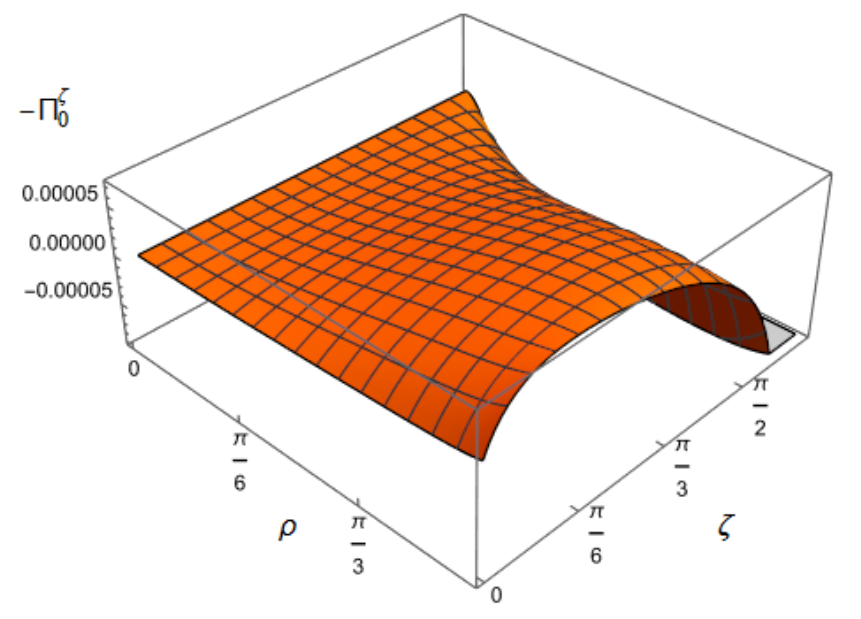}
        \\[0.3cm]
       \includegraphics[width=0.50\textwidth]{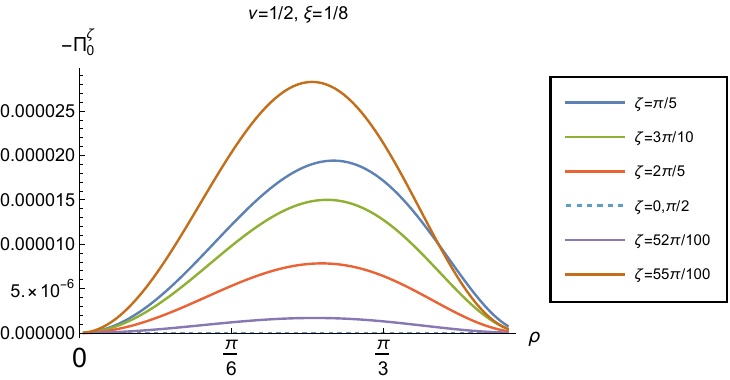} &
       \includegraphics[width=0.40\textwidth]{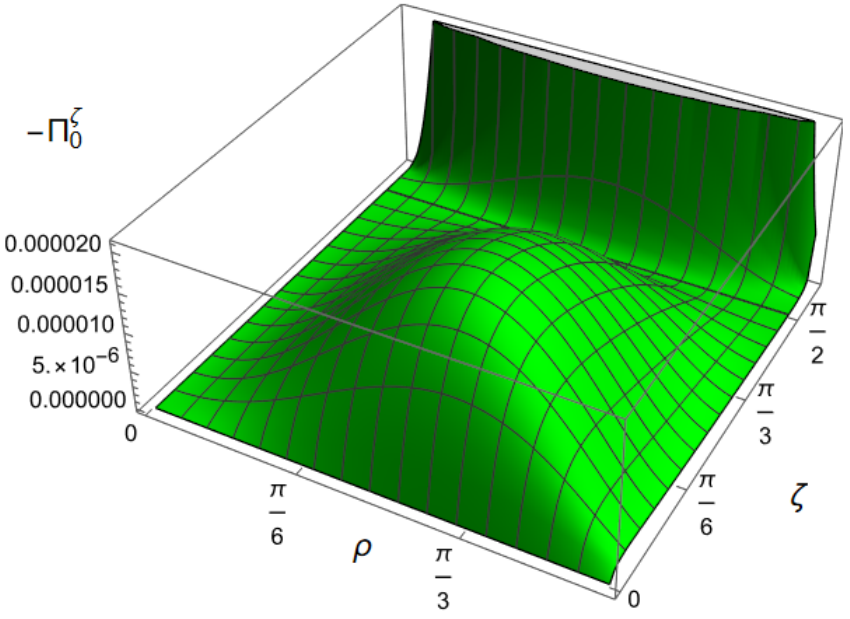}
       \\[0.3cm]
       \includegraphics[width=0.50\textwidth]{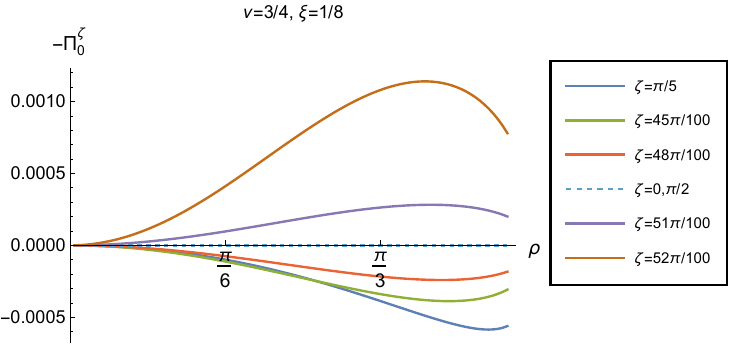} &
       \includegraphics[width=0.40\textwidth]{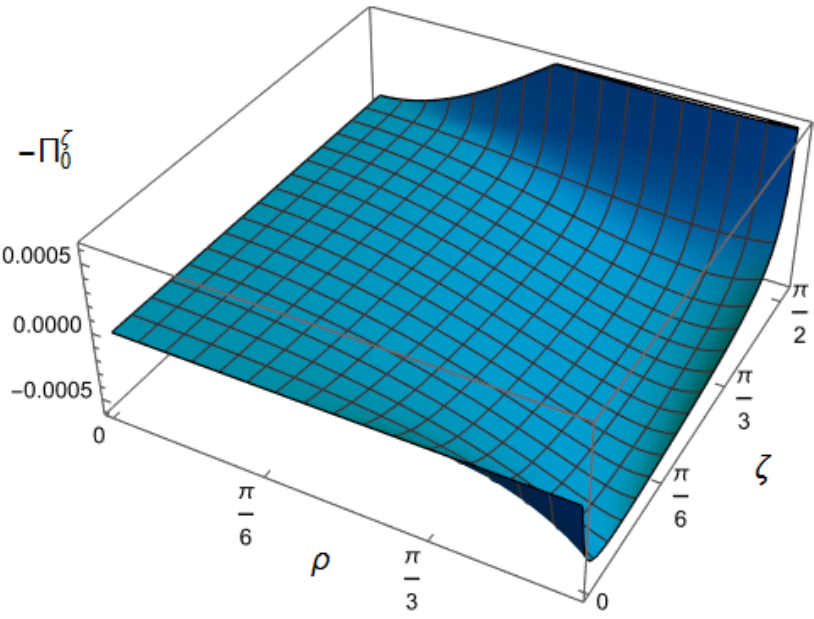}
       \end{tabular}
       \caption{Negative vacuum pressure deviators, $-\Pi_0^\zeta$, with Robin boundary conditions and different values of $\nu$. The left column shows  $-\Pi_0^\zeta$ as a function of the radial coordinate $\rho$ for a selection  of Robin parameters $\zeta$. The right column shows the 3D surface plots  of $-\Pi_0^\zeta$ as functions of $\rho$ and  $\zeta$. For all plots, $\xi=1/8$.}
       \label{fig:ads3_press_dev_vacR}
          \end{center}
   \end{figure}

We begin the discussion of our numerical results by examining, in Figs.~\ref{fig:vacR_tt_nus}, \ref{fig:vacR22} and \ref{fig:ads3_press_dev_vacR} respectively, the v.e.v.s $-\langle \hat{T}_\tau^\tau \rangle _{0}^{\zeta }$, $\langle \hat{T}_\rho^\rho \rangle _{0}^{\zeta }$ and minus the vacuum pressure deviator $-\Pi _{0}^{\zeta }$ with Robin boundary conditions applied to the quantum scalar field (additional plots and discussion can be found in \cite{NamasivayamPhD}).
We take the field to be conformally coupled, $\xi = 1/8$, and consider three values of the parameter $\nu $ (\ref{eq:nu}), namely $\nu = 1/4$, $\nu = 1/2$ and $\nu =3/4$.
When $\nu =1/4$, the scalar field has negative squared mass $m^{2} <0 $; for $\nu = 1/2$ the scalar field is massless and conformally coupled, while for $\nu = 3/4$ we have $m^{2}>0$.
In adS, a negative squared mass does not necessarily indicate an instability; we only consider values of $\nu $ for which the Breitenlohner-Freedman bound is satisfied \cite{Breitenlohner:1982bm,Breitenlohner:1982jf}, and Robin boundary conditions for which the scalar field is classically stable. 

First consider the energy density component, shown in Fig.~\ref{fig:vacR_tt_nus}. 
For all values of $\nu $, when either Dirichlet ($\zeta = 0$) or Neumann ($\zeta = \pi /2$) boundary conditions are applied, this does not depend on the radial coordinate $\rho $ (\ref{eq:SETvacDN}) (note that for a massless, conformally coupled field with $\nu = 1/2$ the v.e.v. vanishes for Dirichlet/Neumann boundary conditions). 
We can see that this is no longer the case when Robin boundary conditions are applied; the maximal symmetry of the underlying space-time is broken \cite{Pitelli:2019svx}.
When the field is massless ($\nu = 1/2$, middle row), the profiles of the energy density have similar qualitative features to those seen for a massless, conformally coupled scalar field in four space-time dimensions \cite{Morley:2023exv}.
In particular, the energy density is positive throughout the space-time. 
The energy density is a maximum at the origin and monotonically decreasing towards the space-time boundary. 
The value of the energy density at the origin is zero for Dirichlet boundary conditions, then increases as the Robin 
parameter $\zeta $ increases, reaches a maximum, then decreases with increasing $\zeta $ until it is zero again for Neumann boundary conditions. 
When $\zeta > \pi /2$, the magnitude of the v.e.v. of the energy density at the space-time origin increases rapidly as $\zeta $ increases, up to the critical value $\zeta _{\mathrm {crit}}$ (\ref{eq:v_values}), beyond which the field is classically unstable.

The energy density profiles for massive fields ($\nu = 1/4$ and $\nu = 3/4$) are qualitatively very different. 
When $m^{2}\neq 0 $, the v.e.v.s for Dirichlet and Neumann boundary conditions (\ref{eq:SETvacDN}) are no longer the same. 
As a function of $\rho $, the v.e.v.~of the energy density is either a monotonically increasing or monotonically decreasing function. 
As $\rho $ increases towards the space-time boundary, it seems to be the case that the v.e.v.~is approaching the v.e.v.~when Neumann boundary conditions are applied.
This suggests that it is Neumann rather than Dirichlet boundary conditions which give the generic behaviour on the boundary, as observed previously for the renormalized vacuum polarization \cite{Namasivayam:2022bky}. 
When $\nu = 1/4$ and the scalar field has negative squared mass, the v.e.v.~of the energy density is monotonically decreasing when $0<\zeta < \pi /2$ and monotonically increasing when $\zeta > \pi /2$.
The opposite is the case when $\nu =3/4$ and the scalar field has positive squared mass; in this case $-\langle \hat{T}_\tau^{\tau} \rangle _0^\zeta$ is monotonically increasing for $0<\zeta < \pi /2$ and monotonically decreasing for $\zeta > \pi /2$.

When $-\langle \hat{T}_\tau^{\tau} \rangle _0^\zeta<0$, the WEC is violated. 
For $\nu =1/4$, this is the case close to the boundary for nearly all $\zeta $ studied, and for sufficiently large $\zeta $ it is the case throughout the space-time.
When $\nu = 3/4$ the energy density is positive (a necessary but not sufficient condition for the WEC to be satisfied) throughout the space-time for sufficiently large $\zeta$, but for $\zeta $ close to zero, the WEC is violated in a neighbourhood of the origin.
Since the energy density depends on $\rho $ for $\zeta \neq 0$, $\pi /2$, when Robin boundary conditions are applied it is not possible to absorb the v.e.v.~of the RSET into a renormalization of the cosmological constant, as in section \ref{sec:tevDN}. 
However, we could decide to take the vacuum state with either Dirichlet or Neumann boundary conditions applied as a reference state, and, since the RSET for these boundary conditions is a constant multiple of the a metric (\ref{eq:SETvacDN}), absorb this into a renormalization of the cosmological constant, so that we are interested in an effective stress-energy tensor $\Delta {\hat {T}}_{\alpha }^{\gamma }$ (as in (\ref{eq:SCEEalt1})), where now
\begin{equation}
    \Delta {\hat {T}}_{\alpha }^{\gamma } = \langle {\hat {T}}_{\alpha }^{\gamma} \rangle _{0}^{\zeta }- \langle {\hat {T}}_{\alpha }^{\gamma} \rangle _{0}^{{{D}}/{{N}}} . 
\end{equation}
Since it is Neumann boundary conditions which give the generic behaviour of the energy density at the space-time boundary, it would be natural to use these boundary conditions in defining the reference state.
With this choice, the effective RSET will violate the WEC everywhere in the space-time for $\zeta > \pi /2$ for $\nu = 1/4$ and for $0<\zeta < \pi /2$ for $\nu = 3/4$.
Alternatively, if we choose Dirichlet boundary conditions to give the reference state, the effective RSET violates the WEC for all $0<\zeta <\zeta _{\mathrm {crit}}$ when $\nu = 1/4$ but the WEC can potentially be satisfied for all $\zeta  $ and $\nu = 3/4$ (whether or not the WEC is satisfied depends also on the sign of $\langle \hat{T}_\rho^{\rho} \rangle _0^\zeta - \langle \hat{T}_\tau^{\tau} \rangle _0^\zeta$, which we consider later in this subsection).

The v.e.v.s of the component $\langle \hat{T}_\rho^{\rho} \rangle _0^\zeta$ of the RSET, shown in Fig.~\ref{fig:vacR22}, have similar properties to those of the energy density.
In particular, as $\rho $ increases towards the space-time boundary, they seem to  approach the v.e.v.~when Neumann boundary conditions are applied, and the profiles as functions of $\rho $ are always either monotonically increasing or decreasing. 
The (negative) vacuum pressure deviators $-\Pi_0^\zeta$ (Fig.~\ref{fig:ads3_press_dev_vacR}) also show significant differences between the different values of $\nu $ that we consider.
For all $\nu $ studied, the pressure deviator is zero at the space-time origin at $\rho = 0$, and has a maximum magnitude at a point between the origin and the space-time boundary at $\rho = \pi /2$.
For $ \nu=1/2$, the pressure deviator $-\Pi_0^\zeta$ also becomes zero at the space-time boundary (similar behaviour is found for a massless, conformally coupled scalar field in four space-time dimensions \cite{Morley:2023exv}). However, for the other values of $\nu $ studied ($\nu= 1/4$, $3/4$),  while the pressure deviators are decreasing in magnitude as the space-time boundary is approached, our numerical results suggest that they are no longer zero at the boundary.

\begin{figure}
     \begin{center}
     \begin{tabular}{cc}
        \includegraphics[width=0.46\textwidth]{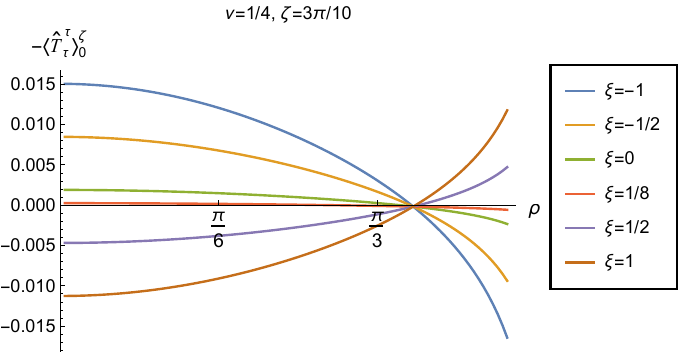} & 
      \includegraphics[width=0.46\textwidth]{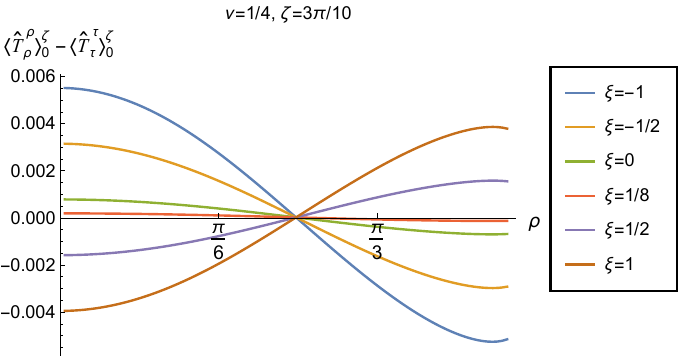}
        \\[0.3cm]
       \includegraphics[width=0.46\textwidth]{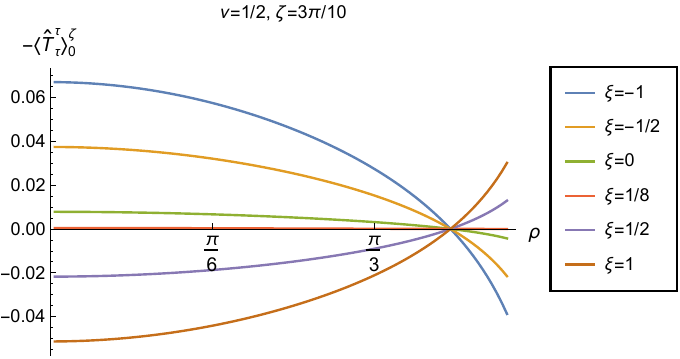} & 
        \includegraphics[width=0.46\textwidth]{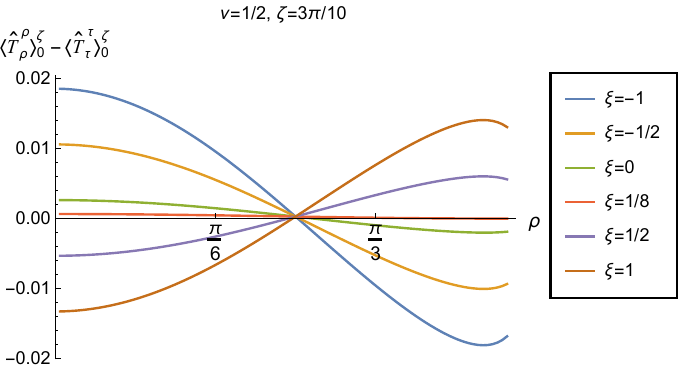}
       
       \\[0.3cm]
       \includegraphics[width=0.46\textwidth]{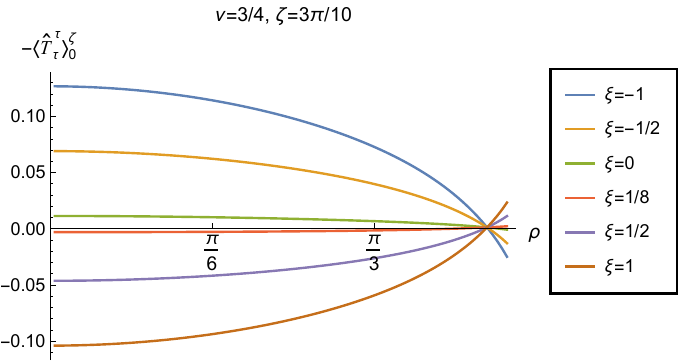} & 
       \includegraphics[width=0.46\textwidth]{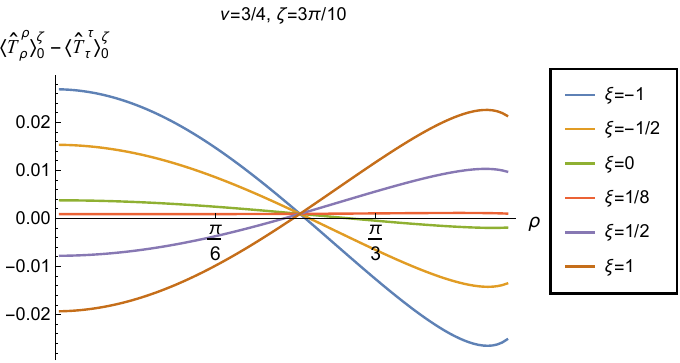}
      
       \end{tabular}
       \caption{Left-hand column: Renormalized v.e.v.s of  the energy density component of the RSET,  $-\langle \hat{T}_\tau^{\tau} \rangle_0^\zeta$.
       Right-hand column: Renormalized v.e.v.s of the combination of RSET components $\langle {\hat {T}}_\rho^\rho\rangle_{0}^{\zeta }  -\langle {\hat {T}}_\tau ^\tau \rangle_{0}^{\zeta }$.  In the top row we fix $\nu=1/4$,  in the middle row $\nu=1/2$ and in the bottom row $\nu=3/4$. For all plots, the Robin parameter $\zeta=3 \pi/10$, and the results are shown for a selection of values of the coupling constant $\xi $.}
       \label{fig:vacECfixedzeta}
          \end{center}
   \end{figure} 

 \begin{figure}
     \begin{center}
     \begin{tabular}{cc}
    
       \includegraphics[width=0.46\textwidth]{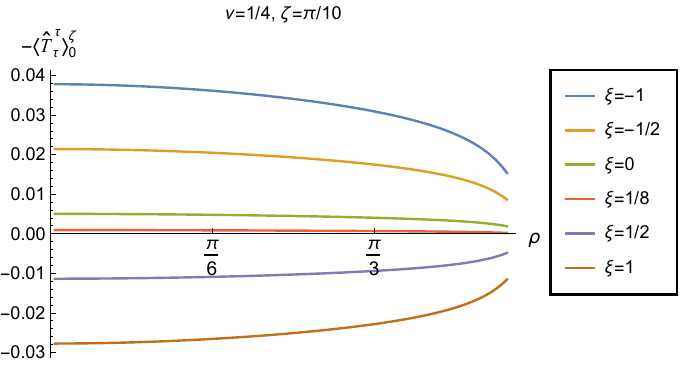} & 
        \includegraphics[width=0.46\textwidth]{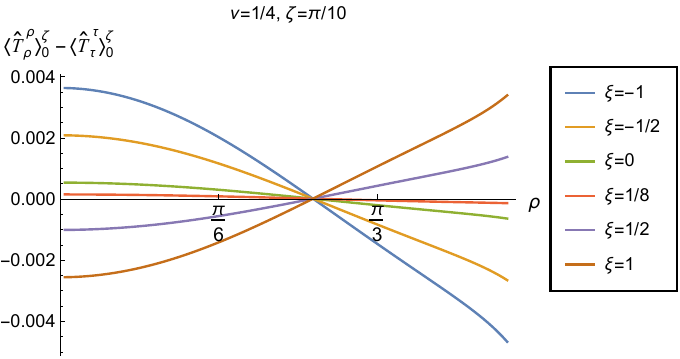}
        \\[0.3cm]
        \includegraphics[width=0.46\textwidth]{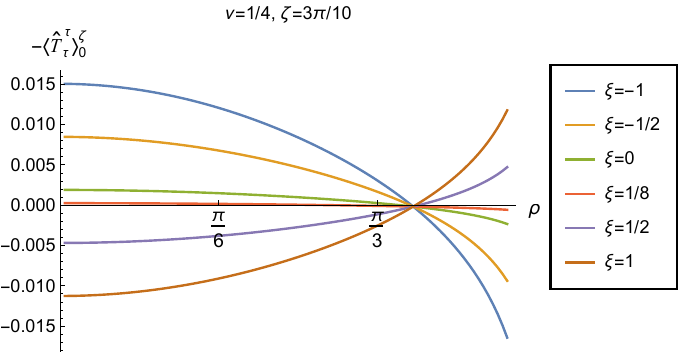} & 
        \includegraphics[width=0.46\textwidth]{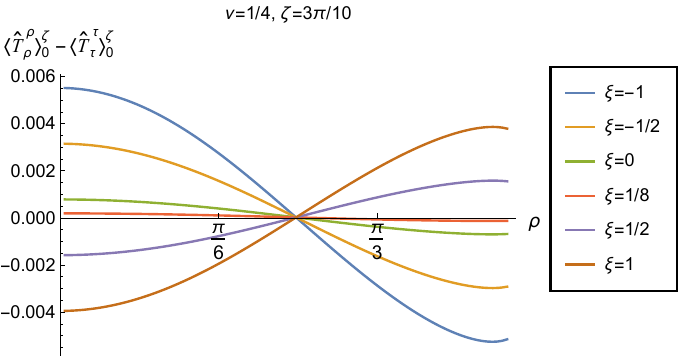}
        \\[0.3cm]
       \includegraphics[width=0.46\textwidth]{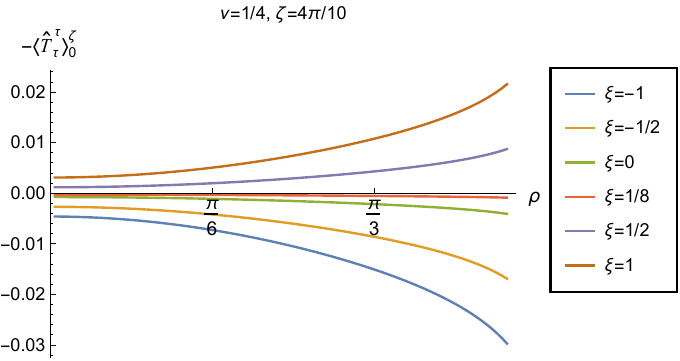} & 
       \includegraphics[width=0.46\textwidth]{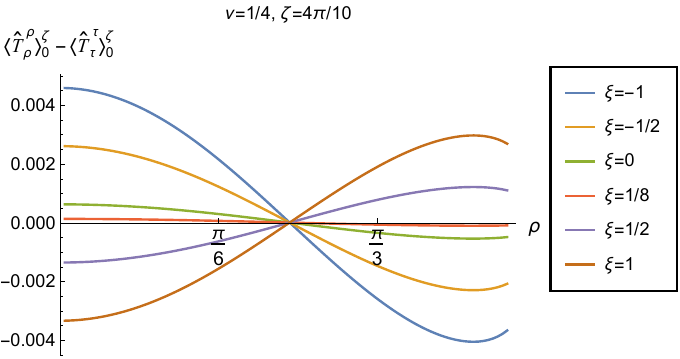}
       
       \\[0.3cm]
       \includegraphics[width=0.46\textwidth]{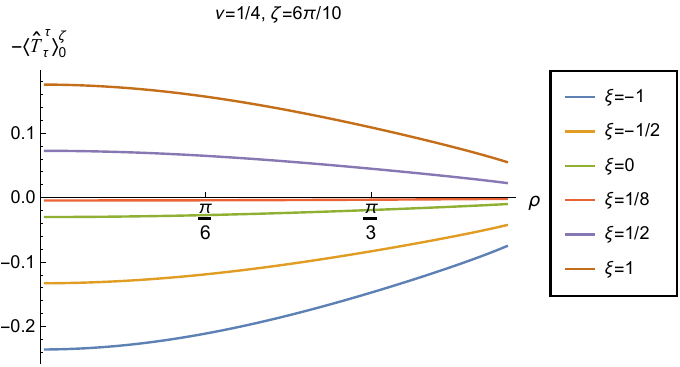} & 
       \includegraphics[width=0.46\textwidth]{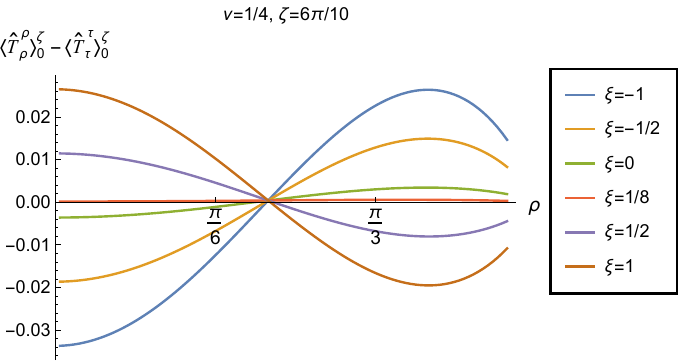}
       
       \end{tabular}
       \caption{
       Left-hand column: Renormalized v.e.v.s of  the energy density component of the RSET,  $-\langle \hat{T}_\tau^{\tau} \rangle_0^\zeta$.
       Right-hand column: Renormalized v.e.v.s of the combination of RSET components $\langle {\hat {T}}_\rho^\rho\rangle_{0}^{\zeta }  -\langle {\hat {T}}_\tau ^\tau \rangle_{0}^{\zeta }$.  In the top row we fix 
       the Robin parameter $\zeta =\pi /10$,  in the second row $\zeta = 3\pi /10$, in the third row $\zeta = 4\pi /10$ and in the bottom row $\zeta =6\pi /10$. For all plots, we fix $\nu = 1/4$, and the results are shown for a selection of values of the coupling constant $\xi $.}
       \label{fig:vacECfixednu}
          \end{center}
   \end{figure}

In Figs.~\ref{fig:vacECfixedzeta} and \ref{fig:vacECfixednu}  we explore the consequences for the WEC and NEC on varying the coupling constant $\xi $ and Robin parameter $\zeta $.
Given that the pressure deviator (see Fig.~\ref{fig:ads3_press_dev_vacR}) is roughly an order of magnitude smaller than the RSET components, we consider only the combination $\langle {\hat {T}}_\rho^\rho\rangle_{0}^{\zeta }  -\langle {\hat {T}}_\tau ^\tau \rangle_{0}^{\zeta  }$ in our study of the NEC.
Recall, as discussed in section~\ref{sec:tevDN}, whether the NEC is satisfied or violated is independent of the choice of reference state to use in renormalizing the cosmological constant. 
In Fig.~\ref{fig:vacECfixedzeta}, we fix the Robin parameter $\zeta = 3\pi /10$, consider three values of the parameter $\nu $ (\ref{eq:nu}), namely $\nu = 1/4$, $\nu = 1/2$ and $\nu = 3/4$ and, in each plot, a selection of values of the coupling constant $\xi $.
The WEC is violated if the energy density $-\langle \hat{T}_\tau^{\tau} \rangle_0^\zeta<0$ (left-hand column),
while the NEC  (and hence also the WEC) is violated if $\langle {\hat {T}}_\rho^\rho\rangle_{0}^{\zeta }  -\langle {\hat {T}}_\tau ^\tau \rangle_{0}^{\zeta  }<0$ (right-hand column).
The WEC can be satisfied only if both quantities in the left- and right-hand columns are positive.

The plots in Fig.~\ref{fig:vacECfixedzeta} have qualitative features in common.
As seen previously in Figs.~\ref{fig:thermNECD} and \ref{fig:thermNECN}, in each plot
the curves for different values of the coupling constant $\xi $ intersect at one value of the radial coordinate $\rho $ (where the terms proportional to  $\xi$ in (\ref{eq:diff_operator3}) vanish). 
The values of $-\langle \hat{T}_\tau^{\tau} \rangle_0^\zeta$ and $\langle {\hat {T}}_\rho^\rho\rangle_{0}^{\zeta }  -\langle {\hat {T}}_\tau ^\tau \rangle_{0}^{\zeta  }$ at the intersection point are not exactly zero; by definition they are equal to their values for a conformally coupled scalar field which are small but nonzero (as can be seen in Figs.~\ref{fig:vacR_tt_nus} and \ref{fig:vacR22}).
For nearly all values of $\xi $ studied, both the WEC and the NEC are violated in some region of the space-time, either in a neighbourhood of the origin (for sufficiently large and positive $\xi $) or in a neighbourhood of the space-time boundary (for sufficiently large and negative $\xi $). 
For a minimally coupled scalar field with $\xi = 0$, the NEC (and hence also the WEC) is violated in a neighbourhood of the space-time boundary. 
The degree to which the NEC is violated increases with increasing $|\xi |$.

We explore the effect of changing the Robin parameter $\zeta $ (as well as the coupling constant $\xi $) in Fig.~\ref{fig:vacECfixednu}, where we have fixed $\nu = 1/4$. 
Examining first the energy density $-\langle \hat{T}_\tau^{\tau} \rangle_0^\zeta$ (left-hand column), the qualitative features of the profiles of this quantity as a function of the radial coordinate $\rho $ depend strongly on the value of the Robin parameter $\zeta $. 
When $\zeta = \pi/10$, the energy density has the same sign for all $\rho $, and is positive for sufficiently large and negative $\xi$, but negative for sufficiently large and positive $\xi $. 
When $\zeta = 3\pi /10$, the curves for different values of $\xi $ intersect at a particular value of the radial coordinate $\rho$, as seen in Fig.~\ref{fig:vacECfixedzeta}. 
For $\zeta = 4\pi /10 $ and $\zeta = 6\pi /10 $, the sign of the energy density is again the same for all values of $\rho $, but is now positive for sufficiently large and positive $\xi $ but negative for sufficiently large and negative $\xi $.
For both $\zeta = \pi /10 $ and $\zeta = 6\pi /10$, the magnitude of the energy density is monotonically decreasing as the radial coordinate $\rho $ increases; in contrast, for $\zeta = 4\pi /10$, the magnitude of the energy density is monotonically increasing as $\rho $ increases.
From the plots in the left-hand column of Fig.~\ref{fig:vacECfixednu}, we see that the WEC is violated (due to the energy density being negative) throughout the space-time for some combinations of the parameters $\zeta $ and $\xi $, in particular for $\xi $ sufficiently large and positive when $\zeta = \pi /10$, and for $\xi $ sufficiently large and negative for $\zeta = 4\pi /10$, $6\pi /10$. 

Turning now to the plots in the right-hand column of Fig.~\ref{fig:vacECfixednu}, the profiles of the quantity 
$\langle {\hat {T}}_\rho^\rho\rangle_{0}^{\zeta }  -\langle {\hat {T}}_\tau ^\tau \rangle_{0}^{\zeta }$ share qualitative features with those in Fig.~\ref{fig:vacECfixedzeta}. 
For all fixed values of $\zeta $ studied, the curves for the different values of the coupling constant $\xi $ intersect at a particular value of the radial coordinate $\rho $ (the precise value depending on $\zeta $). 
We find violations of the NEC (corresponding to $\langle {\hat {T}}_\rho^\rho\rangle_{0}^{\zeta }  -\langle {\hat {T}}_\tau ^\tau \rangle_{0}^{\zeta }<0$) in some region of the space-time for almost all values of $\zeta $ and $\xi $ shown in Fig.~\ref{fig:vacECfixednu}. 
When $0<\zeta < \pi /2$ (top three rows in Fig.~\ref{fig:vacECfixednu}), the NEC is violated in a neighbourhood of the origin for $\xi $ sufficiently large and positive, and in a neighbourhood of the space-time boundary for $\xi $ sufficiently large and negative.
When $\zeta > \pi /2$ (bottom row in Fig.~\ref{fig:vacECfixednu}), the NEC is violated in a neighbourhood of the origin for $\xi $ sufficiently large and negative, and in a neighbourhood of the space-time boundary for $\xi $ 
sufficiently large and positive.

\subsection{Thermal RSET with Robin boundary conditions}
\label{sec:thermalRobin}

Substituting the biscalar (\ref{eq:Euclid_Thermal_Green_diff}) into (\ref{eq:diff_operator3}) gives an expression for the difference in t.e.v.s~of the RSET with Robin and Neumann boundary conditions, involving infinite sums over the quantum numbers $n$ and $\ell $. 
The lengthy expressions for the summands can be found in appendix~\ref{sec:terms_Robin_thermal}.
Again we follow \cite{Namasivayam:2022bky} and our numerical computations are performed in {\tt {Mathematica}}. 
We sum over $n$ first, and then $\ell $.
In order to have a reasonable computation time, we sum over $|n|\le 20 $ and $|\ell |\le 50$. 
Taking a representative value of the Robin parameter, $\zeta = \pi /10$, with $\nu = 1/2$ and $\beta = 1$, for $\rho = 94\pi /200$ we find that the relative error in the RSET components is of order between $10^{-5}$ and $10^{-7}$ compared to summing over $|n|, |\ell | \le 150$.
As in the vacuum case, the relative error is substantially smaller further away from the space-time boundary.

For t.e.v.s with Robin boundary conditions applied to the scalar field, we have a considerable number of parameters on which the RSET depends: the inverse temperature $\beta $, the parameter $\nu $ in the scalar field equation (\ref{eq:nu}), the coupling constant $\xi $ and the Robin parameter $\zeta $.
It is not practical to perform an exhaustive analysis of the t.e.v.s of all components of the RSET for the entire parameter space; instead we present, in this section, a selection of plots illustrating the key qualitative features of both the RSET and the consequences for the NEC and WEC. 
Further plots (and discussion thereof) can be found in \cite{NamasivayamPhD}.

\begin{figure}[t!]
     \begin{center}
     \begin{tabular}{cc}
        \includegraphics[width=0.50\textwidth]{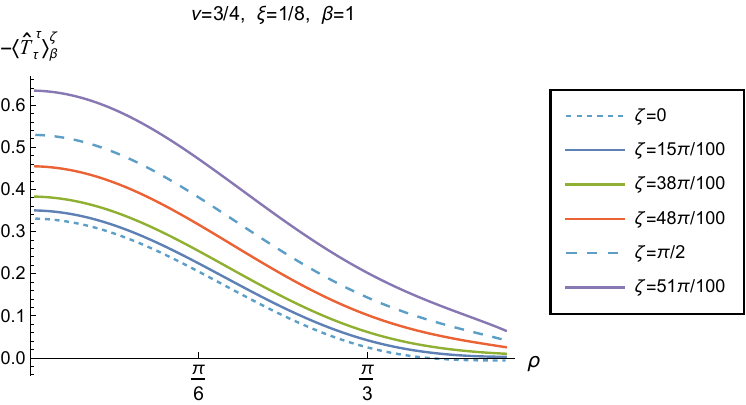}&
       \includegraphics[width=0.40\textwidth]{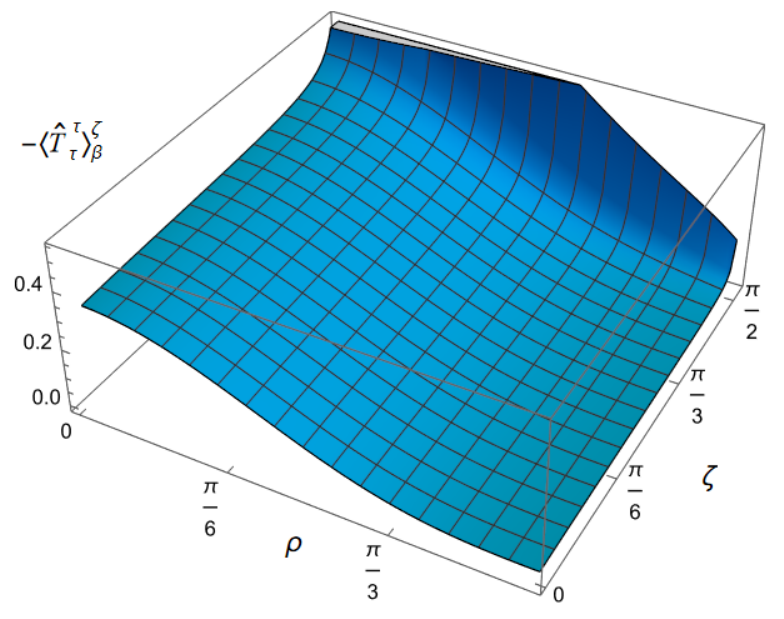}
        \\[0.3cm]
      \includegraphics[width=0.50\textwidth]{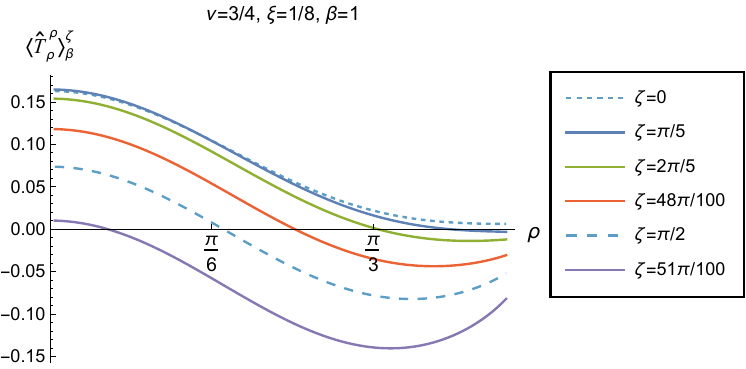}&
       \includegraphics[width=0.40\textwidth]{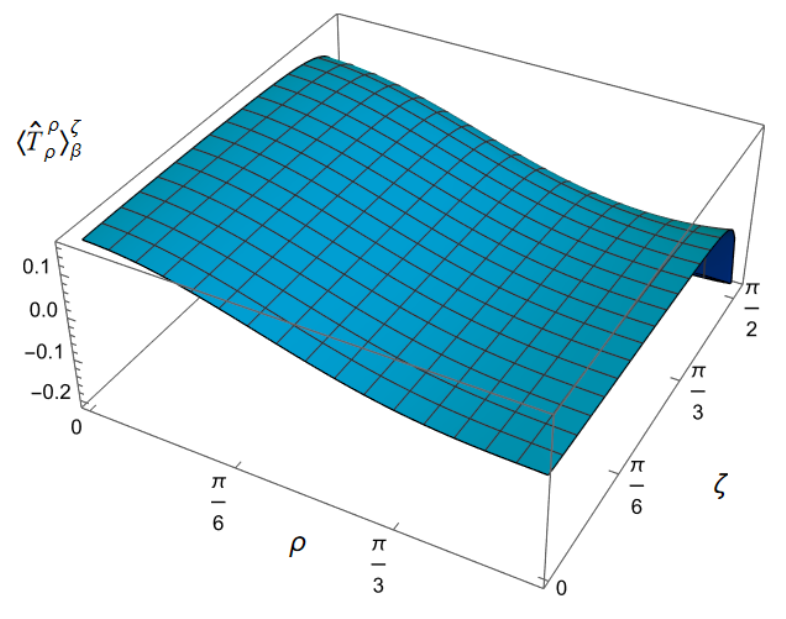}
       \\[0.3cm]
        \includegraphics[width=0.50\textwidth]{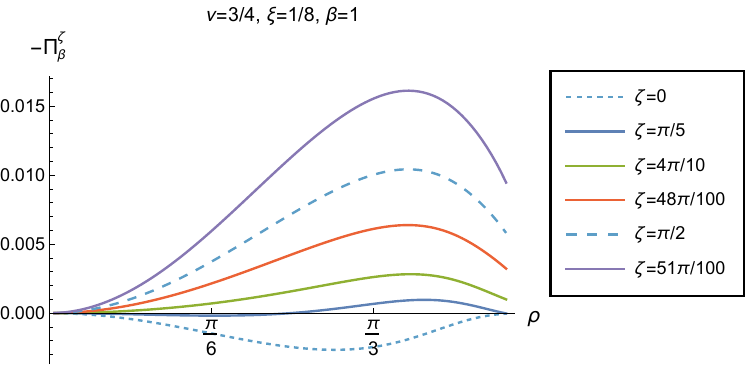}&
       \includegraphics[width=0.40\textwidth]{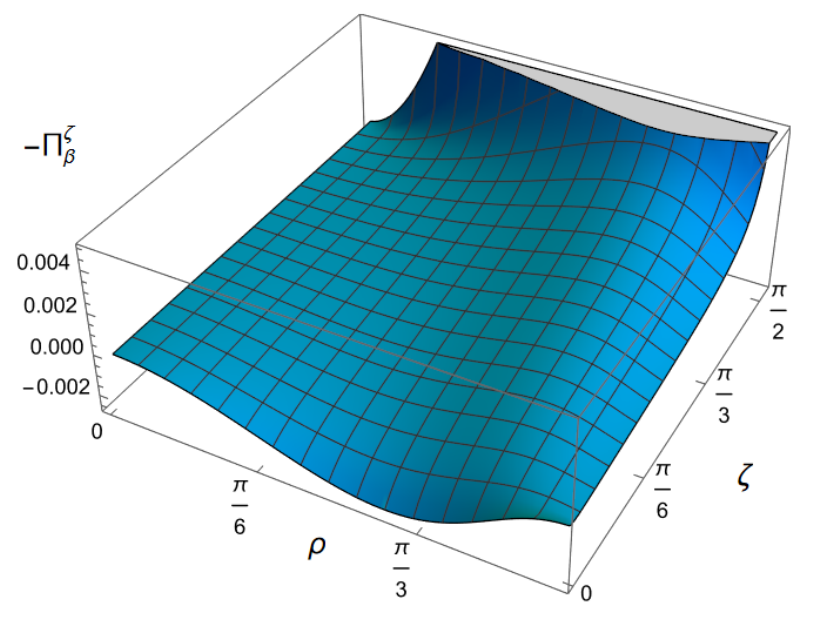}
       \end{tabular}
       \caption{Renormalized t.e.v.s of the RSET components $-\langle \hat{T}_\tau ^\tau \rangle _{\beta }^{\zeta }$ (top row), $\langle \hat{T}_\rho^\rho \rangle _{\beta }^{\zeta }$ (middle row) and negative pressure deviator $-\Pi_\beta^\zeta$ (bottom row), for   a quantum scalar field subject to Robin boundary conditions.   
       Plots in the left column show how these quantities depend on the radial coordinate $\rho $ for a selection of values of the Robin parameter $\zeta $,  whilst the right column shows the 3D surface plots of the quantities as functions of $\rho $ and $\zeta $. For all plots we have fixed the parameter $\nu=3/4$ (\ref{eq:nu}),  the coupling constant  $\xi=1/8$ (conformal coupling), and the inverse temperature $\beta=1$.}
       \label{fig:thermalRobinRSET}
          \end{center}
   \end{figure}

\begin{figure}[t]
     \begin{center}
     \begin{tabular}{cc}   
       \includegraphics[width=0.45\textwidth]{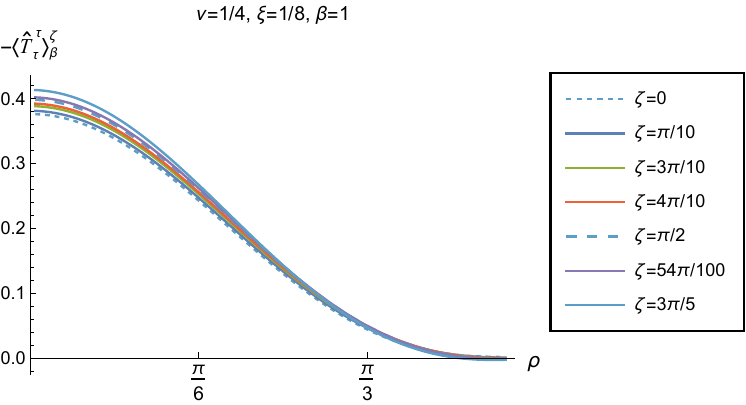} &
       \includegraphics[width=0.42\textwidth]{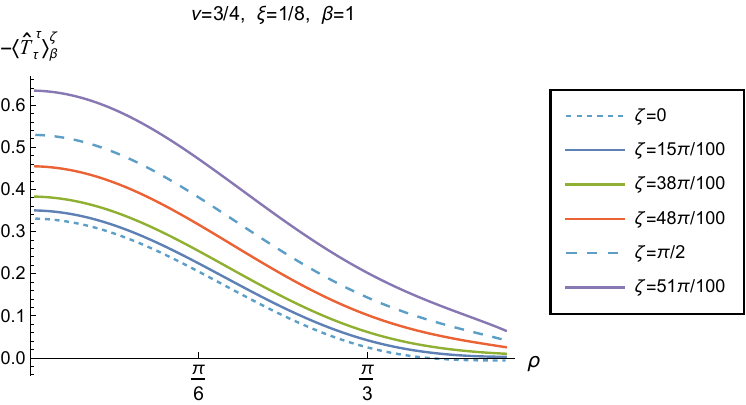}
        \\[0.3cm]
        \includegraphics[width=0.45\textwidth]{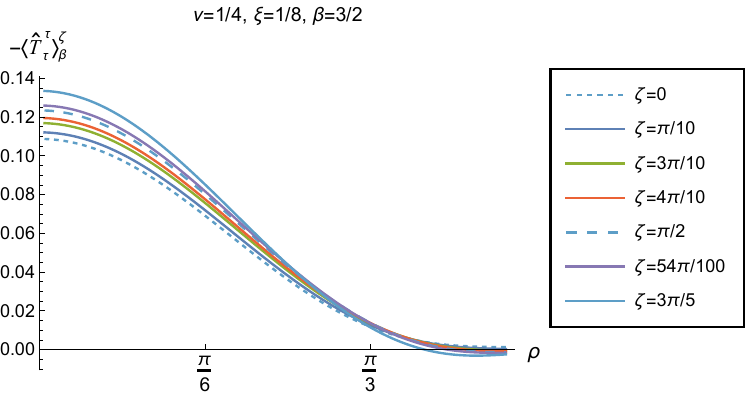} &
        \includegraphics[width=0.42\textwidth]{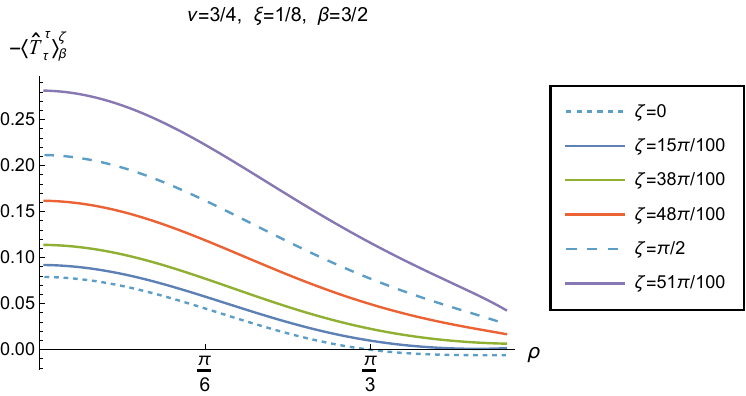}          
        \\[0.3cm]        
       \includegraphics[width=0.45\textwidth]{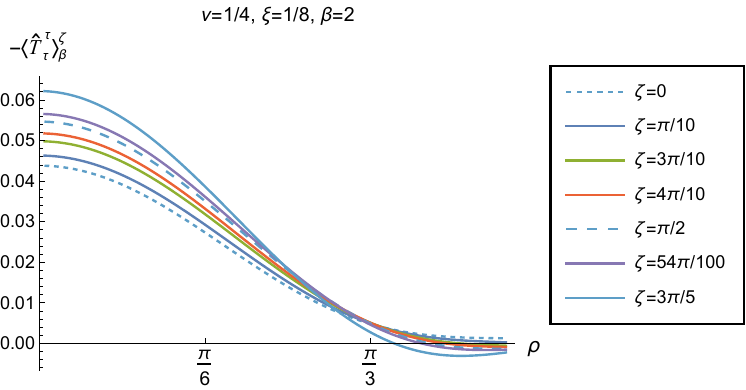} &
       \includegraphics[width=0.42\textwidth]{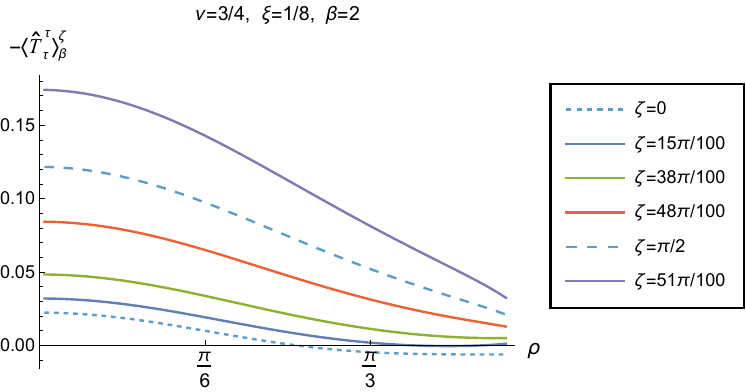}
        \\[0.3cm]
       \includegraphics[width=0.45\textwidth]{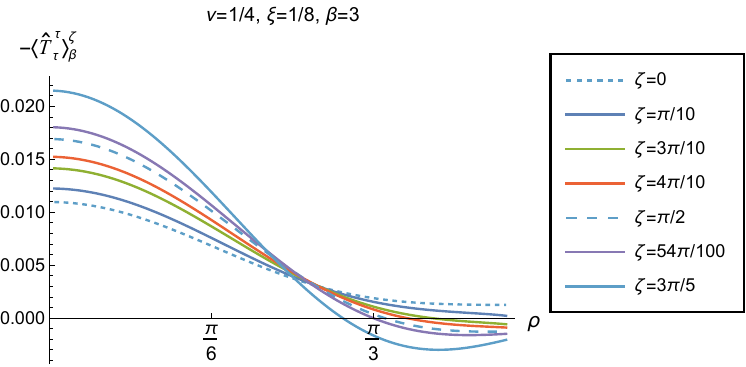} &
        \includegraphics[width=0.42\textwidth]{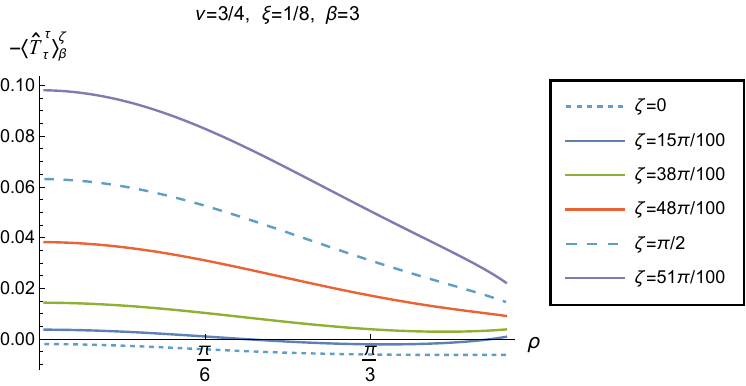}       
       \end{tabular}
       \caption{Renormalized t.e.v.s of  the energy density component of the RSET, $-\langle \hat{T}_\tau^{\tau} \rangle_\beta^\zeta$, for fixed coupling constant $\xi = 1/8$ and fixed parameter  $\nu=1/4$ (left column) and $\nu=3/4$ (right column).
       We consider a selection of values of the inverse temperature, $\beta = 1$ (top row), $\beta =3/2$ (second row), $\beta = 2$ (third row) and $\beta =3$ (bottom row). 
       In each case we show the profiles of the energy density as a function of the radial coordinate $\rho $ for a selection of values of the Robin parameter $\zeta $. }
       \label{fig:therm11_Rv14v34_x18}
          \end{center}
   \end{figure}

 \begin{figure}[t]
     \begin{center}
     \begin{tabular}{cc}
     \includegraphics[width=0.46\textwidth]{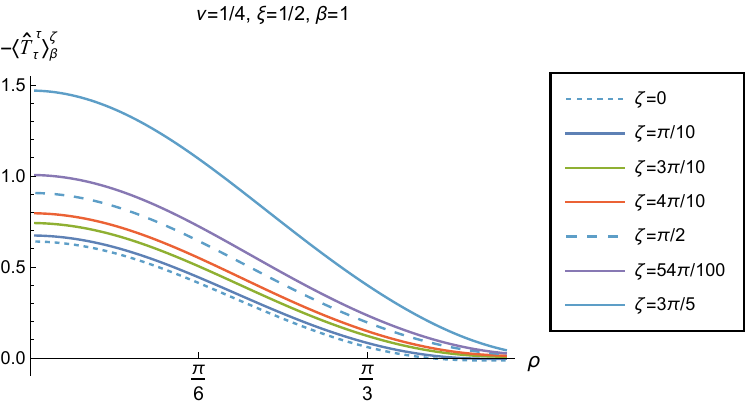} & 
        \includegraphics[width=0.46\textwidth]{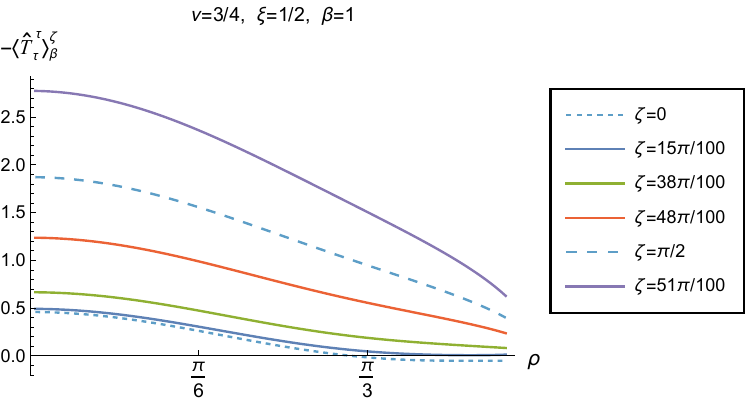}
              \\[0.4cm]
    \includegraphics[width=0.46\textwidth]{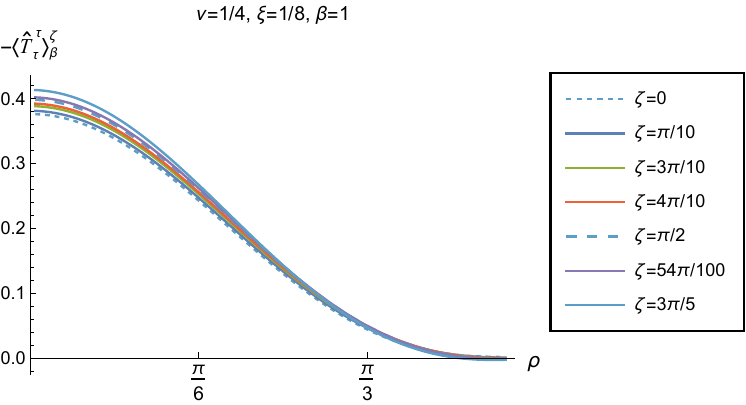} & 
        \includegraphics[width=0.46\textwidth]{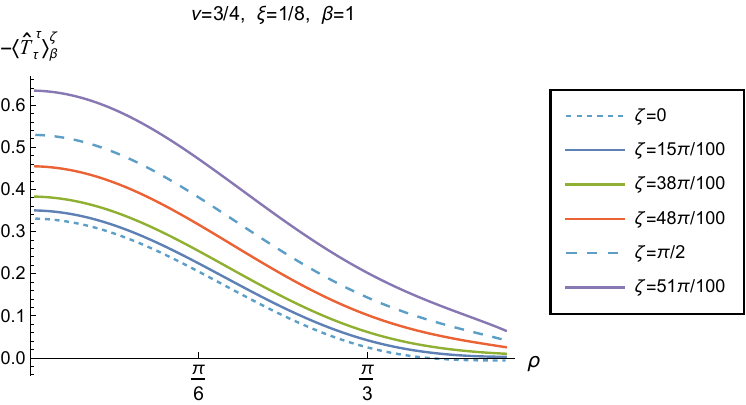}
              \\[0.4cm]
       \includegraphics[width=0.46\textwidth]{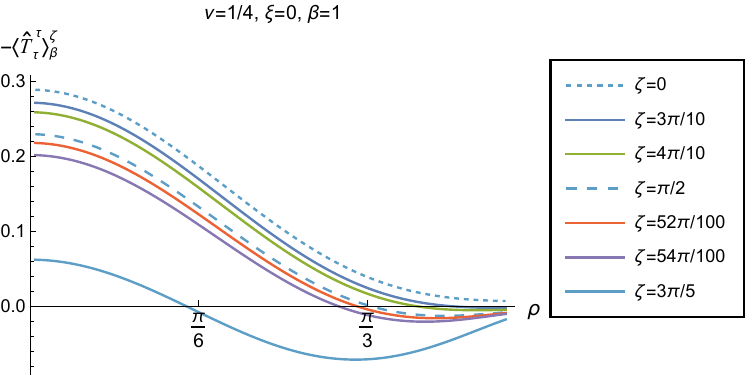} & 
        \includegraphics[width=0.46\textwidth]{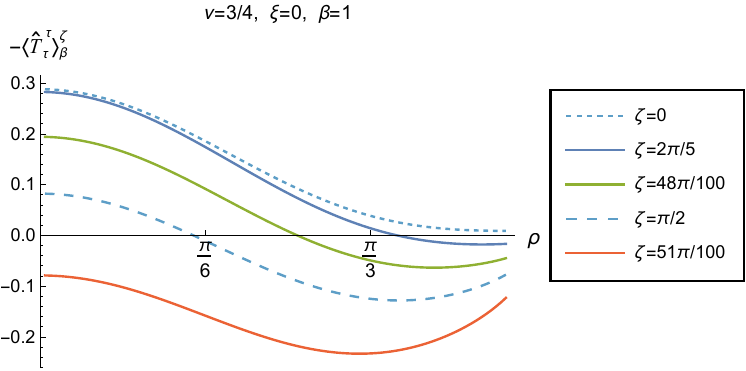}
        \\[0.4cm]
        \includegraphics[width=0.46\textwidth]{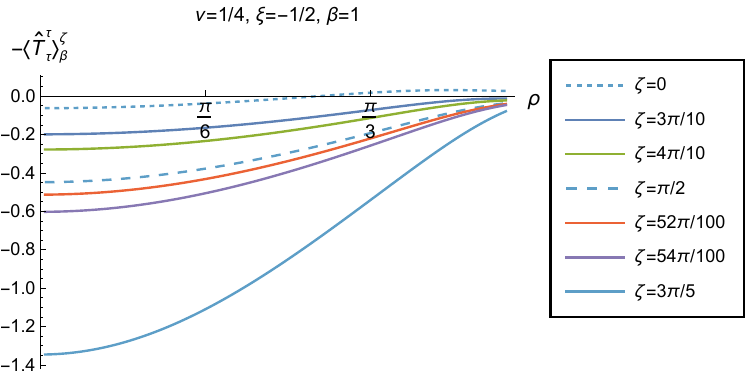} & 
        \includegraphics[width=0.46\textwidth]{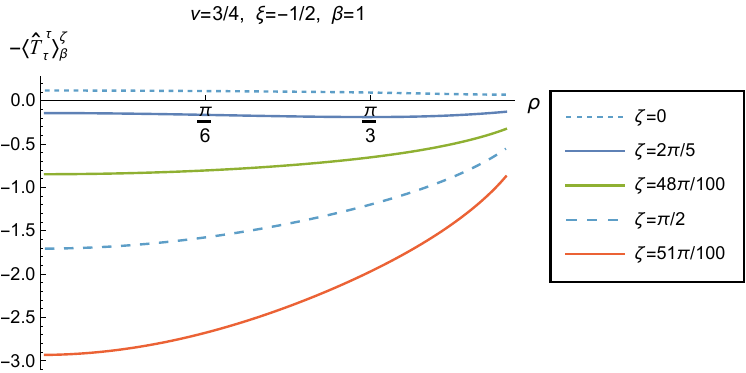}             
       \end{tabular}
       \caption{
       Renormalized t.e.v.s of  the energy density component of the RSET, $-\langle \hat{T}_\tau^{\tau} \rangle_\beta^\zeta$, for fixed inverse temperature $\beta = 1$ and fixed parameter  $\nu=1/4$ (left column) and $\nu=3/4$ (right column).
       We consider a selection of values of the coupling constant, $\xi = 1/2$ (top row), $\xi = 1/8$ (conformal coupling, second row), $\xi =0$ (minimal coupling, third row) and $\xi = -1/2$ (bottom row). 
       In each case we show the profiles of the energy density as a function of the radial coordinate $\rho $ for a selection of values of the Robin parameter $\zeta $.}
       \label{fig:therm11_Rv14_v34_b1}
          \end{center}
   \end{figure}
   
 \begin{figure}[t]
     \begin{center}
     \begin{tabular}{cc}    
       \includegraphics[width=0.46\textwidth]{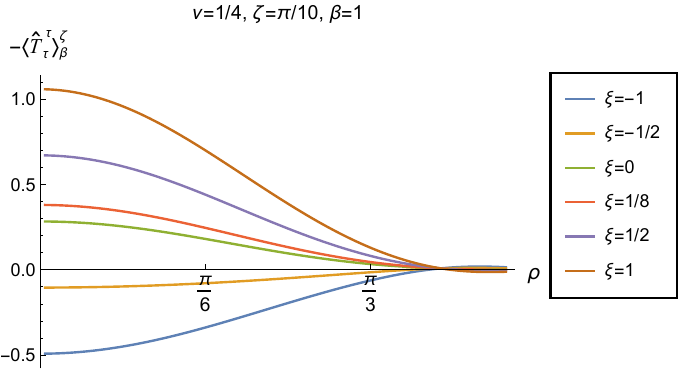} &
       \includegraphics[width=0.46\textwidth]{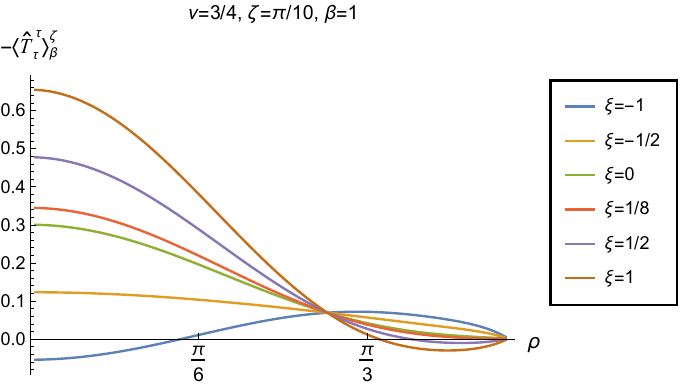}
       \\[0.3cm]
       \includegraphics[width=0.46\textwidth]{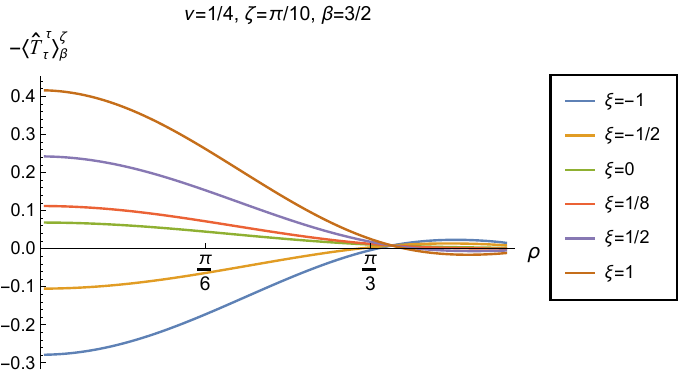} & 
       \includegraphics[width=0.46\textwidth]{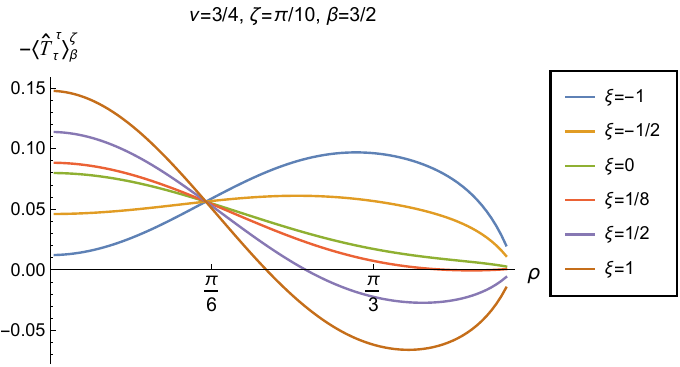}
        \\[0.3cm]              
     \includegraphics[width=0.46\textwidth]{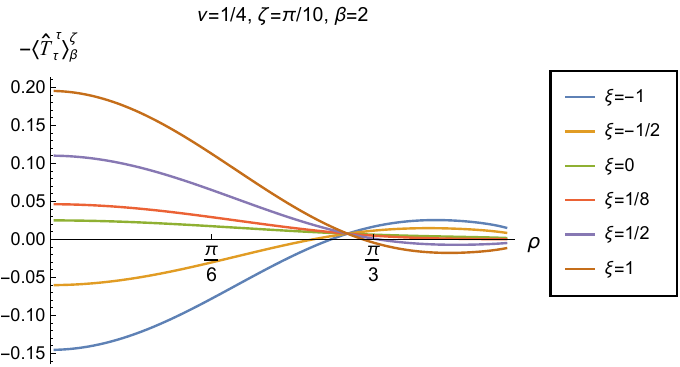} &
       \includegraphics[width=0.46\textwidth]{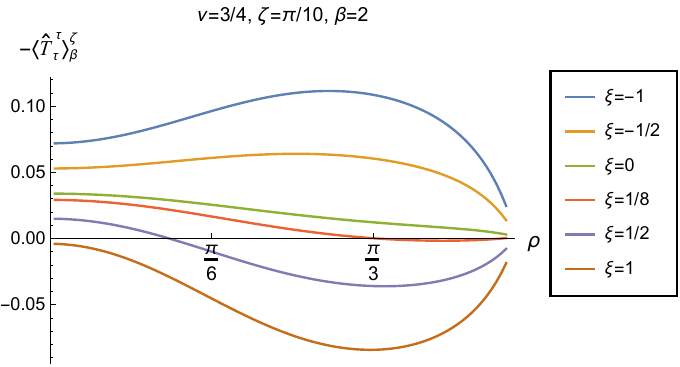}
      \\[0.3cm]
       \includegraphics[width=0.46\textwidth]{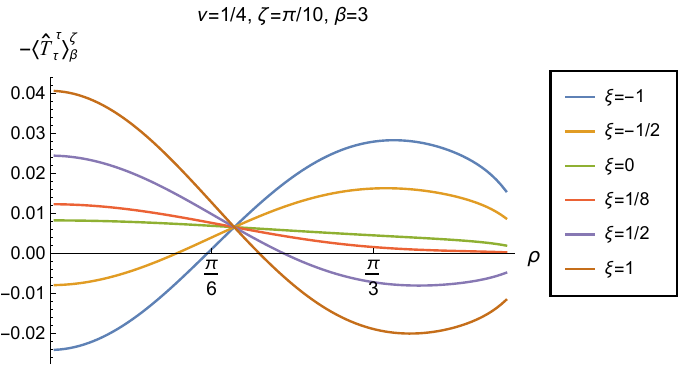} & 
       \includegraphics[width=0.46\textwidth]{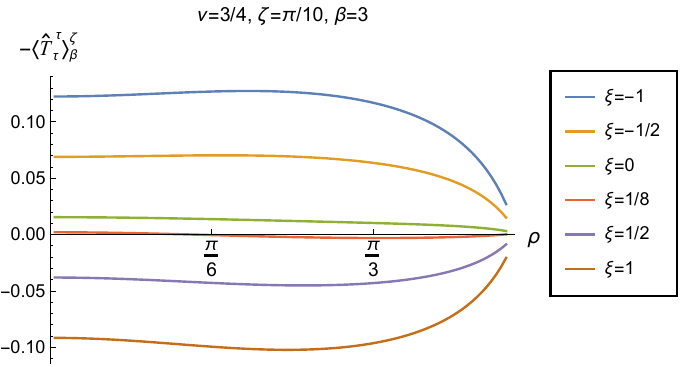}          
              \end{tabular}
       \caption{
       Renormalized t.e.v.s of  the energy density component of the RSET, $-\langle \hat{T}_\tau^{\tau} \rangle_\beta^\zeta$, for fixed Robin parameter $\zeta = \pi /10$ and fixed parameter  $\nu=1/4$ (left column) and $\nu=3/4$ (right column).
       We consider a selection of values of the inverse temperature, $\beta = 1$ (top row), $\beta =3/2$ (second row), $\beta = 2$ (third row) and $\beta =3$ (bottom row). 
       In each case we show the profiles of the energy density as a function of the radial coordinate $\rho $ for a selection of values of the coupling constant $\xi $.}
       \label{fig:therm11_Rv14_v34_z10}
          \end{center}
   \end{figure}
   
\begin{figure}[t!]
\begin{center}
     \begin{tabular}{cc}
    
       \includegraphics[width=0.47\textwidth]{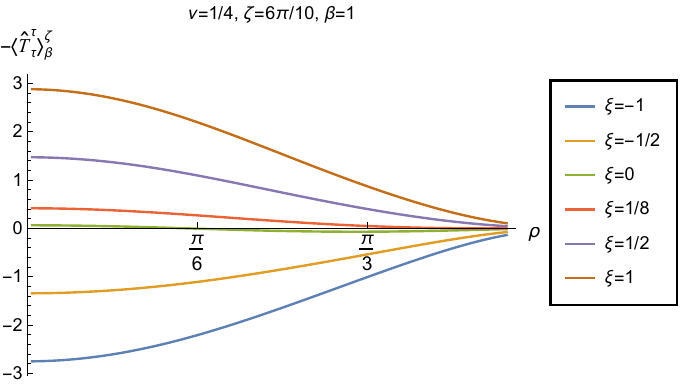} &
       \includegraphics[width=0.45\textwidth]{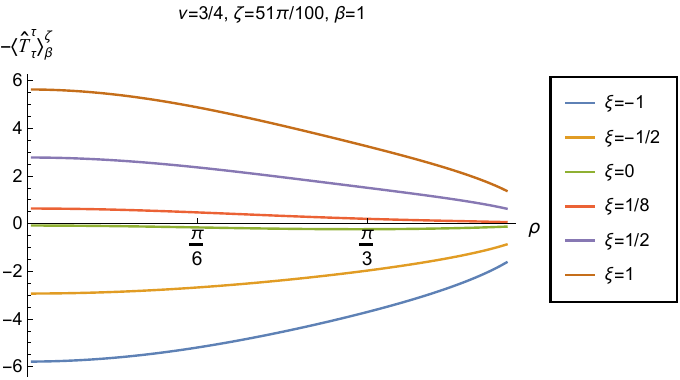}
       
        \\[0.3cm]
           \includegraphics[width=0.47\textwidth]{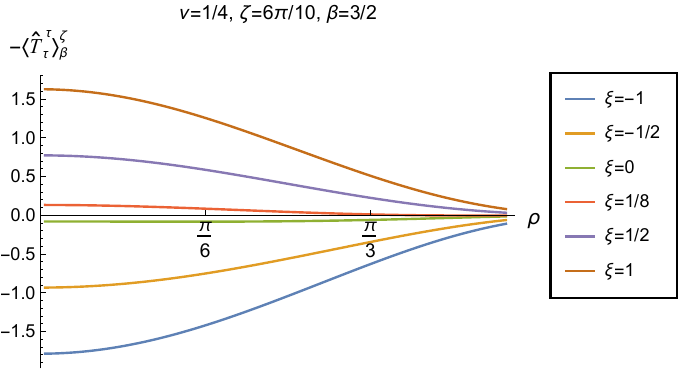} &
       \includegraphics[width=0.45\textwidth]{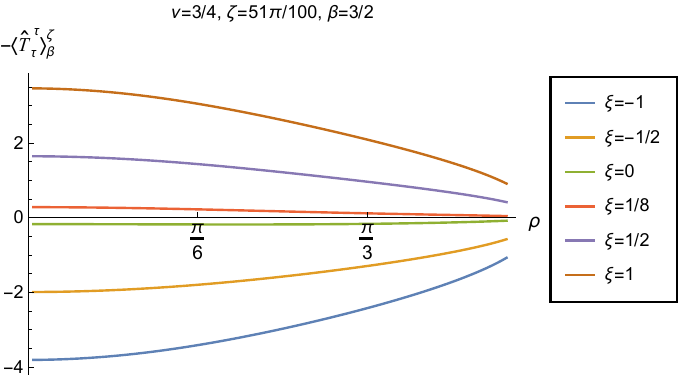}
       
        \\[0.3cm]    
     \includegraphics[width=0.47\textwidth]{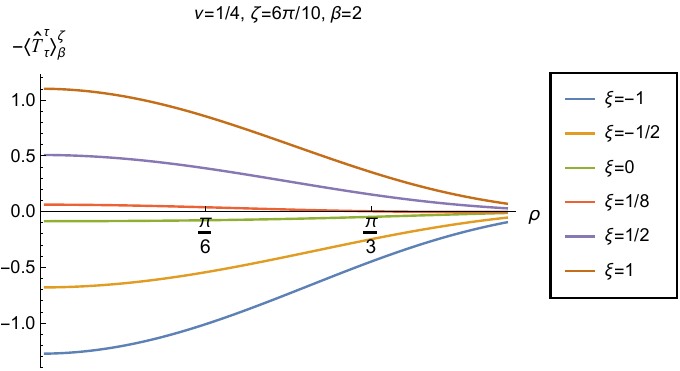} &
       \includegraphics[width=0.45\textwidth]{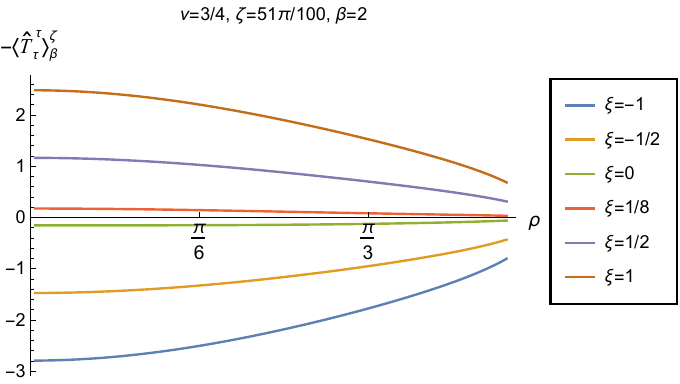}
      \\[0.3cm]
       \includegraphics[width=0.47\textwidth]{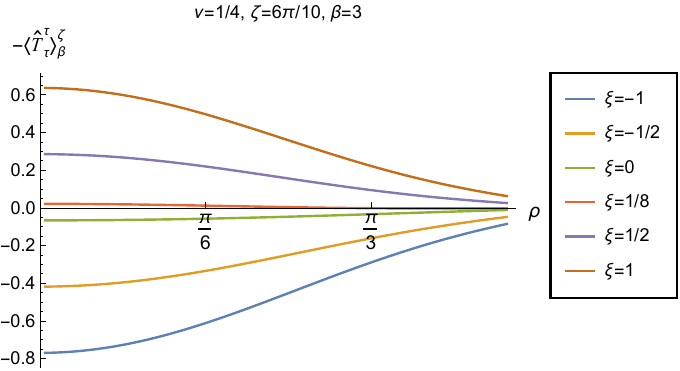} & \includegraphics[width=0.45\textwidth]{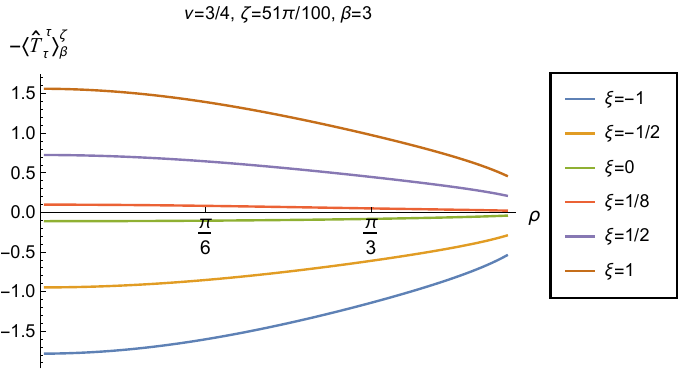}
          
              \end{tabular}
    \caption{
     Renormalized t.e.v.s of  the energy density component of the RSET, $-\langle \hat{T}_\tau^{\tau} \rangle_\beta^\zeta$, for fixed parameter  $\nu=1/4$ with Robin parameter $\zeta = 6\pi /10$ (left column) and $\nu=3/4$ with Robin parameter $\zeta = 51\pi /100$ (right column).
       We consider a selection of values of the inverse temperature, $\beta = 1$ (top row), $\beta =3/2$ (second row), $\beta = 2$ (third row) and $\beta =3$ (bottom row). 
       In each case we show the profiles of the energy density as a function of the radial coordinate $\rho $ for a selection of values of the coupling constant $\xi $.}
    \label{fig:therm11_Rv14_v34_zlarge}
    \end{center}
\end{figure}

 \begin{figure}[t!]
     \begin{center}
     \begin{tabular}{cc}    
       \includegraphics[width=0.47\textwidth]{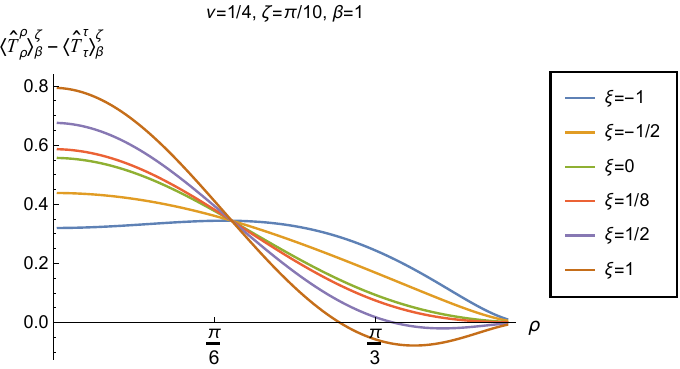} &
       \includegraphics[width=0.47\textwidth]{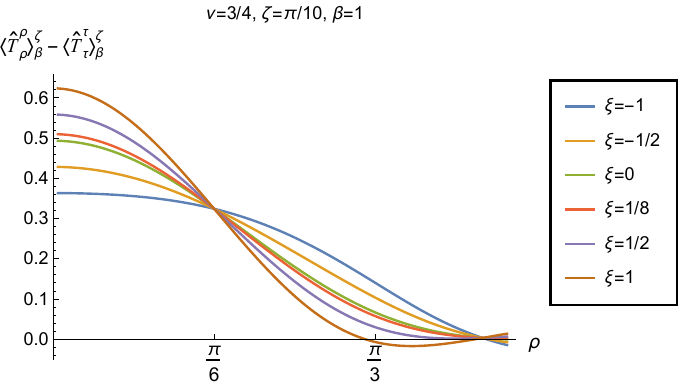}      
        \\[0.3cm]        
         \includegraphics[width=0.47\textwidth]{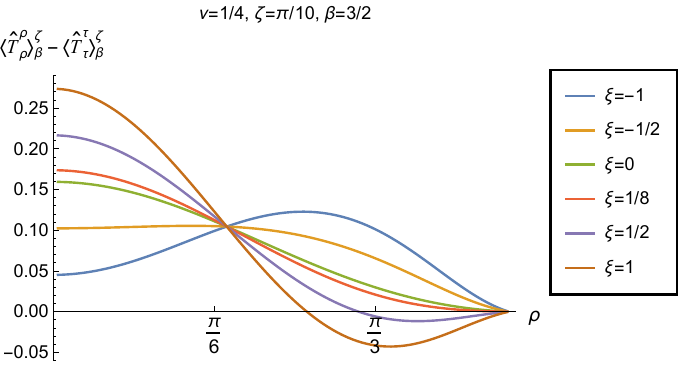} &
       \includegraphics[width=0.47\textwidth]{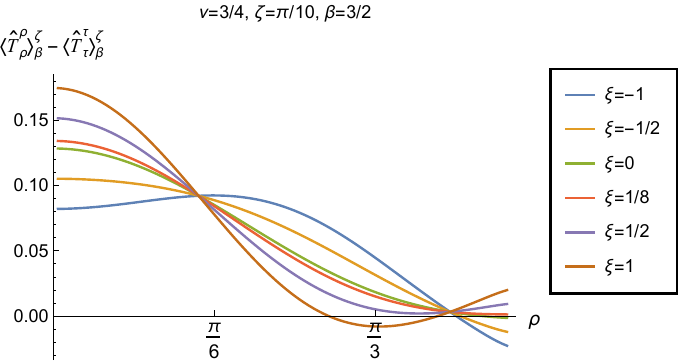}      
        \\[0.3cm]             
     \includegraphics[width=0.47\textwidth]{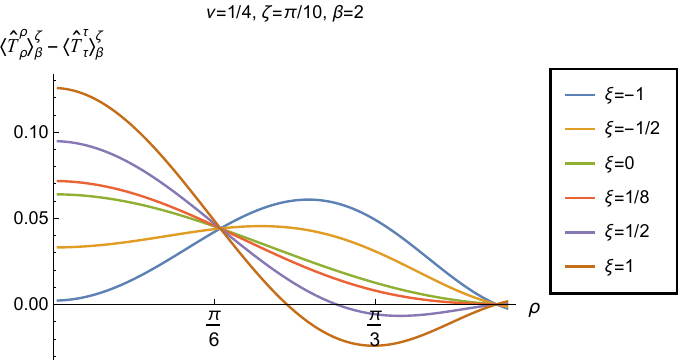} &
       \includegraphics[width=0.47\textwidth]{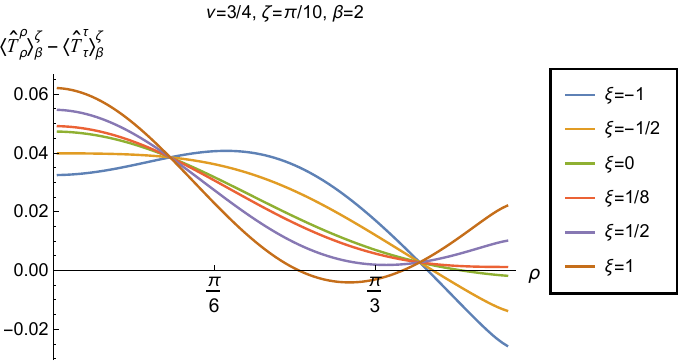}
      \\[0.3cm]
       \includegraphics[width=0.45\textwidth]{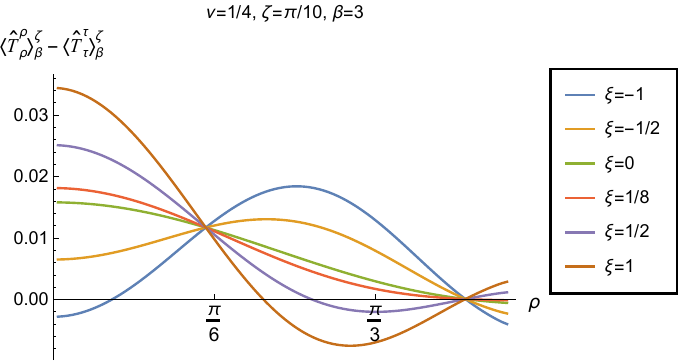} & \includegraphics[width=0.45\textwidth]{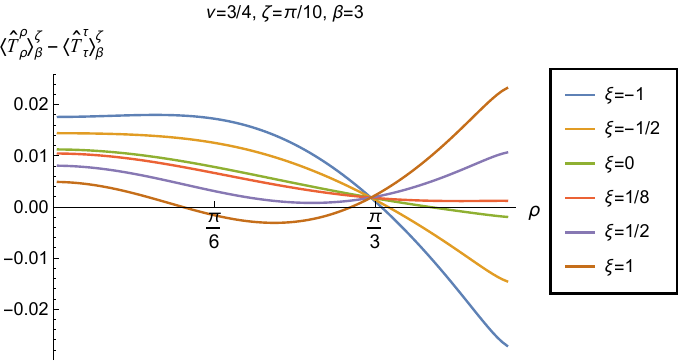}         
              \end{tabular}
       \caption{
       Renormalized t.e.v.s of the combination of RSET components $\langle {\hat {T}}_\rho^\rho\rangle_{\beta }^{\zeta }  -\langle {\hat {T}}_\tau ^\tau \rangle_{\beta }^{\zeta }$,  
       for fixed Robin parameter $\zeta = \pi /10$ and fixed parameter  $\nu=1/4$ (left column) and $\nu=3/4$ (right column).
       We consider a selection of values of the inverse temperature, $\beta = 1$ (top row), $\beta =3/2$ (second row), $\beta = 2$ (third row) and $\beta =3$ (bottom row). 
       In each case we show the profiles as functions of the radial coordinate $\rho $ for a selection of values of the coupling constant $\xi $.}
       \label{fig:NEC_v14v34z10}
          \end{center}
   \end{figure}

\begin{figure}[t!]
     \begin{center}
     \begin{tabular}{cc}
    
       \includegraphics[width=0.47\textwidth]{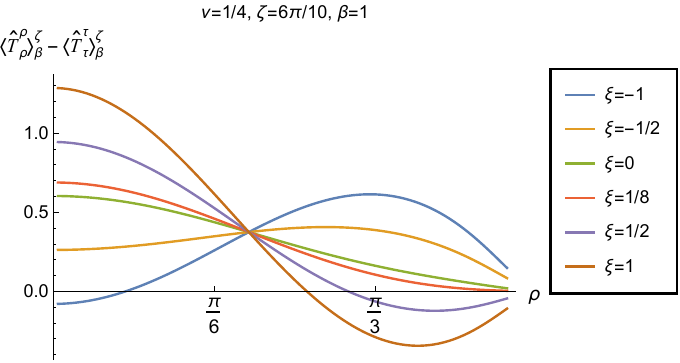} &
       \includegraphics[width=0.47\textwidth]{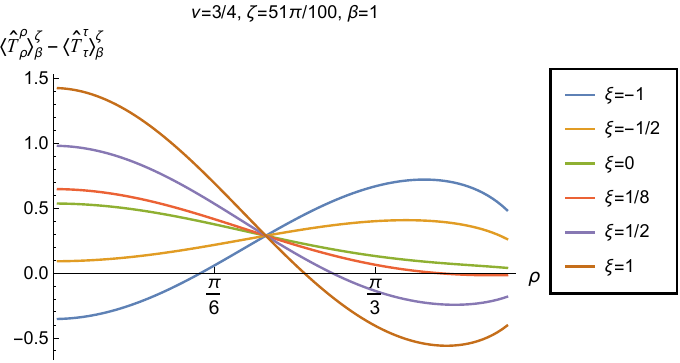}
       
        \\[0.3cm]
           \includegraphics[width=0.47\textwidth]{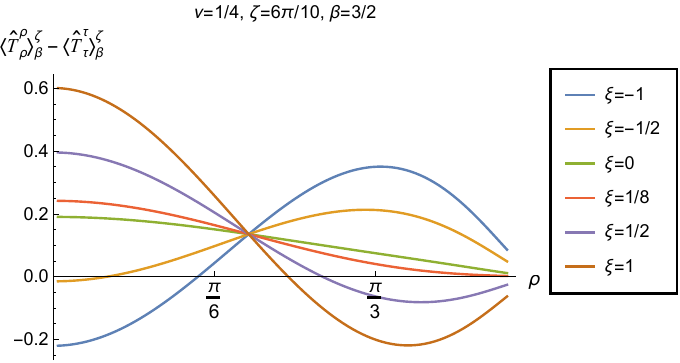} &
       \includegraphics[width=0.47\textwidth]{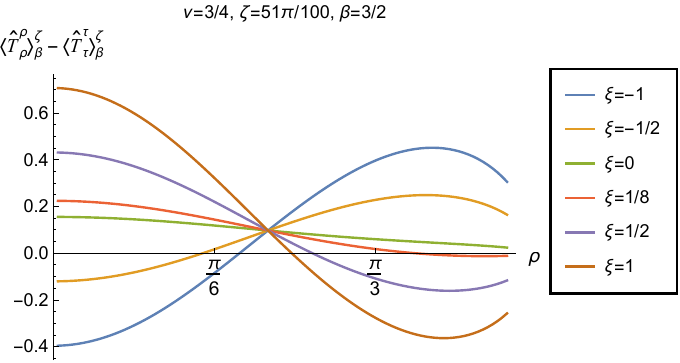}
       
        \\[0.3cm]    
     \includegraphics[width=0.47\textwidth]{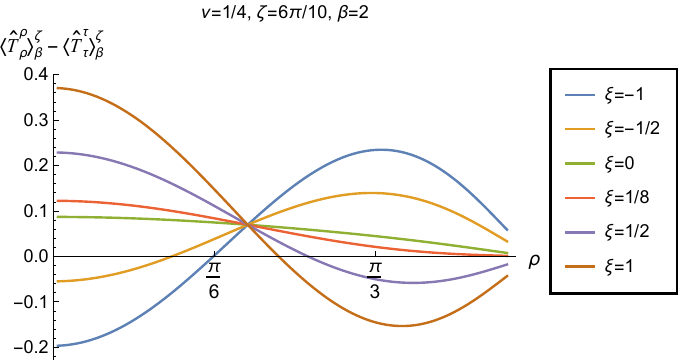} &
       \includegraphics[width=0.47\textwidth]{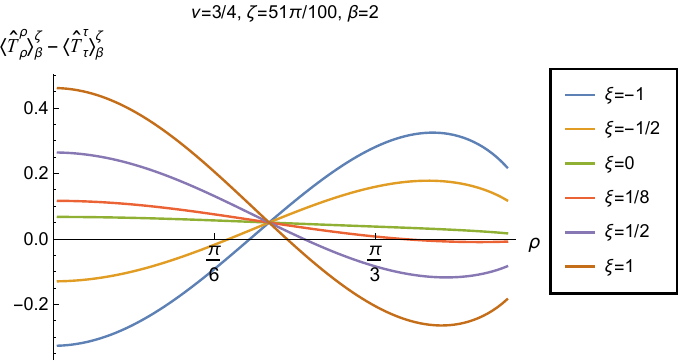}
      \\[0.3cm]
       \includegraphics[width=0.45\textwidth]{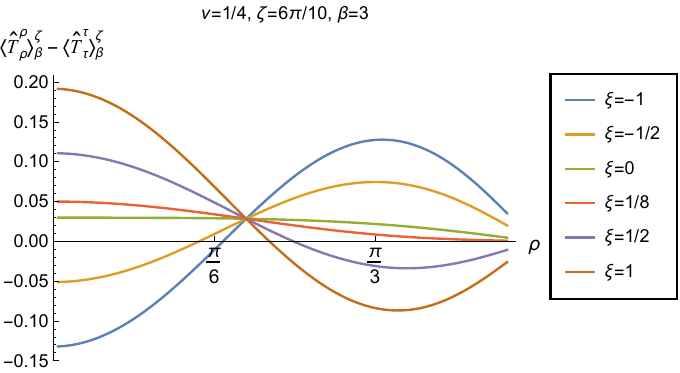} & \includegraphics[width=0.45\textwidth]{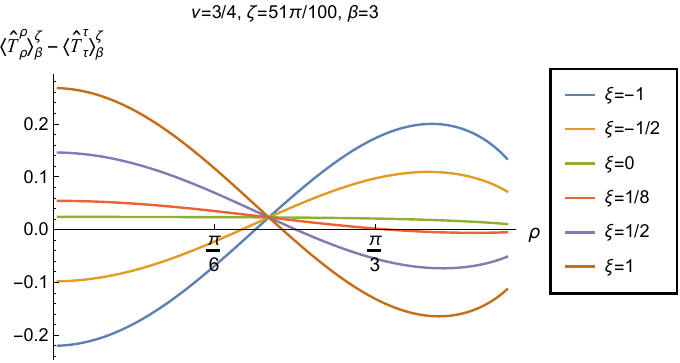}
          
              \end{tabular}
       \caption{
       Renormalized t.e.v.s of the combination of RSET components $\langle {\hat {T}}_\rho^\rho\rangle_{\beta }^{\zeta }  -\langle {\hat {T}}_\tau ^\tau \rangle_{\beta }^{\zeta }$,  for fixed parameter  $\nu=1/4$ with Robin parameter $\zeta = 6\pi /10$ (left column) and $\nu=3/4$ with Robin parameter $\zeta = 51\pi /100$ (right column).
       We consider a selection of values of the inverse temperature, $\beta = 1$ (top row), $\beta =3/2$ (second row), $\beta = 2$ (third row) and $\beta =3$ (bottom row). 
       In each case we show the profiles as functions of the radial coordinate $\rho $ for a selection of values of the coupling constant $\xi $.
       }
       \label{fig:NEC_v14v34zlarge}
          \end{center}
   \end{figure}

We begin, in Fig.~\ref{fig:thermalRobinRSET}, by fixing the inverse temperature $\beta =1$, the parameter $\nu = 3/4$ and the coupling constant $\xi = 1/8$, which corresponds to conformal coupling. 
For this value of $\nu $, the squared mass of the scalar field is positive.
Fig.~\ref{fig:thermalRobinRSET} shows how the t.e.v.s of the RSET components $-\langle \hat{T}_\tau ^\tau \rangle _{\beta }^{\zeta }$ (top row), $\langle \hat{T}_\rho^\rho \rangle _{\beta }^{\zeta }$ (middle row) and negative pressure deviator $-\Pi_\beta^\zeta$  (bottom row) depend on the Robin parameter $\zeta $, as well as the radial coordinate $\rho $.
For these values of the parameters in the theory, the energy density $-\langle \hat{T}_\tau ^\tau  \rangle _{\beta }^{\zeta }$
is positive everywhere throughout the space-time, and monotonically decreasing from its maximum at the origin. 
In contrast, while the component $\langle \hat{T}_\rho^\rho \rangle _{\beta }^{\zeta }$ also has a maximum at the origin (where it is positive), for sufficiently large values of $\zeta $ it becomes negative for sufficiently large $\rho $, beyond which it has a local minimum. 
As for Dirichlet and Neumann boundary conditions (see Fig.~\ref{fig:therm_press_devDN}), the pressure deviator is roughly an order of magnitude smaller than the RSET components. 
It vanishes at the origin and its magnitude has a maximum at some value of the radial coordinate $\rho $.
 
We now explore in some detail how the parameters affect the t.e.v.~of the energy density, $-\langle \hat{T}_\tau^{\,\tau} \rangle_\beta^\zeta$, in Figs.~\ref{fig:therm11_Rv14v34_x18}--\ref{fig:therm11_Rv14_v34_zlarge}.
In Fig.~\ref{fig:therm11_Rv14v34_x18} we fix the coupling constant $\xi = 1/8$ (corresponding to conformal coupling), consider two fixed values of the parameter in the scalar field equation, $\nu = 1/4$ (left-hand column) and $\nu = 3/4$ (right-hand column), and a selection of values of the inverse temperature $\beta $.
In each case we show the energy density profiles as functions of the radial coordinate $\rho $ for a selection of values of the Robin parameter $\zeta $.
The following key features may be gleaned from Fig.~\ref{fig:therm11_Rv14v34_x18}.
First, the curves for the different values of the Robin parameter $\zeta $ are less spread out for smaller values of the inverse temperature $\beta $. 
As observed for a massless, conformally-coupled scalar field on four-dimensional adS \cite{Morley:2023exv}, the effect of varying the boundary conditions is less significant at higher temperatures.
Second, varying the boundary conditions has a rather smaller effect on the energy density when $\nu =1/4$ than it does when $\nu = 3/4$.
Third, for higher temperatures, the energy density is positive throughout the space-time for all values of $\zeta $ considered, and is monotonically decreasing as the radial coordinate $\rho $ increases. 
This ceases to be the case for low temperatures, for which we observe that the energy density profile can become negative close to the boundary (and consequently the WEC is violated close to the boundary). 

Our main focus in this section is the effect of nonminimal coupling on the t.e.v.s of the RSET, and the consequences for the WEC and NEC.
In Fig.~\ref{fig:therm11_Rv14_v34_b1} we fix the inverse temperature $\beta =1$, and consider the same two values of the parameter $\nu $ as in Fig.~\ref{fig:therm11_Rv14v34_x18}, namely $\nu = 1/4$ and $\nu =3/4$.
For four selected values of the coupling constant, $\xi = 1/2$, $\xi =1/8$ (corresponding to conformal coupling), $\xi = 0$ (minimal coupling) and $\xi = -1/2$ we show how the energy density expectation values depend on the radial coordinate $\rho $ as the Robin parameter $\zeta  $ is varied.
We see that changing the Robin parameter has the least impact on the energy density profiles for conformal coupling, $\xi = 1/8$, particularly for the lower value of $\nu $.
For this coupling, and all values of $\zeta $ studied, the energy density is positive for all $\rho $ and monotonically decreasing as $\rho $ increases.
This is not the case for a minimally coupled scalar field.
While the energy density remains positive throughout the space-time for sufficiently small values of the Robin parameter $\zeta $, as $\zeta $ increases we find that the energy density becomes negative in a region close to the space-time boundary, and the size of this region increases as $\zeta $ increases. 
This is particularly marked for the larger value of $\nu $, for which the energy density is negative throughout the space-time for sufficiently large $\zeta $ (with $\xi = 0$).
When $\xi = -1/2$, the energy density is negative throughout the space-time for nearly all values of the Robin parameter, except for $\zeta $ close to zero.
We deduce that, depending on the coupling constant $\xi $ and the Robin boundary conditions applied, the WEC may be violated in a significant region of the adS space-time, and even for the whole of the space-time.

We explore the complex dependence of the energy density on the coupling constant $\xi $ and Robin parameter $\zeta $ further in Figs.~\ref{fig:therm11_Rv14_v34_z10} and \ref{fig:therm11_Rv14_v34_zlarge}.
In Fig.~\ref{fig:therm11_Rv14_v34_z10} we fix the Robin parameter to have a small nonzero value $\zeta = \pi /10$, for both values of the parameter $\nu $, while in Fig.~\ref{fig:therm11_Rv14_v34_zlarge} we choose values of $\zeta $ close to the critical value (\ref{eq:v_values}), namely $\zeta = 6\pi /10$ for $\nu = 1/4$ (left column), and $\zeta = 51 \pi /100$ for $\nu = 3/4$ (right column).
In each case we plot the energy density for a selection of values of the inverse temperature $\beta $, and show how the energy density as a function of the radial coordinate $\rho $ varies as the coupling constant $\xi $ varies. 

Considering first the plots in Fig.~\ref{fig:therm11_Rv14_v34_z10} for $\nu= 1/4$, we see that for all the values of the inverse temperature $\beta $ shown, there is a point where the curves for different $\xi $ intersect. 
This point moves closer to the origin as $\beta $ increases and the temperature decreases.
For nearly every value of $\xi $ studied (the only exceptions being minimal and conformal coupling), there is a region of the space-time for which the energy density is negative (and the WEC is violated). 
This is near the origin for sufficiently large and negative $\xi $, and near the space-time boundary for sufficiently large and positive $\xi $.
For this value  of $\nu $, we do not find any values of the Robin parameter (at the temperatures considered) for which the energy density is negative everywhere in the space-time.

Turning now to the plots in Fig.~\ref{fig:therm11_Rv14_v34_z10} for $\nu = 3/4$, a similar picture emerges at high temperatures (lower values of $\beta $), in particular there is a point at which the curves for different $\xi $ intersect. 
However, the regions of space-time for which the energy density is negative are smaller than those we find for $\nu = 1/4$ at the same inverse temperature, and the energy density is positive everywhere on the space-time except for large $|\xi |$ (larger than for $\nu = 1/4$).
At lower temperatures (higher values of $\beta $), the intersection point disappears and the size of the neighbourhood on which the energy density is negative increases. 
We also find negative energies for smaller values of  $\xi $ than at higher temperatures.
For both $\beta =2$ and $\beta = 3$, and sufficiently large and positive $\xi $, the energy density is negative throughout the space-time, and the WEC is violated.

When the Robin parameter $\zeta $ is close to the critical value (Fig.~\ref{fig:therm11_Rv14_v34_zlarge}) the profiles of the energy density have a much simpler dependence on the coupling constant $\xi $, and a consistent picture emerges.
At least for the values of $\xi $ that we study, it appears to be the case that the energy density is either positive everywhere throughout the space-time (for $\xi $ sufficiently positive) or negative everywhere (for $\xi $ sufficiently large and negative), independent of the value of $\nu $ or the inverse temperature $\beta $.
Comparing the results in Figs.~\ref{fig:therm11_Rv14_v34_z10} and \ref{fig:therm11_Rv14_v34_zlarge}, we also see that the magnitudes of the energy density for large $\zeta $ are considerably larger than for smaller $\zeta $.
This is due to the anticipated breakdown in the semiclassical approximation used here as the scalar field becomes classically unstable in the limit $\zeta \rightarrow \zeta _{\mathrm {crit}}$ (\ref{eq:v_values}).

Having found violations of the WEC in Figs.~\ref{fig:therm11_Rv14_v34_z10} and \ref{fig:therm11_Rv14_v34_zlarge},
we now, in Figs.~\ref{fig:NEC_v14v34z10} and \ref{fig:NEC_v14v34zlarge}, explore whether the NEC is satisfied.
We plot the t.e.v.s of the combination of RSET components $\langle {\hat {T}}_\rho^\rho\rangle_{\beta }^{\zeta }  -\langle {\hat {T}}_\tau ^\tau \rangle_{\beta }^{\zeta }$ (so that the NEC is violated when this is negative), and consider the same parameters in Figs.~\ref{fig:NEC_v14v34z10} and \ref{fig:NEC_v14v34zlarge} as in 
Figs.~\ref{fig:therm11_Rv14_v34_z10} and \ref{fig:therm11_Rv14_v34_zlarge} respectively.
The plots in Figs.~\ref{fig:NEC_v14v34z10} and \ref{fig:NEC_v14v34zlarge} share some common features. 
For example, the curves for different values of $\xi $ in each plot intersect at a particular value of $\rho $, an attribute we have seen in previous figures. 
However, some of the plots in Fig.~\ref{fig:NEC_v14v34z10}, for smaller Robin parameter $\zeta $, feature a second value of $\rho $ at which the curves intersect.
Thus, in this case, there are two points at which the terms in (\ref{eq:diff_operator3}) which are proportional to $\xi $ vanish. 

Notably, for each combination of the fixed parameters $\nu $, $\zeta $ and $\beta $ studied in Figs.~\ref{fig:NEC_v14v34z10} and \ref{fig:NEC_v14v34zlarge}, we find values of the coupling constant $\xi $ for which the NEC is violated. 
For $\nu = 1/4$ and $\zeta = \pi /10$ (left-hand plots in Fig.~\ref{fig:NEC_v14v34z10}), the NEC violations typically occur for large positive $\xi $ in a region close to the space-time boundary, although the situation for low temperature ($\beta = 3$, bottom row) is more complicated, with an additional small region near the origin for which the NEC is violated when $\xi = -1$.
For the larger value of $\nu =3/4$ (right-hand plots in Fig.~\ref{fig:NEC_v14v34z10}), violations of the NEC again occur near the space-time boundary, although there are also violations of the NEC closer to the origin at low temperature.
For larger values of $\beta $ (lower temperatures), the NEC is violated in a neighbourhood of the space-time boundary at $\rho = \pi /2$ for $\xi $ large and negative, whereas for $\xi $ large and positive, small violations of the NEC occur in a region between the origin and the space-time boundary.

When the Robin parameter $\zeta $ is close to its critical value (see Fig.~\ref{fig:NEC_v14v34zlarge}), a simpler picture emerges. 
For both $\nu = 1/4$ and $\nu = 3/4$, and all values of the inverse temperature $\beta $ considered, the NEC is violated in a neighbourhood of the space-time boundary for sufficiently large positive $\xi $, and in a neighbourhood of the origin for sufficiently large and negative $\xi $.
    
\section{Conclusions}
\label{sec:conc}

In this paper we have studied the vacuum and thermal expectation values of the RSET for a quantum scalar field of general mass and curvature coupling on global, three-dimensional adS space-time.
We have examined a region of the (coupling, mass) parameter space in which Robin boundary conditions can be applied to the scalar field. 
Our main focus was to explore whether the RSET satisfies the pointwise energy conditions, in particular the NEC and WEC.

We first reviewed the v.e.v.~of the RSET when either Dirichlet or Neumann boundary conditions are applied to the field. 
In this case the vacuum state preserves the maximal symmetry of the underlying space-time and the RSET is a constant multiple of the metric tensor.
This v.e.v.~can be absorbed into a renormalization of the cosmological constant via the semiclassical Einstein equations.

Introducing a nonzero temperature and/or considering boundary conditions other than Dirichlet or Neumann breaks the maximal symmetry and the RSET is no longer a multiple of the metric tensor. 
We find the v.e.v.s and t.e.v.s in these situations using a state-subtraction technique, since the divergences in the Green functions are state-independent.
Overall a complex picture emerges, with the RSET depending on the temperature of the state, the parameter $\zeta $ which fixes the Robin boundary conditions, as well as the quantity $\nu $ which parameterizes the scalar field equation and the constant $\xi $ describing the coupling between the Ricci scalar curvature and the scalar field.

The WEC is violated whenever the expectation value of the energy density is negative. 
We find this happens in a region of adS space-time for sufficiently large and negative $\xi $ in thermal states with Dirichlet or Neumann boundary conditions. 
When Robin boundary conditions are applied to the scalar field, it is possible for the energy density to be negative throughout the space-time, even in the vacuum state, although this depends strongly on the mass and coupling of the scalar field and the Robin parameter $\zeta $.
Since the RSET in the vacuum state with Robin boundary conditions is not a multiple of the metric, this negative energy density cannot be absorbed into a renormalization of the cosmological constant. 

The NEC can also be violated in a region of the space-time when we have a thermal state, subject to Dirichlet or Neumann boundary conditions, when $|\xi |$ is sufficiently large.  
For Robin boundary conditions, sufficiently large $|\xi |$ can also lead to a violation of the NEC in a region of the space-time in both the vacuum and thermal states.
The size of the region depends strongly on the remaining parameters, namely the inverse temperature, the mass of the scalar field, and the boundary conditions applied.
We have not found any combinations of these parameters for which the NEC is violated on the whole of adS space-time.

In one sense our main result, that the nonminimally-coupled quantum scalar field can violate the NEC (and hence also the WEC) is unsurprising, since there are classical configurations of a nonminimally-coupled scalar field violating the NEC \cite{Kontou:2020bta}.
Furthermore, it is well-known that the NEC is violated in general in quantum field theory \cite{Epstein:1965zza}, even on flat space-time.
However, it is worth emphasizing that the quantum states we consider in this paper are extremely simple, namely global vacuum and thermal equilibrium states on a maximally-symmetric space-time background. 
Such states on Minkowski space-time, even if the scalar field is nonminimally-coupled, satisfy both the WEC and NEC. 

In this paper we have only considered the pointwise NEC and WEC.  
In quantum field theory, it is {\em {averaged}} energy conditions which are more relevant \cite{Kontou:2020bta}. 
These involve integrating the RSET expectation value along a time-like (AWEC) or null (ANEC) geodesic, and state that the resulting integral should be positive.
Averaged energy conditions are much less stringent than the corresponding pointwise energy conditions; indeed, it is shown in \cite{Fewster:2006ti} that the ANEC will be satisfied by a classical scalar field on pure adS space-time if the coupling constant $\xi \in [0, 1/4]$. 
We do find (small) violations of the NEC for values of $\xi $ in this interval for the vacuum state with Robin boundary conditions applied (see Figs.~\ref{fig:vacECfixedzeta} and \ref{fig:vacECfixednu}), as well as larger violations for other values of $\xi $ and thermal states. 
On three-dimensional adS space-time, there are circular null geodesics at any fixed value of the radial coordinate $\rho _{0}$, having momentum/energy ratio equal to $\sin \rho _{0}$. 
Therefore, any violation of the pointwise NEC in a region of adS, as found in this paper, will result in a violation of the ANEC along such a null geodesic.

It would be very interesting to explore the backreaction of the RSET on the space-time geometry when the energy conditions are violated,  in particular to see whether the quantum field can support exotic geometries such as wormholes \cite{Barcelo:1999hq,Barcelo:2000zf}.
The quantum-corrected anti-de Sitter space-time resulting from the solution of the semiclassical Einstein equations (\ref{eq:SCEE}) sourced by the RSET of a quantum scalar field was studied recently in the case of a massless and conformally-coupled scalar field on four-dimensional adS \cite{Thompson:2024vuj}.
In that case the quantum-corrected metrics had a soliton interpretation \cite{Thompson:2024vuj}.
We intend, in future work, to extend the analysis of \cite{Thompson:2024vuj} to the RSET configurations studied in this paper.

\begin{acknowledgements}
We acknowledge IT Services at The University of Sheffield for the provision of services for High Performance Computing.
The work of E.W.~is supported by STFC grant number ST/X000621/1.
Data supporting this publication can be freely downloaded from the University of Sheffield research data repository at {\url {https://doi.org/10.15131/shef.data.28199933}}, under the terms of the Creative Commons Attribution (CC-BY) licence.
This version of the article has been accepted for publication, after peer review but is not the Version of Record and does not
reflect post-acceptance improvements, or any corrections. The Version of Record is available online at:
{\url {https://doi.org/10.1007/s10714-025-03411-3}}.
\end{acknowledgements}

\bibliographystyle{spphys}

\bibliography{rset.bib}

\begin{thebibliography}{10}
\providecommand{\url}[1]{{#1}}
\providecommand{\urlprefix}{URL }
\expandafter\ifx\csname urlstyle\endcsname\relax
  \providecommand{\doi}[1]{DOI \discretionary{}{}{}#1}\else
  \providecommand{\doi}{DOI \discretionary{}{}{}\begingroup
  \urlstyle{rm}\Url}\fi

\bibitem{Kontou:2020bta}
E.A. Kontou, K.~Sanders, {Energy conditions in general relativity and quantum
  field theory}, Class. Quant. Grav. \textbf{37}, 193001 (2020).
\newblock \doi{10.1088/1361-6382/ab8fcf}

\bibitem{Curiel:2014zba}
E.~Curiel, {A primer on energy conditions}, Einstein Stud. \textbf{13}, 43
  (2017).
\newblock \doi{10.1007/978-1-4939-3210-8_3}

\bibitem{Martin-Moruno:2017exc}
P.~Martin-Moruno, M.~Visser, {Classical and semi-classical energy conditions},
  Fundam. Theor. Phys. \textbf{189}, 193 (2017).
\newblock \doi{10.1007/978-3-319-55182-1_9}

\bibitem{Flanagan:1996gw}
E.E. Flanagan, R.M. Wald, {Does back reaction enforce the averaged null energy
  condition in semiclassical gravity?}, Phys. Rev. D \textbf{54}, 6233 (1996).
\newblock \doi{10.1103/PhysRevD.54.6233}

\bibitem{Barcelo:1999hq}
C.~Barcelo, M.~Visser, {Traversable wormholes from massless conformally coupled
  scalar fields}, Phys. Lett. B \textbf{466}, 127 (1999).
\newblock \doi{10.1016/S0370-2693(99)01117-X}

\bibitem{Barcelo:2000zf}
C.~Barcelo, M.~Visser, {Scalar fields, energy conditions, and traversable
  wormholes}, Class. Quant. Grav. \textbf{17}, 3843 (2000).
\newblock \doi{10.1088/0264-9381/17/18/318}

\bibitem{Decanini:2005eg}
Y.~Decanini, A.~Folacci, {Hadamard renormalization of the stress-energy tensor
  for a quantized scalar field in a general spacetime of arbitrary dimension},
  Phys. Rev. D \textbf{78}, 044025 (2008).
\newblock \doi{10.1103/PhysRevD.78.044025}

\bibitem{Magnano:1993bd}
G.~Magnano, L.M. Sokolowski, {On physical equivalence between nonlinear gravity
  theories and a general relativistic selfgravitating scalar field}, Phys. Rev.
  D \textbf{50}, 5039 (1994).
\newblock \doi{10.1103/PhysRevD.50.5039}

\bibitem{Faraoni:1998qx}
V.~Faraoni, E.~Gunzig, P.~Nardone, {Conformal transformations in classical
  gravitational theories and in cosmology}, Fund. Cosmic Phys. \textbf{20}, 121
  (1999)

\bibitem{Fliss:2023rzi}
J.R. Fliss, B.~Freivogel, E.A. Kontou, D.P. Santos, {Non-minimal coupling,
  negative null energy, and effective field theory}, SciPost Phys. \textbf{16},
  119 (2024).
\newblock \doi{10.21468/SciPostPhys.16.5.119}

\bibitem{Epstein:1965zza}
H.~Epstein, V.~Glaser, A.~Jaffe, {Nonpositivity of energy density in quantized
  field theories}, Nuovo Cim. \textbf{36}, 1016 (1965).
\newblock \doi{10.1007/BF02749799}

\bibitem{Avis:1977yn}
S.J. Avis, C.J. Isham, D.~Storey, {Quantum field theory in anti-de Sitter
  space-time}, Phys. Rev. D \textbf{18}, 3565 (1978).
\newblock \doi{10.1103/PhysRevD.18.3565}

\bibitem{Wald:1980jn}
R.M. Wald, {Dynamics in nonglobally hyperbolic, static space-times}, J. Math.
  Phys. \textbf{21}, 2802 (1980).
\newblock \doi{10.1063/1.524403}

\bibitem{Ishibashi:2003jd}
A.~Ishibashi, R.M. Wald, {Dynamics in nonglobally hyperbolic static
  space-times. 2. General analysis of prescriptions for dynamics}, Class.
  Quant. Grav. \textbf{20}, 3815 (2003).
\newblock \doi{10.1088/0264-9381/20/16/318}

\bibitem{Ishibashi:2004wx}
A.~Ishibashi, R.M. Wald, {Dynamics in nonglobally hyperbolic static
  space-times. 3. Anti-de Sitter space-time}, Class. Quant. Grav. \textbf{21},
  2981 (2004).
\newblock \doi{10.1088/0264-9381/21/12/012}

\bibitem{Dappiaggi:2016fwc}
C.~Dappiaggi, H.R.C. Ferreira, {Hadamard states for a scalar field in
  anti\textendash{}de Sitter spacetime with arbitrary boundary conditions},
  Phys. Rev. D \textbf{94}, 125016 (2016).
\newblock \doi{10.1103/PhysRevD.94.125016}

\bibitem{Benini:2017dfw}
M.~Benini, C.~Dappiaggi, A.~Schenkel, {Algebraic quantum field theory on
  spacetimes with timelike boundary}, Annales Henri Poincare \textbf{19}, 2401
  (2018).
\newblock \doi{10.1007/s00023-018-0687-1}

\bibitem{Dappiaggi:2017wvj}
C.~Dappiaggi, H.R.C. Ferreira, {On the algebraic quantization of a massive
  scalar field in anti-de-Sitter spacetime}, Rev. Math. Phys. \textbf{30},
  1850004 (2017).
\newblock \doi{10.1142/S0129055X18500046}

\bibitem{Dappiaggi:2018pju}
C.~Dappiaggi, H.R.C. Ferreira, B.A. Ju\'arez-Aubry, {Mode solutions for a
  Klein-Gordon field in anti\textendash{}de Sitter spacetime with dynamical
  boundary conditions of Wentzell type}, Phys. Rev. D \textbf{97}, 085022
  (2018).
\newblock \doi{10.1103/PhysRevD.97.085022}

\bibitem{Dappiaggi:2018xvw}
C.~Dappiaggi, H.~Ferreira, A.~Marta, {Ground states of a Klein-Gordon field
  with Robin boundary conditions in global anti\textendash{}de Sitter
  spacetime}, Phys. Rev. D \textbf{98}, 025005 (2018).
\newblock \doi{10.1103/PhysRevD.98.025005}

\bibitem{Gannot:2018jkg}
O.~Gannot, M.~Wrochna, {Propagation of singularities on AdS spacetimes for
  general boundary conditions and the holographic Hadamard condition}, J. Inst.
  Math. Jussieu \textbf{21}, 67 (2022).
\newblock \doi{10.1017/S147474802000002X}

\bibitem{Barroso:2019cwp}
V.S. Barroso, J.P.M. Pitelli, {Boundary conditions and vacuum fluctuations in
  ${\mathrm {AdS}}_4$}, Gen. Rel. Grav. \textbf{52}, 29 (2020).
\newblock \doi{10.1007/s10714-020-02672-4}

\bibitem{Morley:2020ayr}
T.~Morley, P.~Taylor, E.~Winstanley, {Quantum field theory on global anti-de
  Sitter space-time with Robin boundary conditions}, Class. Quant. Grav.
  \textbf{38}, 035009 (2021).
\newblock \doi{10.1088/1361-6382/aba58a}

\bibitem{Dappiaggi:2021wtr}
C.~Dappiaggi, A.~Marta, {Fundamental solutions and Hadamard states for a scalar
  field with arbitrary boundary conditions on an asymptotically AdS
  spacetimes}, Math. Phys. Anal. Geom. \textbf{24}, 28 (2021).
\newblock \doi{10.1007/s11040-021-09402-5}

\bibitem{Namasivayam:2022bky}
S.~Namasivayam, E.~Winstanley, {Vacuum polarization on three-dimensional
  anti-de Sitter space-time with Robin boundary conditions}, Gen. Rel. Grav.
  \textbf{55}, 13 (2023).
\newblock \doi{10.1007/s10714-022-03056-6}

\bibitem{Campos:2022byi}
L.d.S. Campos, C.~Dappiaggi, L.~Sinibaldi, {Hidden freedom in the mode
  expansion on static spacetimes}, Gen. Rel. Grav. \textbf{55}, 50 (2023).
\newblock \doi{10.1007/s10714-023-03099-3}

\bibitem{Morley:2023exv}
T.~Morley, S.~Namasivayam, E.~Winstanley, {Renormalized stress-energy tensor on
  global anti-de Sitter space-time with Robin boundary conditions}, Gen. Rel.
  Grav. \textbf{56}, 38 (2024).
\newblock \doi{10.1007/s10714-024-03224-w}

\bibitem{Allen:1985wd}
B.~Allen, T.~Jacobson, {Vector two point functions in maximally symmetric
  spaces}, Commun. Math. Phys. \textbf{103}, 669 (1986).
\newblock \doi{10.1007/BF01211169}

\bibitem{Allen:1986ty}
B.~Allen, A.~Folacci, G.W. Gibbons, {Anti-de Sitter space at finite
  temperature}, Phys. Lett. B \textbf{189}, 304 (1987).
\newblock \doi{10.1016/0370-2693(87)91437-7}

\bibitem{Kent:2014nya}
C.~Kent, E.~Winstanley, {Hadamard renormalized scalar field theory on
  anti\textendash{}de Sitter spacetime}, Phys. Rev. D \textbf{91}, 044044
  (2015).
\newblock \doi{10.1103/PhysRevD.91.044044}

\bibitem{Ambrus:2018olh}
V.E. Ambrus, C.~Kent, E.~Winstanley, {Analysis of scalar and fermion quantum
  field theory on anti-de Sitter spacetime}, Int. J. Mod. Phys. D \textbf{27},
  1843014 (2018).
\newblock \doi{10.1142/S0218271818430149}

\bibitem{Pitelli:2019svx}
J.P.M. Pitelli, {Comment on \textquotedblleft{}Hadamard states for a scalar
  field in anti\textendash{}de Sitter spacetime with arbitrary boundary
  conditions\textquotedblright{}}, Phys. Rev. D \textbf{99}, 108701 (2019).
\newblock \doi{10.1103/PhysRevD.99.108701}

\bibitem{Thompson:2024vuj}
J.C. Thompson, E.~Winstanley, {Quantum-corrected anti\textendash{}de Sitter
  spacetime}, Phys. Rev. D \textbf{110}, 125003 (2024).
\newblock \doi{10.1103/PhysRevD.110.125003}

\bibitem{Birrell:1982ix}
N.D. Birrell, P.C.W. Davies, \emph{{Quantum Fields in Curved Space}}.
\newblock Cambridge Monographs on Mathematical Physics (Cambridge University
  Press, Cambridge, UK, 1982).
\newblock \doi{10.1017/CBO9780511622632}

\bibitem{Tolman:1930zza}
R.C. Tolman, {On the weight of heat and thermal equilibrium in general
  relativity}, Phys. Rev. \textbf{35}, 904 (1930).
\newblock \doi{10.1103/PhysRev.35.904}

\bibitem{Tolman:1930ona}
R.~Tolman, P.~Ehrenfest, {Temperature equilibrium in a static gravitational
  field}, Phys. Rev. \textbf{36}, 1791 (1930).
\newblock \doi{10.1103/PhysRev.36.1791}

\bibitem{NamasivayamPhD}
S.~Namasivayam, Quantum scalar field theory on anti-de {S}itter space-time with
  {R}obin boundary conditions.
\newblock Ph.D. thesis, {School of Mathematical and Physical Sciences, The
  University of Sheffield} (2024).
\newblock \urlprefix\url{{https://etheses.whiterose.ac.uk/35978/}}

\bibitem{Breitenlohner:1982bm}
P.~Breitenlohner, D.Z. Freedman, {Positive energy in anti-de Sitter backgrounds
  and gauged extended supergravity}, Phys. Lett. B \textbf{115}, 197 (1982).
\newblock \doi{10.1016/0370-2693(82)90643-8}

\bibitem{Breitenlohner:1982jf}
P.~Breitenlohner, D.Z. Freedman, {Stability in gauged extended supergravity},
  Annals Phys. \textbf{144}, 249 (1982).
\newblock \doi{10.1016/0003-4916(82)90116-6}

\bibitem{Fewster:2006ti}
C.J. Fewster, L.W. Osterbrink, {Averaged energy inequalities for the
  non-minimally coupled classical scalar field}, Phys. Rev. D \textbf{74},
  044021 (2006).
\newblock \doi{10.1103/PhysRevD.74.044021}

\end{thebibliography}

\appendix
 
\section{Thermal expectation values with  Neumann/Dirichlet boundary conditions}
\label{sec:terms_DN}

In this appendix we explicitly present the terms arising in the RSET (\ref{eq:diff_operator3}) for thermal states with Dirichlet and Neumann boundary conditions applied.
Since the biscalar $W_{\beta }^{D/N}(x,x')$ (\ref{eq:GFoD_renthermal}) involves a sum over $j$, the components of the RSET will also be given as sums over $j$. 
Below we give the summands, with mixed indices.
We first define the following quantities:
\begin{align}
   X & =
    \nu \arccos \left[ 1+2\sec^2\rho\,\sinh^2 \left(\frac{j\beta}{2}\right) \right] ,
    \nonumber \\
Y & = 
\nu \arccos \left[  \cosh(j \beta) \sec^2 \rho -\tan^2\rho\right] ,\nonumber \\
\Upsilon & = 
2 i\sec\rho \left [ 1-2\cos (2\rho) + \cosh(2 j \beta) -4\cosh (j \beta) \sin^2\rho \right]^{1/2} ,
\end{align}
and shall also require the following hypergeometric functions:
\begin{align}
    {\mathcal{F}}_{1} &={}_2F_1 \left(1-\nu,1+\nu,\frac{3}{2};-\sec^2\rho \,\sinh^2\left[\frac{j\beta}{2}\right]\right) , \nonumber \\
{\mathcal {F}}_{2} &={}_2F_1 \left(2-\nu,2+\nu,\frac{5}{2};-\sec^2\rho \,\sinh^2\left[\frac{j\beta}{2}\right]\right) ,\nonumber \\
{\mathcal {F}}_{3} &= {}_2F_1 \left( 3-\nu,3+\nu,\frac{7}{2};-\sec^2 \rho \sinh^2\left[\frac{j\beta}{2}\right] \right) .
\end{align}
In the expressions below, the $+/-$ signs correspond to Neumann and Dirichlet boundary conditions respectively. 
\begin{align}
w  = & ~ \frac{1}{4 \pi L} \left \{ \pm\nu \mathcal{F}_1+ \frac{ \sqrt{2}\cos^2\rho\, \cos X \csch\left( \frac{j\beta}{2}\right)}{2 \sqrt{\cos (2\rho) + \cosh(j\beta)}} \right \},
\label{eq:expr_w}
\\
  w_t^{\,t} = & ~ \frac{i \sin ^2(2\rho) }{192 \pi L^3} 
  \Bigg \{\frac{3i \sqrt{2} \cos Y }{ \{\cos (2\rho) + \cosh (j \beta)\}^{3/2}}\text {cosech} \left( \frac{ j \beta}{2} \right) \left[\cos (2\rho) +2\cosh (j \beta) -1 \right]
  \nonumber \\
    & \qquad  -\frac{12 \nu\,\sin Y}{ [\cos(2\rho)+\cosh (j\beta)]}  \pm 8i \mathcal{F}_2 \nu \, (\nu^2-1)\sec^4\rho\sinh ^2\left( \frac{j \beta}{2}\right )  \Bigg \} 
    \nonumber \\
     & ~+ \frac{i \cos \rho\,\csch \left( \frac{j \beta}{2}\right ) }{768 \pi L^3 \{\cos (2\rho) + \cosh (j\beta)\}^4 } \Big [ 16 \cos^2 \rho + 2 \{2 +3 \cosh (j \beta)\}\cos (4\rho) 
     \nonumber \\
    & \qquad + 13 \cosh (j\beta) +4 \cosh (2j \beta) + 5 \cosh (3j \beta) +16 \{\cosh (j\beta) + \cosh (2 j\beta)\}\cos (2\rho) 
    \Big] 
    \nonumber \\
    & \quad \times \Bigg ( \pm 4 \Upsilon  \nu (\nu^2-1) \mathcal{F}_2 \{\cos (2\rho) +\cosh (j\beta)\}^{3/2} 
    \nonumber \\ & \qquad 
    + 24 \nu \cos^3\rho \,  \text{cosech} \left( \frac{j \beta}{2}\right ) \sin Y \{\cos (2\rho)+\cosh (j\beta)\} \nonumber \\
    & \qquad  + 3 \sqrt{2} \Upsilon \cos Y \cos^4 \rho\, \text{cosech}^3 \left( \frac{j \beta}{2}\right ) [\sin^2\rho - \cosh (j\beta)]
    \Bigg )  
    \nonumber \\
    & ~ - \frac{\cos^5\rho\, \sinh (j \beta)}{960 \pi L^3 \{ \cos (2\rho) + \cosh (j\beta)\}^2 \sinh^4 \left( \frac{j \beta}{2}\right )}  
    \nonumber\\
    &  \quad \times \Bigg \{30 \sqrt{2} \cos Y \sec \rho\, \cosh  \left( \frac{j \beta}{2}\right ) \{\cos (2\rho) + \cosh (j \beta)\}^{1/2} ( \cosh (j\beta)-\sin^2\rho)  
    \nonumber \\
    & \qquad - \frac{15 \Upsilon\, \nu \sin Y \cosh \left( \frac{j \beta}{2}\right )  }{\sqrt{\cos (2\rho) + \cosh (j\beta)}} \Big [ \cos (2\rho) + \cosh (j\beta) + 2 \sinh^2  \left( \frac{j \beta}{2}\right ) \Big ] 
    \nonumber \\
    & \qquad + \frac{60 \sqrt{2} \cos Y \cosh \left( \frac{j \beta}{2}\right )}{\sqrt{\cos (2\rho) + \cosh (j\beta)}}(\cosh (j\beta) -\sin^2\rho) \sec \rho \sinh^2 \left( \frac{j \beta}{2}\right ) 
    \nonumber \\
    & \qquad \pm 80 \mathcal{F}_2 \nu (\nu^2-1) \sec ^5(\rho ) \sinh ^3\left(\frac{j \beta }{2}\right) \cosh \left(\frac{j \beta}{2}\right) [\cosh (j \beta)+\cos (2 \rho )] [ \sin^2\rho-\cosh (j \beta )] 
    \nonumber \\
    & \qquad -15 i \sqrt{2} \Upsilon \cos Y \sinh (j \beta) + 60 i \nu \sec \rho \sin Y \sinh (j \beta) \Big[ \cos (2\rho) + 4 \sinh^2 \left( \frac{j \beta}{2}\right )\Big ] 
    \nonumber \\
    & \qquad \pm 15 i \Upsilon \mathcal{F}_2 \nu(\nu^2-1) \{\cos (2\rho) +\cosh(j \beta)\}^{3/2} \sec^4 \rho \sinh\left( \frac{j \beta}{2}\right ) \sinh(j \beta) 
    \nonumber \\
    & \qquad \pm 30 i \Upsilon \mathcal{F}_2 \nu(\nu^2-1) \sqrt{\cos (2 \rho) + \cosh(j \beta)} \sec^2 \rho \sinh^3\left( \frac{j \beta}{2}\right ) \sinh(j \beta)  
    \nonumber \\
    & \qquad  \pm 4 i \Upsilon \mathcal{F}_3 \nu (\nu^2-4)( \nu^2-1)\{\cos (2 \rho) + \cosh(j \beta)\}^{3/2} \sec^6 \rho\, \sinh^3\left( \frac{j \beta}{2}\right ) \sinh(j \beta)  
    \nonumber \\
& \qquad - \frac{120 i \nu^2 \cos Y[\cos (2\rho) + \cosh (j\beta)]}{\sqrt{1-\{\tan^2\rho-\cosh (j\beta) \sec^2 \rho\}^2}} \sec^3 \rho \sinh^2\left( \frac{j \beta}{2}\right ) \sinh j \beta 
    \Bigg \},
    \label{eq:expr_wuvtt}
\\
 w_\rho^{\, \rho} = & ~ \frac{i\sin^2(2 \rho)}{192 \pi L^3} \Bigg \{ 
 - \frac{3i \sqrt{2} \cos Y }{ \{\cos (2\rho) + \cosh (j \beta)\}^{3/2}}\text {cosech} \left( \frac{ j \beta}{2} \right) \left[\cos (2\rho) +2\cosh (j \beta) -1 \right]  
 \nonumber \\ & \qquad 
     + \frac{12  \nu  \sin Y}{\{\cos (2\rho) + \cosh (j \beta)\}} \pm 8i\mathcal{F}_2 \nu (\nu^2-1) \sec^4 \rho \sinh^2 \left(\frac{ j \beta}{2} \right) \Bigg \} 
     \nonumber \\ & ~
     -\frac{i\, \text{cosech} \left( \frac{j\beta}{2} \right) \cos^3\rho}{6144 \pi L^3 \{\cos (2\rho) + \cosh (j \beta)\}^4} \Bigg (-16[2 + 2\cos (2\rho) + \cos (4\rho)]  
     \nonumber \\
& \qquad      + 2 \cosh (j\beta)\, [17 -36\cos (2\rho)+ 6 \cos (4 \rho)  -\cos (6\rho)]  
\nonumber \\ & \qquad 
      +8 \cos (2 j \beta)\, [2 \cos (2\rho) -\cos (4\rho) -3] + 2 \cosh (3 j \beta)\, [1-3\cos (2\rho)]  \Bigg)  
      \nonumber \\ & \quad 
      \times \Bigg (\pm 8 \Upsilon \mathcal{F}_2 \nu ( \nu^2-1) \sec^4 \rho \, \{\cos (2\rho) + \cosh (j\beta)\}^{3/2} 
      \nonumber \\ & \qquad \qquad
      + 3 \sqrt{2} \Upsilon \cos Y \text{cosech} ^3 \left(\frac{j\beta}{2} \right ) \left [\cos 2\rho +2 \cosh j \beta -1 \right ]
      \nonumber \\ &  \qquad \qquad 
      - 48 \nu \sin Y \sec \rho \{\cos (2\rho) + \cosh (j \beta)\} \text{cosech} \left(\frac{j \beta}{2} \right )\Bigg )
      \nonumber \\ & ~
      - \frac{2i \sin^2 \rho \,\cos^5 \rho \,\, \text{cosech} ^2 \left(\frac{j\beta}{2} \right )}{1920 \pi L^3 \{\cos (2\rho) + \cosh (j\beta)\}^2 } 
      \nonumber \\ & \quad 
       \times \Bigg \{ \pm20 \Upsilon \mathcal{F}_2 \,\nu (\nu^2-1) \sec^6 \rho\, \sinh^3  \left(\frac{j\beta}{2} \right) \{\cos (2 \rho) + \cosh (j \beta)\}^{1/2} \Big [ 5 \cos (2 \rho) + 8 \cosh (j \beta) -3\Big ]  
       \nonumber \\ & \qquad 
      \pm 4 \sqrt{2} \nu ( \nu^2-1) \sec^8 \rho\, \sinh^4\left(\frac{j\beta}{2} \right ) \{\cos (2\rho) + \cosh (j\beta)\} 
      \nonumber \\ & \qquad \qquad \times 
      \Big[ -20i \sqrt{2} \mathcal{F}_2 \cos^3 \rho  
      +2 \sqrt{2} \Upsilon \mathcal{F}_3 (\nu^2-4) \sinh\left(\frac{j\beta}{2} \right ) \{\cos (2\rho) + \cosh (j\beta)\}^{1/2} \Big ] 
      \nonumber \\ & \qquad 
      +  15 \Bigg [ 2 i \Big( 2 \sqrt {2} \cos Y \{\cos (2\rho) + \cosh (j \beta)\}^{1/2} \sec \rho  
      + \Upsilon \nu \{\cos (2\rho) + \cosh (j \beta)\}^{1/2}  \sec^2 \rho \sin Y \Big )
      \nonumber \\ & \qquad \quad 
      + 2 \sec^2 \rho\, \sinh^2\left(\frac{j\beta}{2} \right )(-3 \sqrt{2}\Upsilon \cos Y + 8\nu \cos (2\rho) \sec \rho \sin Y)  
      \nonumber \\ & \qquad \quad 
      + \frac{4 i \sinh^3\left(\frac{j\beta}{2} \right )}{\sqrt{\cos (2\rho) + \cosh (j\beta)}} \Big [ 2\sqrt{2} \cos Y \sec \rho \Big( 1+ 2\nu^2 \cos (2\rho) \sec^2 \rho \Big) 
       + \Upsilon \nu \sec^2 \rho \sin Y \Big ]  
       \nonumber \\ & \qquad \quad 
       +32 \nu \sec^3 \rho \,\sin Y \sinh ^4\left(\frac{j\beta}{2} \right ) -2\sqrt{2} \Upsilon \cos Y \sec^2 \rho\, \sinh ^2\left(\frac{j\beta}{2} \right )  -2\sqrt{2} \Upsilon \cos Y  
       \nonumber \\ & \qquad \quad
      + 16 \nu \sec^3 \rho\, \sin Y \sinh ^2\left(\frac{j\beta}{2} \right ) \cosh (j \beta) + 
      \frac{16 \sqrt{2} i \nu^2 \cosh (j\beta) \sec^3 \rho\,   \cos Y \sinh ^3 \left(\frac{j\beta}{2} \right )}{ \sqrt{\cos (2\rho) + \cosh (j\beta)}}
      \Bigg]
      \Bigg \},
\\
  w_\theta^{\,\theta} = & ~ -\frac{i \cos^2\rho }{48 \pi L^3 } \Bigg \{ - \frac{6 i \sqrt{2} \csch \left(\frac{j\beta}{2}\right) \cos X \cos^2\rho}{\{\cos (2\rho) +\cosh (j\beta)\}^{3/2}} 
  \pm 8 i \mathcal{F}_2 \,\nu(\nu^2-1) \sinh^2\left (\frac{j\beta}{2}\right)\sec^4\rho  
  \nonumber \\ & \qquad 
    + \frac{12 \nu  \sin X}{ \{\cos (2\rho) +\cosh (j\beta )\}}    \Bigg \} 
     \nonumber \\
    & ~ - \frac{i\sin^2 (2 \rho) \csch^3 \left( \frac{ j \beta }{2}\right)}{128 \pi L^3 \{\cos (2\rho) + \cosh (j\beta)\}^{3/2}} \Bigg [ 
\cos X [\cosh (j\beta) -\sin^2 \rho] 
\nonumber \\ & \qquad  
+ \sqrt{2}i \nu \sin X \sinh \left(  \frac{j \beta}{2}\right) \sqrt{ \cos (2\rho) + \cosh (j\beta )} \, \Bigg ]
\pm \frac{ \mathcal{F}_2 \nu(\nu^2-1) }{12 \pi L^3},
    \label{eq:expr_wuvtheta}
\\
   w_{;t}^{\,\,;t} = & ~\frac{\sin (2\rho) }{8\pi L^3}\Bigg \{\frac{\sin (2\rho)}{4(\cos(2\rho)+\cosh (j \beta))^{3/2}}\Bigg[ 2\sqrt{2} 
 \cos X \left(\cosh(j \beta)-\sin^2 \rho \right ) \text{cosech} \left(\frac{j\beta}{2}\right)
 \nonumber \\ & \qquad
    + 4 i \nu \sqrt{\cos (2\rho) + \cosh(j\beta)}\,\sin X \Bigg ] 
  \pm \frac{4}{3}\nu(\nu^2-1) \mathcal{F}_2  \sec^2\rho\, \tan\rho\, \sinh^2 \left (\frac{j\beta}{2} \right ) \Bigg\},
  \label{eq:expr_wcovuvtt}
\\
    w_{;\rho}^{\,; \rho} = & ~
    \frac{\cos^2 \rho}{960\pi L^3} \Bigg \{- \frac{15\sqrt{2}\,\cos X\, \csch \left (\frac{j\beta}{2} \right )  }{(\cos (2\rho) +\cosh (j\beta))^{5/2}}\Big(15\cos (2\rho) -10 +2\cos (4\rho) +\cos (6\rho) \nonumber \\
    & \qquad + 8\{\cosh(2 j \beta)-\cosh(j \beta)\}\cos(2\rho) 
    + 4\{5+ \cos (4\rho)\}\cosh(j \beta) \Big ) 
    \nonumber \\
    & \qquad  - \frac{120 i \nu \Big ( 7+ 4(2\cosh j\beta -1)\cos (2\rho)  + \cos (4\rho) -4\cosh (j\beta)\Big) \sin X  }{[\cos (2\rho) +\cosh (j\beta) ]^2} 
    \nonumber \\ & \qquad 
    \pm 160 \nu (\nu^2-1)\mathcal{F}_2 \,(3\cos (2\rho) -5) \, \sec^4 \rho\, \sinh^2 \left ( \frac{j\beta}{2}\right )  + \frac{960 \sqrt{2}\, \nu ^2 \cos X \sin^2 \rho \,\sinh \left ( \frac{j\beta}{2}\right )}{ [\cos (2\rho) +\cosh (j\beta)]^{3/2}}   
    \nonumber \\
     & \qquad   + \frac{60 \sin (2\rho) \tan \rho}{[\cos (2\rho) + \cosh (j\beta)]^{3/2}} \Bigg[ 2\sqrt{2}  \cos X [\cosh (j\beta)-\sin^2\rho]\text{cosech} \left(\frac{j\beta}{2}\right)
     \nonumber \\ & \qquad \qquad 
     + 4 i \nu \sqrt{\cos (2\rho) + \cosh (j\beta)} \sin X \Bigg] 
     \nonumber \\
     & \qquad  \pm 256 \nu(\nu^2-4)(\nu^2-1)\mathcal{F}_2 \,\sec^4 \rho\,\tan^2\rho\, \sinh^4 \left( \frac{j\beta}{2}\right) 
    \Bigg \},
    \label{eq:expr_wcovuvrr}
\\
    w_{;\theta}^{ \,\,;\theta} = & ~
    \frac{\cot\rho}{4 \pi L^3}\Bigg \{ -\frac{\sin(2\rho)}{4\{\cos (2\rho) + \cosh (j \beta)\}^{3/2}}\Bigg [ 2\sqrt{2} \cos X \{\cosh (j\beta)-\sin^2\rho\}\text{cosech}\left(\frac{j\beta}{2}\right) \nonumber \\
    & \qquad + 4 i \nu \sqrt{\cos (2\rho) + \cosh (j \beta)}\,  \sin X\bigg] \pm \frac{4}{3} \nu (\nu^2-1)\mathcal{F}_2 \sec^2 \rho\, \sinh^2 \left( \frac{j\beta}{2}\right) \tan \rho
    \Bigg \},
    \label{eq:expr_wcovuvtheta}
\\
\Box w = & -\frac{\cos^2\rho\,\cot\rho}{240 \pi  L^3}\Bigg \{\frac{15 \cos X  \sin (2\rho)}{\sqrt{2}\,\{\cos (2\rho) +\cosh (j\beta)\}^{5/2}}   
\bigg(-1 +4\cos(2\rho) + \cos (4\rho)+ 4\cosh (2 j \beta) 
\nonumber \\ & \qquad \qquad 
+ 8\cosh(j\beta) \cos^2\rho \bigg) \nonumber \\
 & ~ \pm 40 \nu(\nu^2-1)\mathcal{F}_2 \sec^5\rho \, \left[ \sin (3\rho)-7 \sin \rho \right] \sinh^2 \bigg( \frac{j\beta}{2}\bigg) 
 \nonumber \\ & \qquad
 + \frac{30 i \nu\, [7+ \cos (4\rho) +8\cosh(j\beta) \cos(2\rho)] \sin X \tan\rho }{\{\cos(2\rho)+\cosh (j\beta)\}^2}\nonumber \\
 & \qquad  -\frac{240 \sqrt{2}\nu^2\,\cos X \sin^2\rho\, \tan\rho}{  \{\cos (2\rho)+\cosh(j\beta) \}^{3/2}} \sinh \left( \frac{j\beta}{2}\right) 
 \nonumber \\ & \qquad
 \pm 64 \nu(\nu^2-4)(\nu^2-1) \mathcal{F}_3 \, \sec^4\rho \,\tan^3\rho\sinh^4 \bigg( \frac{j\beta}{2}\bigg)
 \Bigg \}.
 \label{eq:expr_box}
\end{align}

\section{Vacuum expectation values with Robin boundary conditions}
\label{sec:terms_Robin_vac}

In this appendix we explicitly present the terms arising in the RSET (\ref{eq:diff_operator3}) for vacuum states with Robin boundary conditions applied.
Quantities appearing in the RSET (\ref{eq:diff_operator3}) are derived from the biscalar $W_{0}^{\zeta }(x,x')$ (\ref{eq:Euclid_vac_Green_diff}), which involves an integral over $\omega $ and a sum over $\ell $.
Below, using mixed space-time indices, we give the quantities in the RSET which are to be integrated/summed over. 

We first define the auxiliary quantities
\begin{align}
    \mathcal{A} = & ~ 1+\ell^2 +2\nu +\nu^2 +2|\ell |(1+\nu)+\omega^2 ,
    \nonumber \\
    \mathcal{J} = & ~ -4 \nu F_2 \, \mathcal{N}^\zeta_{\omega \ell}  \cos\zeta (\cos \rho)^{2\nu} -\frac{F_4  \, \mathcal{N}^\zeta_{\omega \ell}}{(1+\nu)}|1+|\ell |+\nu + i \omega|^2 \cos\zeta (\cos\rho)^{2+2\nu} 
    \nonumber \\ & \qquad 
    -\frac{F_5}{(\nu-1)}|1+|\ell |-\nu+ i\omega|^2 (\mathcal{N}^N_{\omega \ell}-\mathcal{N}^\zeta_{\omega \ell}\sin\zeta) \cos^2\rho ,
    \nonumber \\
    \mathcal{C} = & ~ |\ell |-1-\nu +(1+|\ell |+\nu)\cos 2\rho , \nonumber \\
     \mathcal{K}= & ~\mathcal{N}^\zeta_{\omega \ell}\cos\zeta \,\left( \cos\rho \right) ^{2\nu} F_2+F_3\,(\mathcal{N}^\zeta_{\omega \ell}\,\sin\zeta-\mathcal{N}^N_{\omega \ell}),
     \label{eq:vac_auxiliary}
    \end{align}
    where the normalization constant $\mathcal{N}^{\zeta }_{\omega \ell}$ is given in (\ref{eq:Nconstantzeta}) and  $\mathcal{N}^{N}_{\omega \ell}$ is obtained by setting $\zeta = \pi /2$ (corresponding to Neumann boundary conditions) in (\ref{eq:Nconstantzeta}). We will also require the following hypergeometric functions:
    \begin{align}
    F_1 &= {}_2F_1\Big(\frac{1}{2}[1+|\ell |+\nu+i\omega],\frac{1}{2}[1+|\ell |+\nu-i\omega],1+|\ell |;\sin^2\rho  \Big),
    \nonumber \\
    F_2 &={}_2F_1\Big(\frac{1}{2}[1+|\ell |+\nu+i\omega],\frac{1}{2}[1+|\ell |+\nu-i\omega],1+\nu;\cos^2\rho  \Big),
    \nonumber \\
    F_3 &= {}_2F_1\Big(\frac{1}{2}[1+|\ell |-\nu+i\omega],\frac{1}{2}[1+|\ell |-\nu-i\omega],1-\nu;\cos^2\rho  \Big),
    \nonumber \\
     F_4 &= {}_2F_1\Big(\frac{1}{2}[3+|\ell |+\nu+i\omega],\frac{1}{2}[3+|\ell |+\nu-i\omega],2+\nu;\cos^2\rho  \Big),
     \nonumber \\
      F_5 &= {}_2F_1\Big(\frac{1}{2}[3+|\ell |-\nu+i\omega],\frac{1}{2}[3+|\ell |-\nu-i\omega],2-\nu;\cos^2\rho  \Big),
      \nonumber \\
       F_6 &= {}_2F_1\Big(\frac{1}{2}[3+|\ell |+\nu+i\omega],\frac{1}{2}[3+|\ell |+\nu-i\omega],2+|\ell |;\sin^2\rho  \Big), \nonumber \\
       F_7 &= {}_2F_1 \Big(\frac{1}{2}[5 + |\ell | + \nu + i \omega],
       \frac{1}{2}[5 + |\ell | + \nu - i \omega],3+ |\ell |;\sin^2 \rho \Big), 
       \nonumber \\
       F_8 &= {}_2F_1 \Big(\frac{1}{2}[5 + |\ell | - \nu + i \omega ],
       \frac{1}{2}[5 + |\ell | - \nu - i \omega],3- \nu;\cos^2 \rho \Big), 
       \nonumber \\
       F_9 &= {}_2F_1 \Big(\frac{1}{2}[5 + |\ell | + \nu + i \omega],
       \frac{1}{2}[5 + |\ell | + \nu - i \omega],3+ \nu;\cos^2 \rho \Big).
       \label{eq:vacR_HG_expression}
\end{align}
The quantities appearing in the v.e.v.s of the RSET with Robin boundary conditions applied are then:
\begin{align}
    w = & ~ 
    \frac{F_1 \mathcal{K}}{4 \pi^2} \cos^2\rho\, (\sin \rho)^{2|\ell |}, 
    \\
    w_\tau^{\,\tau } = & ~ 
    -\frac{\cos^2\rho\, (\sin\rho)^{2|\ell |}\mathcal{K}}{64 \pi ^2 L^2}\Bigg\{8 F_1 \mathcal{C} 
    + \frac{2 \mathcal{A}}{1+|\ell |}F_6 \sin^2 (2 \rho) -16 F_1\omega^2\,\cos^2\rho \Bigg \}, 
    \\
    w_\rho^{\,\rho} = & ~
    -\frac{\cos^2\rho\,(\sin \rho)^{2|\ell |}\mathcal{K}}{64\pi^2 L^2(1+|\ell |)} \Bigg \{8F_1 (1+|\ell |)[1+3|\ell |+\nu+3(1+|\ell |+\nu)\cos (2\rho)] 
    -4F_6 \mathcal{C} \cos^2\rho\, |1+|\ell |+\nu + i \omega|^2 
    \nonumber \\ &  \qquad
   -\frac{4F_7\,\mathcal{A}}{(2+|\ell |)}|3+|\ell |+\nu + i \omega|^2  \cos^4\rho \sin ^2 \rho 
   -4F_6 \mathcal{A} \cot\rho \sin (4\rho) 
   \nonumber \\ &  \qquad
   +2F_6\mathcal{A} \sin ^2(2\rho)  \Big(  1+\nu +(1-|\ell |) \cot^2\rho\Big) 
    +8 F_1 \mathcal{C}(1+|\ell |) \Big (
   (1-|\ell |) \cot^2\rho +\nu \Big )
    \Bigg \},  
    \\
    w_\theta^{\, \theta } = & ~
    \frac{\cos^2\rho \,(\sin\rho)^{2(|\ell |-1)}\mathcal{K}}{64 \pi ^2 L^2}\Bigg\{8 F_1 \mathcal{C}
    + \frac{2 \mathcal{A}}{(1+|\ell |)}F_6 \sin^2 (2 \rho) - 16 F_1 \ell ^2\,\cos^2\rho \Bigg \},
\\
    w_{;\tau}^{\,;\tau } = & ~ -\frac{\cos^2\rho\,(\sin\rho) ^{2(1+|\ell |)}}{8 \pi^2 L^2} \Bigg \{4F_1 |\ell | \mathcal{K}  \cot^2 \rho +F_1 (\mathcal{J}-4 \mathcal{K}) + \frac{F_6\mathcal{K}}{(1+|\ell |)}\,|1+|\ell |+\nu + i \omega|^2\cos^2\rho
    \Bigg \},
\\
    w_{;\rho}^{\,;\rho}  = & ~
    \frac{ \cos^2 \rho\,(\sin \rho)^{2(1+|\ell |)}}{8 \pi^2 L^2}\Bigg \{  F_1 \cos^2 \rho \Big[ \mathcal{J}(1+ 4|\ell |)-4\mathcal{K}(1+5|\ell |)\Big] 
    \nonumber \\ & \qquad 
    + \frac{F_6}{(1+|\ell |)}|1+|\ell |+\nu +i\omega|^2 \cos^2\rho \Big [ \mathcal{K}\cos^2\rho(1+4|\ell |) + \sin^2\rho(\mathcal{J}-6 \mathcal{K})\Big ] 
    \nonumber \\
    &  \qquad + 4F_1 \Big [ \mathcal{K} |\ell | \,(2 |\ell |\cos^2\rho-1)\cot^2\rho -\sin^2\rho(\mathcal{J}-2\mathcal{K})\Big ] 
    \nonumber \\
    &  \qquad +\frac{F_7 \mathcal{K}}{2(1+|\ell |)(2+|\ell |)}|1+|\ell |+\nu +i \omega|^2\, |3+|\ell |+\nu|^2 \cos^4\rho \sin^2\rho 
    \nonumber \\
   &  \qquad +\frac{F_1 \sin^2\rho}{2} \Bigg ( 16F_2\mathcal{N}^\zeta_{\omega \ell} \nu^2 \cos \zeta (\cos \rho)^{2\nu}+ 4 F_4 \mathcal{N}^\zeta_{\omega \ell} |1+|\ell |+\nu|^2\cos \zeta (\cos \rho)^{2(1 + \nu)} 
   \nonumber \\ &  ~  
   + \frac{\mathcal{N}^\zeta_{\omega \ell }}{(1+\nu)(2+\nu)}|1+|\ell |+\nu +i \omega|^2 \cos \zeta (\cos \rho )^{2(1+\nu)} \Big[ 4F_4 \nu(2+\nu)
   + F_9 |3+|\ell |+\nu +i \omega|^2 \cos^2\rho\Big ] 
   \nonumber\\
   &  ~ + \frac{\cos^2\rho}{(\nu-1)(\nu-2)}|1+|\ell |-\nu +i \omega|^2(\mathcal{N}^N_{\omega \ell}-\mathcal{N}^\zeta_{\omega \ell}\sin\zeta)\Big[ 4F_5 (\nu-2) 
   -F_8 |3+|\ell |-\nu +i \omega|^2 \cos^2 \rho\Big ] \Bigg) 
    \Bigg \}, \\
    w_{;\theta}^{\, ;\theta } = & ~ \frac{ \cos^2 \rho \,(\sin\rho) ^{2|\ell |}}{8 \pi^2 L^2}\Bigg \{4F_1 |\ell | \mathcal{K}  \cot^2 \rho +F_1 (\mathcal{J}-4 \mathcal{K}) 
    + \frac{F_6\mathcal{K}}{1+|\ell |}\,|1+|\ell |+\nu + i \omega|^2\cos^2\rho  
    \Bigg \},
\\
    \Box w = & ~ \frac{\cos^2\rho\, (\sin \rho)^{2|\ell |}}{16 \pi^2 L^2} \Bigg \{F_1 (\mathcal{J}-4\mathcal{K}) \Big [3+4|\ell | +(1+4|\ell |)\cos (2\rho) \Big] 
    \nonumber \\ & ~\qquad 
    +\frac{2F_6}{(1+|\ell |)}|1+|\ell |+\nu +i \omega|^2 \cos^2\rho \Big (
    \mathcal{K} \Big[ 1+\cos^2\rho(1+4|\ell |)\Big] + \mathcal{J} \sin^2 \rho\Big ) 
    \nonumber \\
    & ~ \qquad + 2F_1 \Big[ 8 \mathcal{K} |\ell |^2 \cos^2\rho \cot^2\rho -(4 \mathcal{K}-3\mathcal{J}) \sin^2\rho \Big] 
    - \frac{5F_6 \mathcal{K}}{2(1+|\ell |)}|1+|\ell |+\nu +i \omega|^2 \sin^2 (2\rho)
    \nonumber \\ & ~ \qquad 
    + \frac{F_7 \mathcal{K}}{(1+|\ell |)(2+|\ell |)}|3+|\ell |+\nu +i \omega|^2  |1+|\ell |+\nu +i \omega|^2 \cos^4\rho \sin^2\rho 
    \nonumber \\
    & ~ \qquad + F_1 \sin^2\rho \Bigg ( 16F_2\mathcal{N}^\zeta_{\omega \ell} \nu^2 \cos \zeta (\cos \rho)^{2\nu}+ 4 F_4 \mathcal{N}^\zeta_{\omega \ell}|1+|\ell |+\nu|^2\cos \zeta (\cos \rho)^{2(1 + \nu)} 
    \nonumber \\
& ~ + \frac{\mathcal{N}^\zeta_{\omega \ell}}{(1+\nu)(2+\nu)}|1+|\ell |+\nu +i \omega|^2 \cos \zeta (\cos \rho) ^{2(1+\nu)} \Big[ 4F_4 \nu(2+\nu) + F_9 |3+|\ell |+\nu +i \omega|^2 \cos^2\rho\Big ] 
\nonumber\\
   & ~ + \frac{\cos^2\rho}{(\nu-1)(\nu-2)}|1+|\ell |-\nu +i \omega|^2(\mathcal{N}^N_{\omega \ell}-\mathcal{N}^\zeta_{\omega \ell}\sin\zeta)\Big[ 4F_5 (\nu-2) -F_8 |3+|\ell |-\nu +i \omega|^2 \cos^2 \rho\Big ] \Bigg)
    \Bigg \}. 
\end{align}

\section{Thermal expectation values with Robin boundary conditions}
\label{sec:terms_Robin_thermal}

In this appendix we explicitly present the terms arising in the RSET (\ref{eq:diff_operator3}) for thermal states with Robin boundary conditions applied.
Quantities appearing in the RSET (\ref{eq:diff_operator3}) are derived from the biscalar $W_{\beta }^{\zeta }(x,x')$ (\ref{eq:Euclid_Thermal_Green}), which involves sums over $n$ and $\ell $.
As in appendix~\ref{sec:terms_Robin_vac}, we give the quantities in the RSET which are to be summed over,  using mixed space-time indices. 

We require the auxiliary quantities (\ref{eq:vac_auxiliary}), hypergeometric functions (\ref{eq:vacR_HG_expression}) and normalization constant (\ref{eq:Nconstantzeta}) with $\omega $ replaced by $n\kappa $, as well as the additional quantities
\begin{align}
    \mathcal{R} = & ~ -4 \nu F_2 \mathcal{N}^\zeta_{\omega \ell}  \cos\zeta (\cos \rho)^{2\nu} -\frac{F_4 \mathcal{N}^\zeta_{\omega \ell}|(1+|\ell |+\nu)\beta+2 i n \pi|^2 \cos\zeta (\cos\rho)^{2(1+\nu)}}{(1+\nu)\beta^2} \nonumber \\
    & ~\qquad -\frac{F_5|(1+|\ell |-\nu)\beta +2 i n \pi|^2\cos^2\rho (\mathcal{N}^N_{\omega \ell}-\mathcal{N}^\zeta_{\omega \ell}\sin\zeta)}{(\nu-1)\beta^2},
    \nonumber \\
    \mathcal{D} = & ~ (1+|\ell |+\nu)^2\beta^2 + 4 \pi^2 n^2.
\end{align}
The quantities appearing in the t.e.v.s of the RSET with Robin boundary conditions applied are then:
\begin{align}
    w = & ~ \frac{F_1 \mathcal{K} \cos^2\rho\,(\sin \rho)^{2|\ell |}}{2 \pi \beta}, 
    \\
     w_{\tau }^{\, \tau} = & ~   -\frac{ \cos^2\rho\,(\sin\rho)^{2|\ell |}\mathcal{K}}{32 \pi L^2 \beta^3} \Bigg \{ -64 F_1 \pi^2 n^2 \cos^2\rho 
    +   8\beta^2 \mathcal{C} F_1 
     + \frac{2 \mathcal{D} F_6 \sin^2 (2\rho)}{(1+|\ell |)}
        \Bigg \},
        \\
    w_{\rho }^{\, \rho } = & ~ 
    -\frac{ \cos^2\rho (\sin\rho)^{2|\ell |} \mathcal{K}}{32(1+|\ell |)\pi L^2\beta^3} \Bigg \{
    32F_1(1+|\ell |)(1+|\ell |+\nu)\beta^2 \cos^2\rho - 4F_6 \mathcal{D} \cot\rho \sin (4\rho) 
     + 2F_6 \mathcal{D} \sin^2 (2\rho) 
    \nonumber \\
    & ~ \qquad + 8F_1\,\mathcal{C}(1+|\ell |)\beta^2 
    -4F_6\, \mathcal{C}\beta^2 \left | 1+|\ell |+\nu +\frac{2 i \pi n}{\beta} \right |^2 \cos^2\rho\,  
    \nonumber \\ & ~ \qquad
    - \frac{4 F_7 \mathcal{D}\cos^4 \rho \sin^2 \rho}{(2+|\ell |)\beta^2} \left | 
  (3+|\ell |+\nu)\beta +2 i \pi n\right |^2 
+\nu \left [ 8F_1(1+|\ell |)\beta^2 \mathcal{C} +2F_6 \mathcal{D} \sin^2 (2\rho) \right ] 
\nonumber \\ & ~ \qquad
 + (1-|\ell |)\cot^2\rho \Big[ 8F_1 (1+|\ell |)\beta^2 \mathcal{C}  + 2F_6 \mathcal{D} \sin^2 (2\rho)\Big ]  \Bigg \},
 \\
    w_{\theta }^{\,\theta } = & ~ 
    -\frac{\cot^2\rho\, (\sin\rho)^{2|\ell |} \mathcal{K}}{32 \pi L^2 \beta^3} \Bigg \{ 16 F_1 |\ell |^2 \beta^2 \cos^2\rho \, -8\beta^2 \mathcal{C} F_1 
     - \frac{2 \mathcal{D} F_6 \sin^2 (2\rho)}{(1+|\ell |)}  
    \Bigg \}, 
    \\
    w_{;\tau }^{\,;\tau} = & ~
    - \frac{\cos^2\rho\,(\sin \rho)^{2(1+ |\ell |)}}{ 4 \pi L^2\beta} \Bigg\{ 4 |\ell | F_1 \mathcal{K} \cot^2\rho   
    + F_1 ( \mathcal{R}  - 4 \mathcal{K}) 
       + \frac{F_6\, \mathcal{K}\, \cos^2\rho}{(1+|\ell |)\beta^2}  |(1+|\ell |+\nu)\beta +2i n \pi|^2  
    \Bigg \},
\\
    w_{;\rho}^{\,;\rho } = & ~ 
    \frac{ \cos^2 \rho \,(\sin \rho) ^{2 |\ell |}}{8 \pi L^2 \beta} \Bigg \{F_1\cos^2\rho  \left[ -8 \mathcal{K}(1+5|\ell |) + 2 \mathcal{R} (1+4|\ell |)\right] 
    \nonumber \\ & ~ \qquad
    + \frac{2\cos ^4 \rho}{(1+|\ell |)\beta^2} \left | (1+|\ell |+\nu)\beta + 2 i n \pi \right |^2 F_6 \mathcal{K} \left ( 1+ 4 |\ell |\right)  
     +8 F_1 \mathcal{K} |\ell | ( 2|\ell | \cos^2\rho-1) \cot^2 \rho 
     \nonumber \\ &  ~ \qquad
     + 8 F_1 (2 \mathcal{K}-\mathcal{R})\sin^2\rho 
    + \frac{F_6  \sin^2(2\rho)}{2(1+|\ell |)\beta^2}\left |  (1+|\ell |+\nu)\beta + 2 i n \pi\right |^2 (\mathcal{R}- 6 \mathcal{K}) 
    \nonumber \\
     & ~ \qquad + \frac{F_7\, \mathcal{K}\,\cos^2\rho \sin^2(2\rho)}{4(1+|\ell |)(2+|\ell |)\beta^4}\left |  (1+|\ell |+\nu)\beta + 2 i n \pi \right |^2 \left |  (3+|\ell |+\nu)\beta + 2 i n \pi \right |^2 \nonumber \\
    & ~ \qquad  + F_1 \sin ^2\rho\Bigg (16 \nu^2 F_2 \mathcal{N}^\zeta_{\omega \ell} \cos \zeta (\cos \rho)^{2\nu} 
    \nonumber \\ & ~ \qquad \quad 
    + \frac{4F_4 \mathcal{N}^\zeta_{\omega \ell}\,\cos \zeta (\cos \rho) ^{2(1+\nu)}}{(1+\nu)\beta^2} \left | (1+|\ell |+\nu)\beta + 2 i n \pi \right |^2(1+2\nu) 
    \nonumber \\
    & ~ \qquad \quad + \frac{F_9 \mathcal{N}^\zeta_{\omega \ell}\cos \zeta (\cos \rho)^{2(2+\nu)}}{(1+\nu)(2+\nu)\beta^4}\left | (1+|\ell |+\nu)\beta + 2 i n \pi \right |^2\left | (3+|\ell |+\nu)\beta + 2 i n \pi\right |^2 
    \nonumber \\
     &  ~ \qquad \quad + \frac{4F_5\,\cos^2\rho}{(\nu-1)\beta^2}\left |  (1+|\ell |-\nu)\beta + 2 i n \pi \right |^2   \left (\mathcal{N}^N_{\omega \ell} - \mathcal{N}^\zeta_{\omega \ell} \sin\zeta\right ) \nonumber \\
     & ~ \qquad \quad - \frac{F_8\,\cos^4 \rho}{(\nu-2)(\nu-1)\beta^4}\left | (1+|\ell |-\nu)\beta + 2 i n \pi \right |^2 \left |  (3+|\ell |-\nu)\beta + 2 i n \pi \right |^2  \left ( \mathcal{N}^N_{\omega \ell}- \mathcal{N}^\zeta_{\omega \ell} \sin\zeta\right )
     \Bigg )
     \Bigg \},
     \\
    w_{;\theta }^{\, ;\theta } = & ~ 
    \frac{ \cos^2 \rho \,(\sin\rho)^{2|\ell |}}{4 \pi L^2\beta}\Bigg \{ 4F_1 \mathcal{K} |\ell |  \cot^2 \rho + F_1  (\mathcal{R}- 4 \mathcal{K}) 
    + \frac{F_6\mathcal{K}\cos^2\rho}{(1+|\ell |)\beta^2}\left |  (1+|\ell |+\nu)\beta + 2 i n \pi \right |^2   \Bigg \},
\\
    \Box w  = & ~ 
    \frac{\cos^2\rho (\sin\rho)^{2|\ell |}}{8 \pi L^2 \beta} \Bigg \{2F_1(\mathcal{R}-4 \mathcal{K})[1+\cos^2\rho (1+4 |\ell |)] 
    \nonumber \\ & ~ \qquad
    + \frac{2F_6\mathcal{K}\cos^2\rho}{(1+|\ell |)\beta^2} \left|  (1+|\ell |+\nu)\beta + 2 i n \pi \right |^2  \left [ 1+\cos^2\rho(1+4|\ell |)\right ] 
    \nonumber \\
   & ~ \qquad + 2F_1 \left [8 |\ell |^2 \mathcal{K} \cos^2\rho \cot^2\rho +(4\mathcal{K}-3\mathcal{R})\sin^2\rho \right ] 
   \nonumber \\ & ~ \qquad
   +  \frac{F_6 \sin^2 (2\rho)}{(1+|\ell |)\beta^2}
   \left|  (1+|\ell |+\nu)\beta + 2 i n \pi \right |^2(\mathcal{R}-10\mathcal{K}) 
   +\frac{F_6 \mathcal{R}\sin^2 (2\rho) }{(1+|\ell |)}\left|  (1+|\ell |+\nu)\beta + 2 i n \pi \right |^2 
   \nonumber \\ & ~ \qquad
   + \frac{F_7\mathcal{K}\cos^2 \rho \,\sin^2 (2\rho )}{(1+|\ell |)(2+|\ell |)\beta^4}\left| (1+|\ell |+\nu)\beta + 2 i n \pi \right |^2\left|  (3+|\ell |+\nu)\beta + 2 i n \pi \right |^2 
   \nonumber \\
    & ~ \qquad  + F_1 \sin ^2\rho\Bigg (16 \nu^2 F_2 \mathcal{N}^\zeta_{\omega \ell} \cos \zeta (\cos \rho)^{2\nu} 
    \nonumber \\ & ~ \qquad
\quad    + \frac{4F_4 \mathcal{N}^\zeta_{\omega \ell} \,\cos \zeta (\cos \rho )^{2(1+\nu)}}{(1+\nu)\beta^2} \left |  (1+|\ell |+\nu)\beta + 2 i n \pi \right |^2(1+2\nu) 
    \nonumber \\
    & ~ \qquad \quad + \frac{F_9 \mathcal{N}^\zeta_{\omega \ell}\cos \zeta (\cos \rho)^{2(2+\nu)}}{(1+\nu)(2+\nu)\beta^4}\left |  (1+|\ell |+\nu)\beta + 2 i n \pi \right |^2\left |  (3+|\ell |+\nu)\beta + 2 i n \pi \right |^2 \nonumber \\
     & ~ \qquad \quad + \frac{4F_5\,\cos^2\rho}{(\nu-1)\beta^2}\left |  (1+|\ell |-\nu)\beta + 2 i n \pi \right |^2   \left ( \mathcal{N}^N_{\omega \ell}- \mathcal{N}^\zeta_{\omega \ell} \sin\zeta\right ) \nonumber \\
     & ~ \qquad \quad - \frac{F_8\, \cos^4 \rho}{(\nu-2)(\nu-1)\beta^4}\left |  (1+|\ell |-\nu)\beta + 2 i n \pi \right |^2 \left |  (3+|\ell |-\nu)\beta + 2 i n \pi \right |^2 \left (\mathcal{N}^N_{\omega \ell} - \mathcal{N}^\zeta_{\omega \ell} \sin\zeta\right )
     \Bigg )
  \Bigg \}.
\end{align}

\end{document}